\documentclass[twocolumn,twocolappendix]{aastex63}
\usepackage{amsmath}
\usepackage{amssymb}
\usepackage{bm}
\usepackage{wasysym}
\usepackage{hyperref}
\hypersetup{colorlinks,breaklinks,allcolors=blue}
\usepackage[super]{nth}
\usepackage{wrapfig}
\usepackage{graphicx}
\usepackage{grffile}
\usepackage{longtable}
\usepackage{multirow}
\usepackage{xspace}

\newcommand{\swift}{{\sc SWIFT}\xspace}
\newcommand{\woma}{{\sc WoMa}\xspace}
\newcommand{\seagen}{{\sc SEAGen}\xspace}

\newcommand{\Sa}{{\scriptsize\saturn{}}}

\defcitealias{Cuk+2016b}{{\'C}16}
\newcommand{\tCuk}{\citetalias{Cuk+2016b}\xspace}
\newcommand{\pCuk}{\citepalias{Cuk+2016b}\xspace}

\defcitealias{Hyodo+Charnoz2017}{HC17}

\shorttitle{A recent impact origin of Saturn's rings}
\shortauthors{Teodoro \& Kegerreis et al.}

\begin{document}

\title{\Large A recent impact origin of Saturn's rings and mid-sized moons}

\author[0000-0002-8346-0138]{L. F. A. Teodoro}
\affiliation{School of Physics and Astronomy, University of Glasgow, G12 8QQ, Scotland, UK}
\affiliation{Centre for Space Sensors and Systems, Faculty of Mathematics and Natural Sciences, University of Oslo, 2007 Kjeller, Norway}

\correspondingauthor{Jacob Kegerreis}
\email{jacob.kegerreis@durham.ac.uk}
\author[0000-0001-5383-236X]{J. A. Kegerreis}
\affiliation{NASA Ames Research Center, Moffett Field, CA 94035, USA}

\author[0000-0002-1132-5594]{P. R. Estrada}
\affiliation{NASA Ames Research Center, Moffett Field, CA 94035, USA}
\author[0000-0003-1226-7960]{M. {\'C}uk}
\affiliation{Carl Sagan Center, SETI Institute, Mountain View, CA 94043, USA}
\author[0000-0001-5416-8675]{V. R. Eke}
\affiliation{Physics Department, Institute for Computational Cosmology, Durham University, Durham, DH1 3LE, UK}
\author[0000-0003-0553-1436]{J. N. Cuzzi}
\affiliation{NASA Ames Research Center, Moffett Field, CA 94035, USA}
\author[0000-0002-6085-3780]{R. J. Massey}
\affiliation{Physics Department, Institute for Computational Cosmology, Durham University, Durham, DH1 3LE, UK}
\author[0000-0002-4630-1840]{T. D. Sandnes}
\affiliation{Physics Department, Institute for Computational Cosmology, Durham University, Durham, DH1 3LE, UK}



\begin{abstract}

We simulate the collision of precursor icy moons analogous to Dione and Rhea
as a possible origin for Saturn's remarkably young rings.
Such an event could have been triggered a few hundred million years ago
by resonant instabilities in a previous satellite system.
Using high-resolution smoothed particle hydrodynamics simulations,
we find that this kind of impact can produce a wide distribution of massive objects
and scatter material throughout the system.
This includes the direct placement of pure-ice ejecta
onto orbits that enter Saturn's Roche limit,
which could form or rejuvenate rings.
In addition, fragments and debris of rock and ice
totalling more than the mass of Enceladus
can be placed onto highly eccentric orbits that would intersect
with any precursor moons orbiting in the vicinity of Mimas, Enceladus, or Tethys.
This could prompt further disruption and facilitate
a collisional cascade to distribute more debris
for potential ring formation,
the re-formation of the present-day moons,
and evolution into an eventual cratering population of planeto-centric impactors.

\end{abstract}



\section{Introduction} \label{sec:introduction}

Whatever their origin, Saturn's rings appear to be young.
Key observations made during the Cassini mission
provided new measurements of the ring mass \citep{Iess+2019},
the fraction of non-icy material in the rings \citep{Zhang+2017a,Zhang+2017b},
and the extrinsic micrometeoroid flux at Saturn \citep{Kempf+2023}.
Together, these three factors constrain the ring age to be less
than a few 100~Myr \citep{Estrada+2018,Kempf+2023,Durisen+Estrada2023}.
This is because the primarily icy rings
\citep[$>$$95$\% by mass,][]{Doyle+1989,Cuzzi+Estrada1998,Zhang+2017a,Zhang+2017b}
are continuously subjected to micrometeoroid bombardment and are darkened over time,
so an upper bound on the rings' exposure age can be determined
by assuming that they began as pure ice \citep{Cuzzi+Estrada1998,Estrada+2015}.
The present mass of the main rings is $\sim$0.4~Mimas masses
($\sim$$1.5 \times 10^{19}$ kg).

Moreover, the rings appear to be losing mass at a remarkable rate
\citep{Waite+2018,Hsu+2018,ODonoghue+2019},
suggesting they are not only young, but ephemeral as well.
\citet{Durisen+Estrada2023} showed that the process of
direct deposition of micrometeoroids and the ballistic transport
of their impact ejecta can account for these mass-loss rates,
and \citet{Estrada+Durisen2023} used these results to demonstrate that
mass loading and ballistic transport due to micrometeoroid bombardment
continue to drive the rings' dynamical evolution once viscosity becomes too weak.
These studies concluded that if the rings have been losing mass
at more or less their current rate,
then they may have begun with a mass of
one to a few times the mass of Mimas a few 100~Myr ago.
They also find that, when ballistic transport is included,
the rings' dynamical age and exposure age \emph{are} similar
-- contrary to the view that these timescales can be vastly different,
which would otherwise allow for the rings to look young
but be dynamically old \citep{Crida+2019}.

The rings' geological youth and apparent transience are at odds
with most current ideas about their origin,
in which they are either primordial \citep{Pollack+1976,Canup2010},
or derived from the remnants of either
a collisionally disrupted, Mimas-mass moon
\citep{Harris1984,Charnoz+2009,Dubinski2019}
or a tidally split comet \citep{Dones1991,Dones+2007,Hyodo+2017c}
-- catastrophes that are highly unlikely in the current solar system
and hence presumed ancient.
In the latter tidal disruption case, a comet or Centaur
a few hundreds of kilometres in size
is required to pass close enough to Saturn to be tidally disrupted,
with the rings forming from some of the debris.
The collisional disruption of a Mimas-mass moon
requires a body tens of kilometres in size.
Furthermore, the precursor moon must be kept close enough to Saturn
long enough for the impact event to occur,
although this could be facilitated by mean-motion resonances
with Enceladus and Dione \citep{Dubinski2019}.
However, the flux of such bodies decreases sharply
after the Late Heavy Bombardment \citep{Zahnle+2003},
giving only a probability of occurrence of $\sim$$10^{-4}$ in the past few 100~Myr
\citep{Ip1988,Lissauer+1988,Dones1991,Dones+2007,Charnoz+2009}.

\citet{Wisdom+2022} recently explored a novel scenario for recent ring formation
in which Saturn once had an additional moon between Titan and Iapetus.
This Iapetus-mass satellite helped to maintain a secular resonance with Neptune
that drove up Saturn's obliquity.
Subsequent destabilisation of the moon's orbit by resonances with Titan
could have led to a grazing encounter with Saturn
and a recent ring origin $\sim$$100$ Myr ago,
in addition to kicking Saturn out of the resonance.
This would imply that Saturn is no longer in the resonance with Neptune,
but confirming this may require more accurate measurements
of Saturn's pole precession rate \citep{Jacobson2022}.
Detailed predictions are also yet to be made
for the fate of the destabilised moon's rock and ice material,
if it were sent on an elliptical orbit
that crosses into the Roche limit \citep{Dones1991},
or for its potentially disruptive effects on the rest of the satellite system
-- perhaps along similar lines to the collisional scenario we consider below.

In addition to the rings,
Saturn's current system of icy moons (interior to Titan)
also presents some evidence of recent formation.
\citet[][hereafter \citetalias{Cuk+2016b}]{Cuk+2016b}
demonstrated that if the present-day mid-sized satellite system were ancient,
then recent tidal evolution through dynamical resonances
would have excited the moons' orbits beyond what is seen today.
In particular, the ``smoking gun'' for the moons not being ancient is
Rhea's present inclination, which is an order of magnitude lower today
than if it had survived crossing an evection resonance%
\footnote{
  The evection resonance occurs when a satellite's orbital precession period
  becomes commensurate with the planet's orbital period around the Sun.
}
at around 8.3~$R_{\Sa}$,
where $R_{\Sa} = 60,268$~km is the equatorial radius of Saturn
\citep[at 1 bar,][]{Lindal+1985},
and such inclinations are not readily damped \citep{Murray+Dermott1999,Chen+2014}.
Rhea today is located outside evection, at 8.7~$R_{\Sa}$,
with an inclination of only $0.35^\circ$,
so it cannot have formed long ago and migrated outwards.
A very slow tidal evolution rate might also
have been able to explain the present-day moons' un-excited orbits,
but this would conflict with the apparent
fast orbital evolution of Rhea \citep{Lainey+2020,Jacobson2022}
and the intense tidal heating of Enceladus \citep{Meyer+Wisdom2007}.
The alternative is that Rhea formed (or re-formed) outside the resonance \pCuk.
This implies a similarly recent, and thus perhaps shared, formation
to the rings for at least some of these icy satellites.

More recent work has shown that at least some moons of Saturn evolve much faster
than predicted by the equilibrium tides assumed by \tCuk
\citep{Lainey+2020,Jacobson2022},
possibly indicating resonance locking of the moons
with gravity modes or inertial waves within Saturn \citep{Fuller+2016}.
The apparent presence of resonant tides
weakens the constraints on the age of the system from
the lack of a past Tethys--Dione 3:2 mean-motion resonance found by \tCuk,
as it is no longer clear that these moons are currently on converging orbits.
However, the observed fast migration of Rhea,
if driven by resonance locking \citep{Lainey+2020},
still implies that an ancient Rhea
would have crossed and had its orbit excited by
the solar evection resonance only a few hundred Myr ago.

Here we examine a new scenario based on the work of \tCuk,
which hypothesizes a recent dynamical instability
in a precursor satellite system $\sim$$100$~Myr ago.
Any pair of moons with orbits and masses
loosely comparable with present-day Dione and Rhea
will be suddenly excited when the Rhea-analogue outer moon encounters
the evection resonance during migration.
Both moons have their orbits excited
to significant eccentricities ($\sim$$0.1$--$0.2$)
and inclinations of several degrees.
This paired excitation can naturally arise
from the coupled migration of moons in mean-motion resonance;
from the inner moon catching up owing to the stronger tidal effects nearer Saturn;
and from an initially separate outer moon
being trapped just interior to evection for a long time,
by a cycle of temporary inwards migration caused by the damping
of evection-driven eccentricity \pCuk.
Regardless of the precise evolution to this point,
the common outcome of large eccentricities for both moons
leads to orbit crossing and high-velocity collisions.
Fragments and debris ejected by this impact
could then re-accrete into the present-day moons,
significantly erode other precursor moons leading to a collisional cascade,
and perhaps even deliver material directly inside Saturn's Roche limit
to form or rejuvenate rings.

This scenario does not rule out previous hypotheses
for ancient satellite formation around Saturn,
such as by accretion from primordial rings \citep{Crida+Charnoz2012,Crida+Charnoz2014}.
Such mechanisms would now simply apply
to the formation of the precursor satellite system
that was then recently destabilised and disrupted
to (re-)form the moons we see today.

Some aspects of the \tCuk scenario were investigated using
smoothed particle hydrodynamics (SPH) and $N$-body simulations
of two colliding Rhea-mass moons \citep{Hyodo+Charnoz2017},
with results that suggested that the debris
would re-accrete into a satellite without forming rings.
However, this conclusion could be premature for a few reasons, including:
only a single impact angle of $45^\circ$ was considered
as an example input for the longer-term models;
the SPH simulations used a low resolution of $2 \times 10^5$ particles;
the subsequent evolution assumed a simplified hard-sphere
and dissipative coefficient-of-restitution model
for the re-accretion of fragments,
-- without the possibility of further fragmentation;
and the presence of other precursor moons in the system
and potential cascade interactions with them were neglected.

Regarding the last issue of other satellites, the \tCuk scenario is based
on a multiple-moon primordial system comparable to the current one.
Such a precursor satellite system would likely have a broadly similar mass
and architecture to the present system
-- perhaps with a somewhat larger total mass.
The current systems of both Saturn and Uranus
(and probably at one time, Neptune; \citealt{Goldreich+1989})
have families of mid-sized moons that extend inward close to their parent planets,
so this expectation appears well justified.
Furthermore, the initial distribution of moons cannot be random,
since their radial locations must satisfy
orbital stability criteria \citep[e.g.][]{Wisdom1980,Lissauer1995}
that restrict the regions in which moons can orbit stably
\citep[e.g. Fig.~1b,][]{Mosqueira+Estrada2003b}.
We also can expect size sorting (increasing radius with distance),
as it is a natural outcome of models that form mid-sized moons
in a gaseous circumplanetary disk \citep{Mosqueira+Estrada2003}
or if the moons were ``spun out'' from viscously evolving,
massive, primordial rings after planet formation
\citep{Crida+Charnoz2012,Crida+Charnoz2014},
while a collisional origin might allow some variation
in size and composition \citep{Asphaug+Reufer2013}.
We therefore take the current Saturn system as a guide
for the context and implications of impacts in a precursor one.

Here we use SPH simulations with over two orders of magnitude
higher resolution than \citet{Hyodo+Charnoz2017}
to model collisions based on \tCuk integrations.
We find that such an impact can send significant debris throughout the system,
including interior to Saturn's Roche limit,
and can produce an extended distribution of massive fragments.
These objects are likely to continue to fragment or erode
and circularise with one another on future orbits,
and they also have semi-major axes and eccentricities
that will lead to crossing orbits, close encounters,
and possible collisions with other precursor moons.
This opens the possibility of a collisional cascade
that leads to additional rubble piles and debris crossing within the Roche limit,
which may be tidally disrupted or collisionally captured
(if some material is already present) into rings.

\section{Methods} \label{sec:methods}

\subsection{Smoothed particle hydrodynamics simulations} \label{sec:methods:sph_sims}

Smoothed particle hydrodynamics \citep[SPH,][]{Lucy1977,Gingold+Monaghan1977}
is a Lagrangian method widely used
to model planetary impacts and diverse astrophysical and other systems.
Materials are represented by many interpolation points or `particles'
that are evolved under hydrodynamical forces,
with pressures computed via an equation of state (EoS), and gravity.
SPH is particularly well suited for modelling asymmetric geometries
and/or large dynamic ranges in density and distribution of material,
such as the messy impacts considered here.
However, standard resolutions in planetary simulations,
using $10^{5}$--$10^{6}$ SPH particles,
can fail to converge on even large-scale outcomes of giant impacts
\citep{Genda+2015,Hosono+2017,Kegerreis+2019,Kegerreis+2022}.

Here we run simulations with up to $\sim$$10^{7.5}$ SPH particles,
to resolve the detailed outcomes of collisions
and to test for numerical convergence,
using the open-source hydrodynamics and gravity code \swift%
\footnote{
  \swift \citep{Schaller+2018,Schaller+2023}
  is publicly available at \url{www.swiftsim.com}.
}
\citep{Kegerreis+2019}.
For this study, we use a `vanilla' form of SPH
plus the \citet{Balsara1995} switch for the artificial viscosity.
Future work will include more sophisticated modifications
recently added to \swift that mitigate issues that can arise
from density discontinuities in SPH \citep{RuizBonilla+2022,Sandnes+2023}.
As in previous works, we also neglect material strength,
as gravitational stresses are expected to dominate \citep{Asphaug+Reufer2013},
though as smaller bodies are considered
this assumption will be valuable to test in future work.
Nevertheless, we tentatively expect these standard simplifications
to be comparatively innocuous for the high-speed collisions
(relative to the mutual escape speed) considered here
\citep{Asphaug+Reufer2013,Kegerreis+2020}.

\subsection{Initial conditions} \label{sec:methods:init_cond}

To distinguish the Dione-/Rhea-like moons we collide from
their present-day analogues,
we refer to the precursor satellites as `p-Dione' and `p-Rhea'.
While a precursor system may have been larger in mass,
we start by considering moons like those today,
with masses of $1.10$ and $2.31 \times 10^{21}$~kg, respectively.
Their detailed internal state cannot readily be predicted, so,
for simplicity and in line with previous works,
we model them both as differentiated into silicate cores and ice mantles,
at an isothermal 100~K,
with core mass fractions of 60\% and 40\%, respectively.
These materials are modelled with
\citet{Tillotson1962} basalt and water EoS \citep{Melosh2007},
or with ANEOS forsterite and AQUA water \citep{Stewart+2020,Haldemann+2020}
for a small subset of comparison simulations.

We generate the moons' internal profiles by integrating inwards
while maintaining hydrostatic equilibrium%
\footnote{
  The \woma code \citep{RuizBonilla+2021}
  for producing spherical and spinning planetary profiles and
  initial conditions is publicly available with documentation and examples at
  \href{https://github.com/srbonilla/WoMa}{github.com/srbonilla/WoMa},
  and the python module \texttt{woma} can be installed directly with
  \href{https://pypi.org/project/woma/}{pip}.
},
then place the roughly equal-mass SPH particles
to match the resulting density profiles
using the stretched equal-area method%
\footnote{
  \seagen \citep{Kegerreis+2019} is publicly available at
  \href{https://github.com/jkeger/seagen}{github.com/jkeger/seagen},
  or as part of \woma.
}.
Before simulating the impact, we first run a brief 2~h settling simulation
for each body in isolation, to allow any final settling to occur.
The specific entropies of the particles are kept fixed,
enforcing that the particles relax themselves adiabatically.
The impact simulations are then evolved from 1~h before contact to 9~h after.
As this is only a modest fraction of the orbital period around Saturn ($\sim$82~h),
we neglect the effects of Saturn's gravity during the SPH simulations,
as discussed in \S\ref{sec:methods:frame_trans}.

\subsection{Impact scenarios} \label{sec:methods:impact_scenarios}

As detailed in \tCuk,
a pair of precursor moons can be suddenly excited
by p-Rhea encountering the evection resonance,
leading to high eccentricities, inclinations, and crossing orbits.
We take two specific example outcomes of \tCuk models,
which result in collisions with relative velocities of 2 and 3~km~s$^{-1}$.
The orbital parameters are listed in \S\ref{sec:methods:frame_trans}.
For comparison, the mutual escape speed is $v_{\rm esc} \approx 580$~m~s$^{-1}$,
so the impacts are at $\sim$$3.4$ and $5.1$~$v_{\rm esc}$.
These two orbit scenarios set the velocity at infinity, $v_\infty$,
for the impact initial conditions.
Given the moons' small masses and high relative velocities,
little additional speed ($\sim$20~m~s$^{-1}$)
is gained from self-gravity as the moons approach.

\begin{figure}[t]
  \centering
  \includegraphics[
  width=\columnwidth, trim={40mm 8mm 47mm 6mm}, clip]{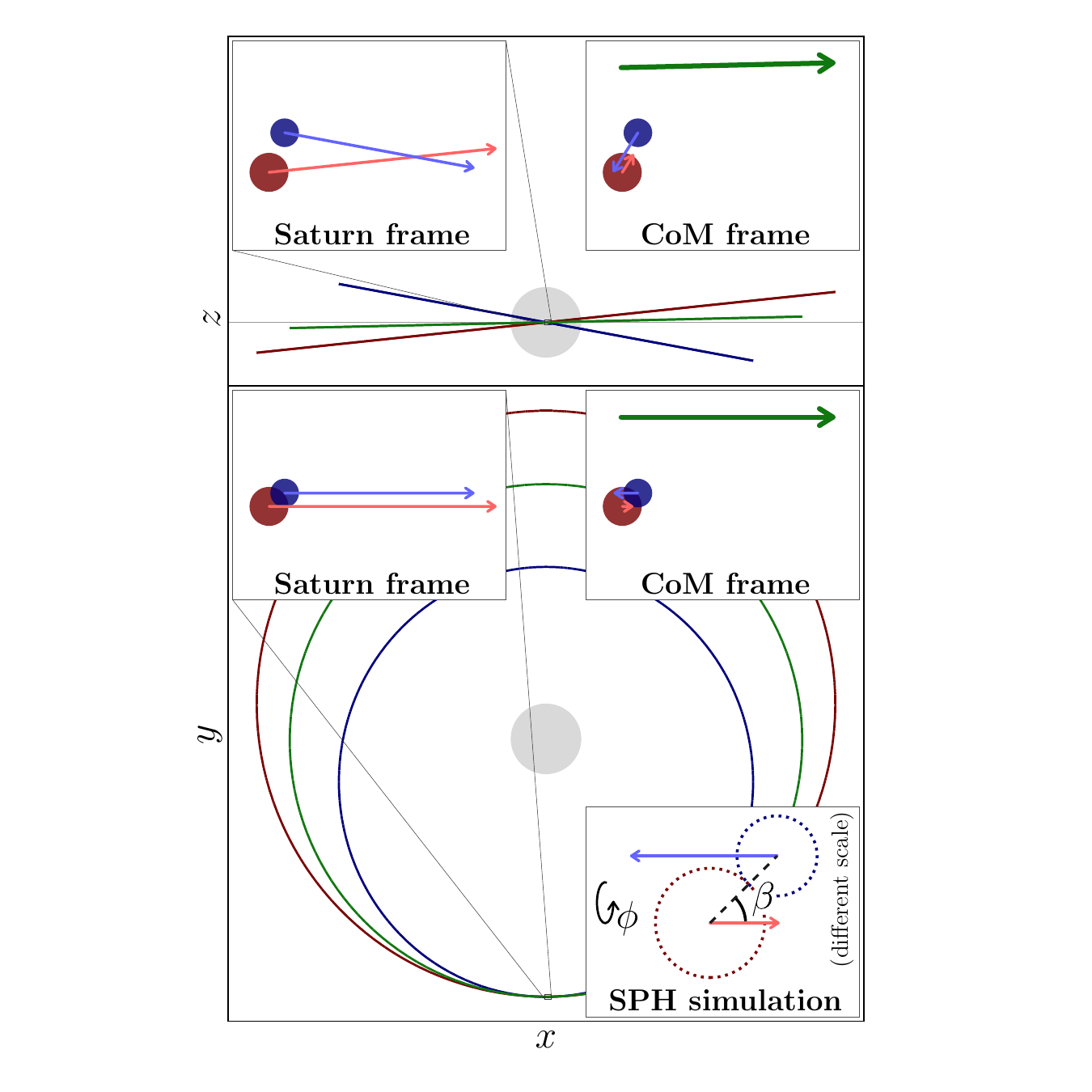}
  \\\vspace{-0.6em}
  \caption{
  The pre-impact orbits of p-Rhea, p-Dione, and their centre of mass (CoM)
  around Saturn in dark red, blue, and green, respectively,
  in $x$--$y$ and $x$--$z$ projections
  (where $z=0$ is Saturn's equatorial plane),
  for the 3~km~s$^{-1}$ scenario
  detailed in \S\ref{sec:methods:frame_trans}.
  The grey circle shows Saturn, to scale.
  The inset zoom-in panels show, as labelled,
  the moons and their velocities shortly before impact in Saturn's frame
  and in the centre of mass frame.
  The bottom-right panel shows the orientation for the SPH simulation,
  with the velocity at contact in the $x$ direction
  and an impact angle $\beta$ (ignoring any tidal deformation),
  plus a possible extra rotation about the $x$ axis in this frame
  by an angle $\phi$, to change the orientation of the collision
  once placed in Saturn's frame.
  The initial separation for the simulation is set
  such that the time to impact is 1~hour.
  \label{fig:frames_diagram}}
\end{figure}

The moons' orbits are rapidly and chaotically varying at this time,
and their radii are a negligible fraction of their orbital semi-major axes,
so we freely set a variety of impact angles
to examine the range of possible outcomes at these speeds.
For each of the two speed scenarios,
we simulate collisions at eight impact angles (see Fig.~\ref{fig:frames_diagram})
from nearly head-on to highly grazing,
with $\beta = 5^\circ$--$75^\circ$ in steps of $10^\circ$.
This angle positions the point of contact along a line
from the centre to the edge of the target moon.
The impact scenario can also be rotated
by an angle $\phi$ (default $0^\circ$)
about the $x$ axis in the SPH frame
to position the point of contact anywhere on the target.
While a $\phi \neq 0$ simulation is identical in the SPH frame,
the orientation in Saturn's frame
and thus the resulting distribution of orbiting debris are affected.
Therefore, we repeat the analysis of the debris with
$\phi = 90^\circ, 180^\circ, 270^\circ$ for the primary suite of simulations.

For each angle and speed scenario,
the SPH particle mass, which controls the resolution,
is set such that the number of particles per
$M_{\rm Rh} + M_{\rm Di} = 3.4 \times 10^{21}$~kg
is $10^{5}$ up to $10^{7.5}$ in steps of 0.5 dex.
In addition to the primary suite of p-Dione--p-Rhea impacts,
we repeat these same simulations for p-Rhea--p-Rhea
as considered by \citet{Hyodo+Charnoz2017},
although they used non-equal ratios between rock and ice.

For a cleaner side test of how the results might scale
with the precursor moons' masses,
we also run a subset of impacts with p-Dione masses of
$1$, $2$, $4$, and $8 \times 10^{21}$~kg
and corresponding p-Rhea masses of double these values.
These moons are given a core mass fraction of $\tfrac{1}{3}$.
We simulate these collisions for $\beta = 15^\circ$, $25^\circ$, $35^\circ$
for the four mass pairs.

Finally, to further examine the numerical effects of our finite-particle models,
each scenario at our second-highest resolution of $10^{7}$ particles
is repeated four more times using a rotated orientation
for the settled p-Rhea and p-Dione.
Particularly for low-speed collisions,
this can yield non-negligible effects
even at the high resolutions used here \citep{Kegerreis+2020,Kegerreis+2022}
-- an uncertainty that these tests will help to constrain.

\subsection{Transforming SPH outputs into Saturn's frame} \label{sec:methods:frame_trans}

The SPH simulations are run in the centre-of-mass (CoM) frame
of the colliding moons,
with the relative velocity at contact in the simulation-box $x$ direction,
as illustrated in Fig.~\ref{fig:frames_diagram}.
The simulation results must therefore be rotated and translated into Saturn's frame
to provide physically realistic pre-impact orbit conditions.
The orbits of the resulting debris can then be computed,
as detailed in Appx.~\ref{appx:orbits}.

The pre-impact orbital elements of p-Rhea and p-Dione are,
for the 2~km~s$^{-1}$ collision:
\begin{align}
  a_{\rm pR} &= 8.3~R_{\Sa}, \;\; e_{\rm pR} = 0.12, \;\; i_{\rm pR} = 3^\circ, \nonumber\\
  a_{\rm pD} &= 6.09~R_{\Sa}, \;\; e_{\rm pD} = 0.2, \;\; i_{\rm pD} = 5.25^\circ, \nonumber
\end{align}
where $a$, $e$, and $i$ are the semi-major axis, eccentricity, and inclination, respectively.
The orbits intersect at the periapsis and ascending node of p-Rhea
and the apoapsis and descending node of p-Dione.
The relative velocity at this point of impact is:
\begin{equation}
  \vec{v}_{\rm rel} = [1.544, \, 0, \, 1.273]~{\rm km~s}^{-1}, \nonumber
\end{equation}
and the centre of mass has orbital elements:
\begin{equation}
  a_{\rm com} = 7.356~R_{\Sa}, \;\; e_{\rm com} = 0.0068, \;\; i_{\rm com} = 0.64^\circ. \nonumber
\end{equation}
For the 3~km~s$^{-1}$ collision, the corresponding values are:
\begin{align}
  a_{\rm pR} &= 8.3~R_{\Sa}, \;\; e_{\rm pR} = 0.12, \;\; i_{\rm pR} = 6^\circ, \nonumber\\
  a_{\rm pD} &= 6.09~R_{\Sa}, \;\; e_{\rm pD} = 0.2, \;\; i_{\rm pD} = 10.5^\circ, \nonumber\\
  \vec{v}_{\rm rel} &= [1.608, \, 0, \, 2.538]~{\rm km~s}^{-1}, \nonumber\\
  a_{\rm com} &= 7.246~R_{\Sa}, \;\; e_{\rm com} = 0.0059, \;\; i_{\rm com} = 1.30^\circ. \nonumber
\end{align}
The centre of mass of the particles in the SPH simulation
is at the origin at rest,
and the relative velocity at impact is in the $x$ direction.
The particles' positions and velocities are transformed
into Saturn's rest frame by the following steps:
\begin{enumerate}
  \item Rotate about the $z$ axis such that $\vec{v}_\infty$
    rather than $\vec{v}_{t=0}$ is in the $x$ direction.
    This rotation is by $<$$1^\circ$ and has a negligible effect,
    but we include it for completeness.
  \item Rotate about the $y$ axis such that the direction
    of the relative velocity changes from $x$ to that of $\vec{v}_{\rm rel}$.
    This corresponds to $39.5^\circ$ and $58.6^\circ$
    for the two scenarios.
  \item Compute the position and velocity of the centre of mass
    at the appropriate time from the input orbital elements,
    then add them to the particles' positions and velocities.
\end{enumerate}

\begin{figure*}[t]
  \centering
  \includegraphics[
  width=0.495\textwidth, trim={18mm 8mm 25mm 7mm}, clip]{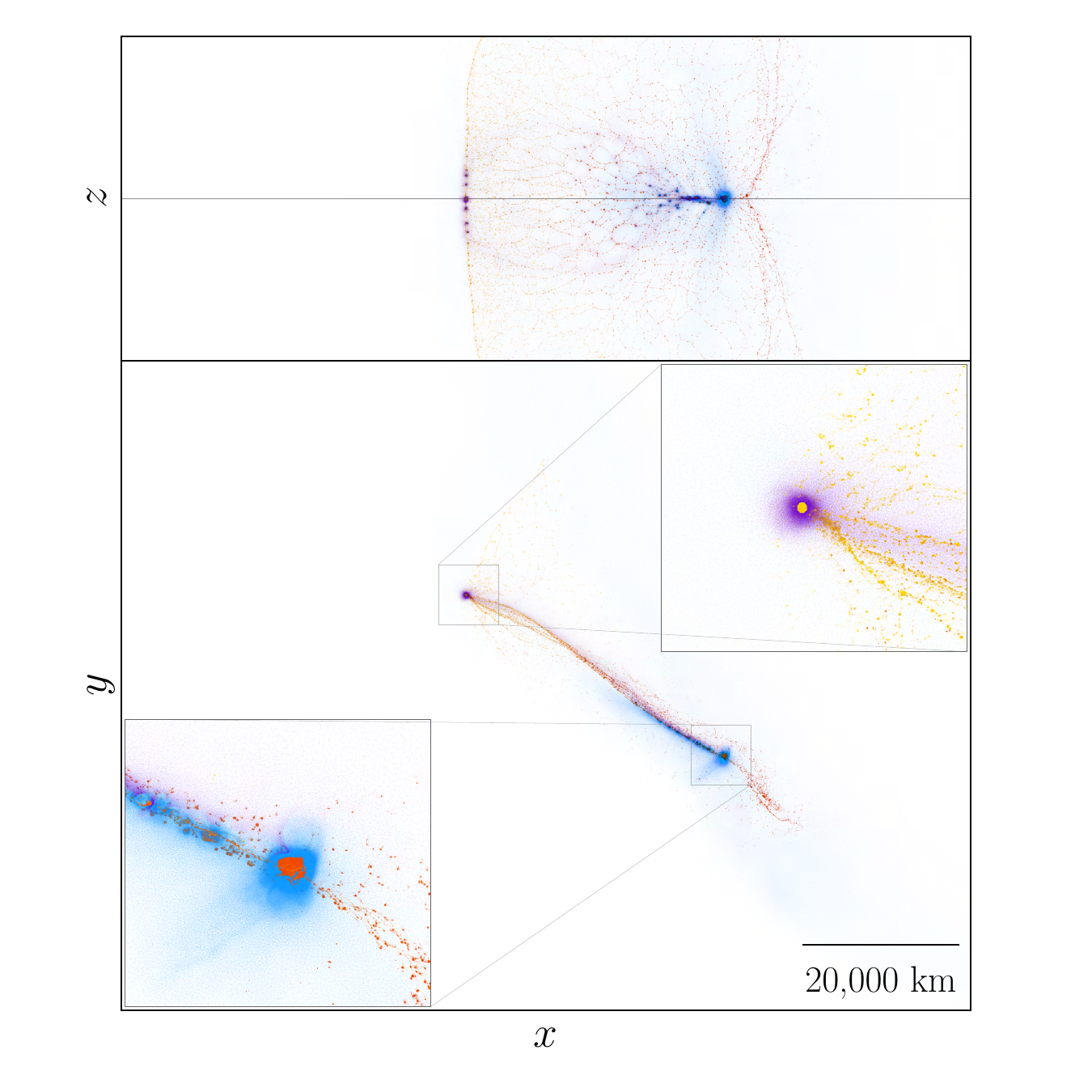}
  \includegraphics[
  width=0.495\textwidth, trim={18mm 8mm 25mm 7mm}, clip]{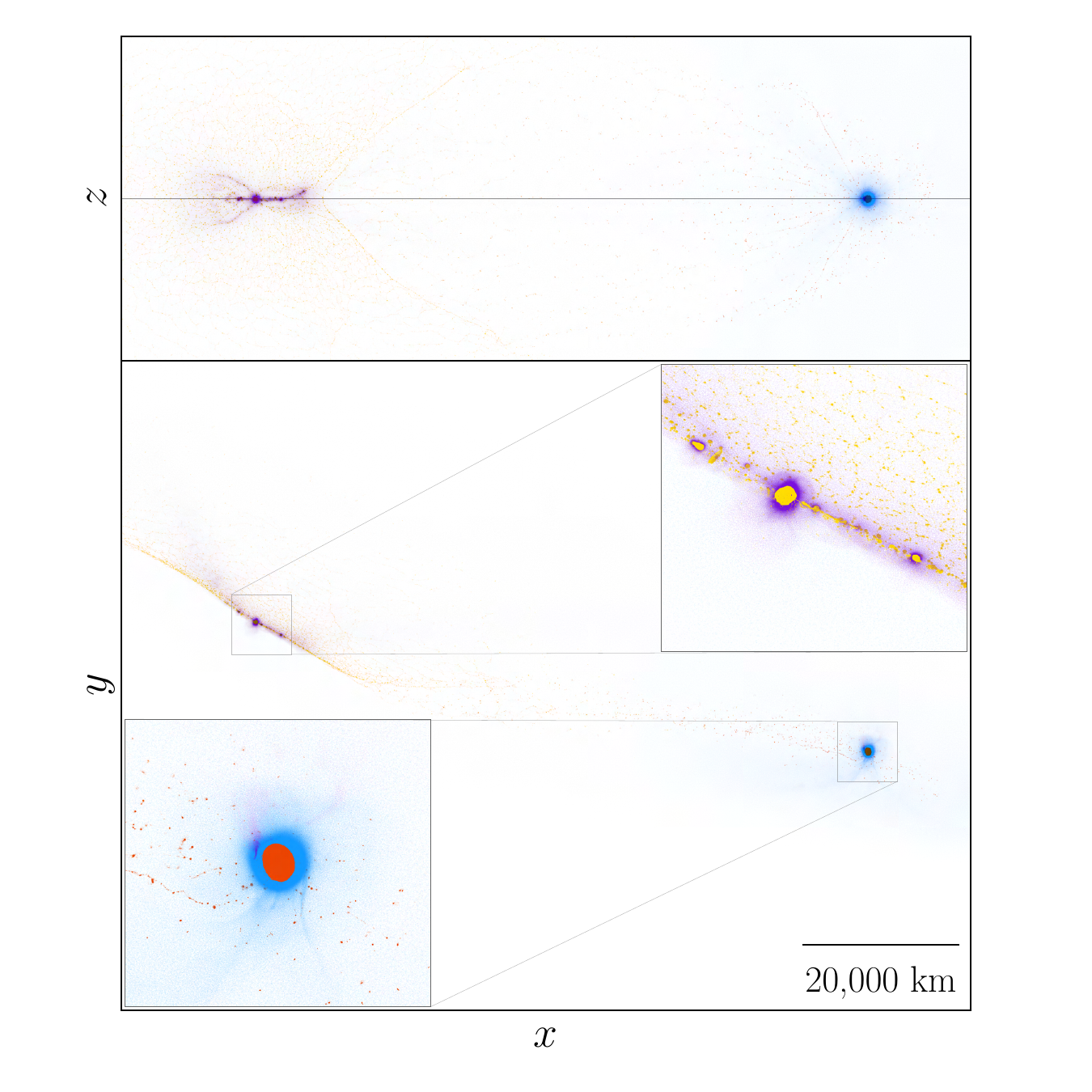}
  \\\vspace{-0.5em}
  \caption{
  Illustrative final snapshots in the SPH box frame
  from simulations with $10^{7.5}$ particles
  for impact angles and speeds at infinity of
  (left) $\beta = 15^\circ$, $v_{\infty} = 2.0$~km~s$^{-1}$ and
  (right) $\beta = 35^\circ$, $v_{\infty} = 3.0$~km~s$^{-1}$.
  The lower and upper panels show the $x$--$y$ and $x$--$z$ planes, respectively.
  Orange and blue show p-Rhea's
  core-rock and mantle-ice material, respectively,
  and yellow and purple the same for p-Dione.
  Animations are available at
  \href{http://icc.dur.ac.uk/giant_impacts/icy_moons_2kms_py.mp4}{icc.dur.ac.uk/giant\_impacts/icy\_moons\_2kms\_py.mp4}
  and \href{http://icc.dur.ac.uk/giant_impacts/icy_moons_3kms_py.mp4}{...\_3kms\_py.mp4},
  and with the same data rendered in 3D at
  \href{https://youtu.be/OWaMeUF4enU}{icc.dur.ac.uk/giant\_impacts/icy\_moons\_2kms.mp4}
  and \href{https://youtu.be/q8FYMhOE6TE}{...\_3kms.mp4}.
  \label{fig:final_snaps}}
\end{figure*}

As a sanity check, we confirm that these transformations are correct
by analysing the SPH initial snapshots to estimate the SPH moons' orbits,
using the equations in Appx.~\ref{appx:orbits}.
These indeed reproduce the expected orbital elements for each moon,
with slight variations caused by the shifted positions and velocities
we imposed to yield the different impact angles.
When the SPH simulation evolves further,
the orbital estimates become less accurate,
depending on how far material travels away from the centre of mass.
Using the final snapshots from our highly grazing impacts as examples,
the semi-major axes remain close to the initial values,
but by 9~h after impact the extracted eccentricities can vary by $\sim$10\%.
This, alongside the less predictable effects on the debris evolution
from tidal forces from Saturn,
results in up to tens of percent uncertainty
in the mass of debris that crosses different orbits.
However, this remains a smaller potential variation
than that between different impact angles
and moon masses (\S\ref{sec:results:orbits}),
and regardless would not affect the overall conclusions
of this proof-of-concept study,
since significant masses of debris can still be delivered throughout the system.
Nevertheless, this simplification should be redressed in future work.

\subsection{Identifying debris objects} \label{sec:methods:fof}

\begin{figure}[t]
	\centering
	\includegraphics[
    width=\columnwidth, trim={7mm 7mm 6mm 6mm}, clip]{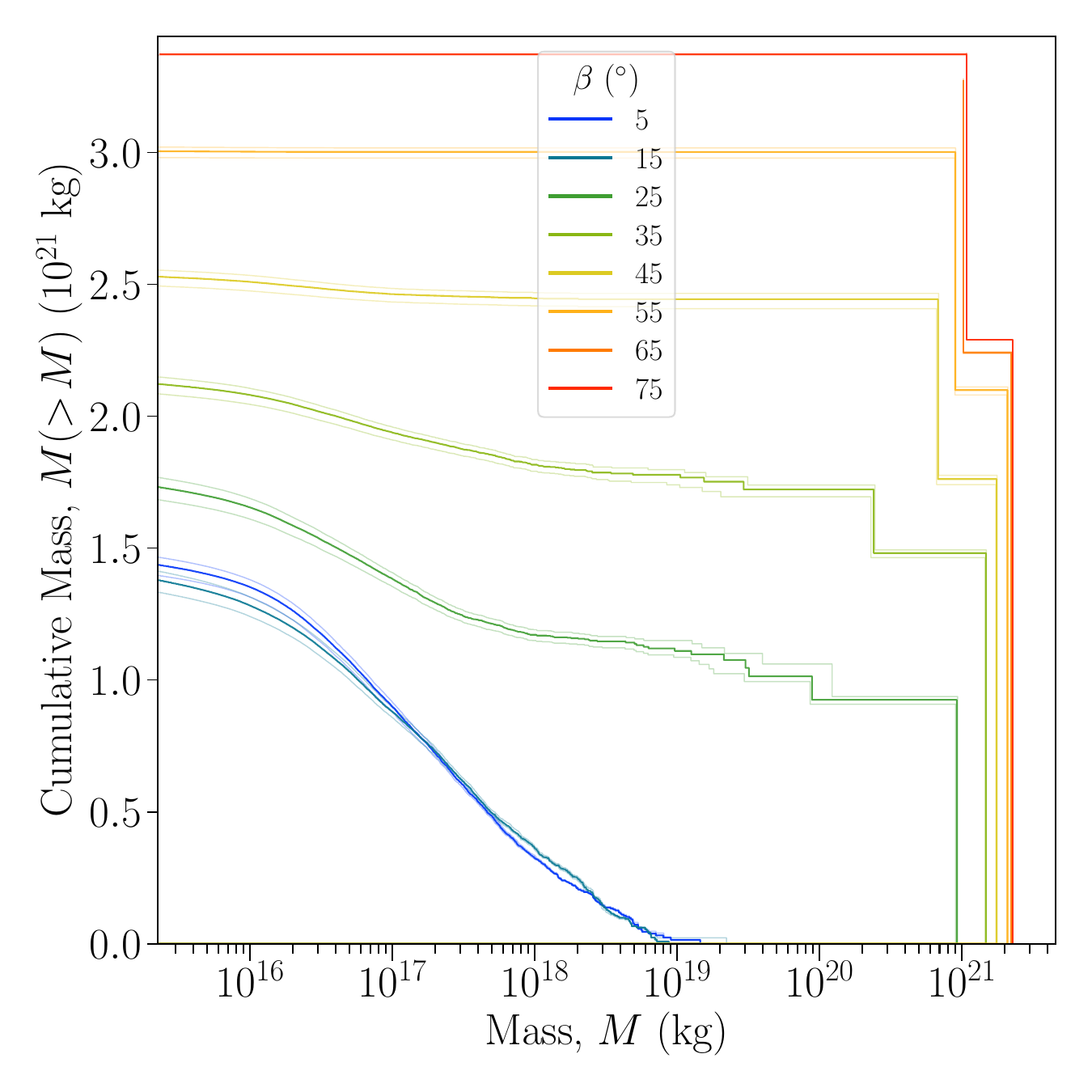}
  \\\vspace{-0.5em}
  \caption{
    The cumulative mass distributions of objects
    produced by impacts with $v_{\infty} = 3$~km~s$^{-1}$,
    after 9~h, simulated with $10^{7.5}$ SPH particles,
    coloured by the impact angle.
    Additional material is ejected as diffuse debris
    or smaller, unresolved objects that bring
    the cumulative total mass in all cases to the same value
    beyond the low end of the horizontal axis shown.
    The faint lines above and below each primary result
    show the distributions for 20\% smaller and larger linking lengths.
	\label{fig:mass_function}}
  \vspace{-1em}
\end{figure}

The surviving remnants and accumulated fragments produced by the SPH simulations
are identified using a standard friends-of-friends (FoF) algorithm,
where particles within a certain distance of each other
-- the `linking length', $l_{\rm link}$ -- are grouped together.
At the relatively short time after impact of our analysis,
most of these objects are still embedded in a debris field.
This makes the precise choice of linking length somewhat arbitrary,
because small changes in the linking length yield slightly different groups,
in contrast with bodies in a vacuum with clear surfaces,
which would be identified much the same for a wider range of linking lengths.
For our primary resolution of $10^{7.5}$ particles, we set $l_{\rm link} = 15$~km,
which scales for other resolutions with the cube root of the particle mass.
We test the sensitivity of our results to this choice
by repeating the analysis for significantly lower and higher
$l_{\rm link}$ of $\pm$20\% (12 and 18~km for $10^{7.5}$ particles
-- see Fig.~\ref{fig:mass_function}).
Even smaller or even larger linking lengths start to
no longer include the ice mantles and rocky cores of bodies in the same group,
or connect together many incohesive fragments into the same group, respectively.

\section{Results and discussion} \label{sec:results}

We first examine the general outcomes of collisions between precursor moons
in the center-of-mass frame in which the SPH simulations were run
in \S\ref{sec:results:sph},
before then considering the context of the full Saturn system
in \S\ref{sec:results:orbits}
(see Fig.~\ref{fig:frames_diagram}).
We examine numerical and other-scenario comparisons
for the primary-simulation results in \S\ref{sec:results:comparisons},
with a discussion of the limitations and speculative implications
of the results in \S\ref{sec:results:implications}.

\subsection{Impact outcomes}  \label{sec:results:sph}

For head-on to mid-angle impacts,
a significant amount of ice and rock is ejected from both bodies,
as illustrated for two example scenarios in Fig.~\ref{fig:final_snaps}.
The faster or closer to head-on the collision,
the greater the disruption,
the smaller the surviving remnants of the two original bodies (if any),
and the further the debris is spread away from the impact point
-- both in and out of the orbital plane.

Much of the ice is dispersed into diffuse debris,
while the rock remains or more rapidly re-accretes into
a large number of cohesive fragments.
These disruptive impacts produce a relatively smooth mass distribution
of many low-mass and fewer high-mass objects,
as shown by Fig.~\ref{fig:mass_function}.
Smaller and diffuse debris are not reliably resolved
into discrete objects by the SPH simulations,
so are not included in the figure,
and the detailed results below $\sim$$10^{16}$~kg
should be interpreted with caution.
As such, although the total mass remains the same in all simulations,
the cumulative distributions in Fig.~\ref{fig:mass_function}
show only the mass of resolved fragments.
The corresponding results for 2~km~s$^{-1}$ impacts are qualitatively similar,
but with a more rapid trend to fewer large fragments at grazing angles
(see Appx.~\ref{appx:extended_results}).

For more grazing collisions,
the pre-impact moons are incrementally less disrupted.
Less ice and much less rock is ejected,
particularly for impact angles at which the moons' cores do not intersect.
The debris becomes dominated by the remnants of the two pre-impact bodies,
with a smaller number of other fragments and diffuse ejecta (Fig.~\ref{fig:mass_function}).
Although grazing collisions may not immediately deliver
ring-forming or moon-disrupting material across the system,
the orbits of the remnant moons are negligibly changed\footnote{
  The post-impact semi-major axes, eccentricities, and inclinations
  of p-Rhea and p-Dione are changed by less than $\sim$1\% for $\beta > 45^\circ$.
},
and they will likely collide again in a future orbit \pCuk,
at perhaps more head-on angles.

Many fragments can be produced with masses
above $10^{18}$ and even above $10^{19}$~kg (i.e.\ order 0.1--1 Mimas masses),
which could readily disrupt other precursor moons
analogous to Tethys, Enceladus, and Mimas on crossing orbits
\citep{Leinhardt+Stewart2012},
as discussed further in \S\ref{sec:results:orbits}.
The ice fractions of these larger fragments are typically
a few tens of percent by mass,
as more massive objects are better able to hold on to and re-accrete
ice in addition to their rocky cores.
Less ice is also ejected into distant debris
by slower and/or more grazing impacts.
For close to head-on collisions,
fragments can be composed of up to equal mixes of p-Rhea and p-Dione material,
while for mid-to-grazing angles,
most large fragments are built from almost purely p-Rhea or p-Dione.

These results are similar for the different linking lengths
discussed in \S\ref{sec:methods:fof},
and any minor differences are negligible compared with
those between impact angles and speeds,
as illustrated in Fig.~\ref{fig:mass_function}.
The ice fractions are similarly insensitive to the linking length,
with the larger and smaller values typically yielding a few percent
more and less ice by mass, respectively.

\begin{figure}[t]
	\centering
	\includegraphics[
    width=\columnwidth, trim={7.5mm 8mm 7mm 17mm}, clip]{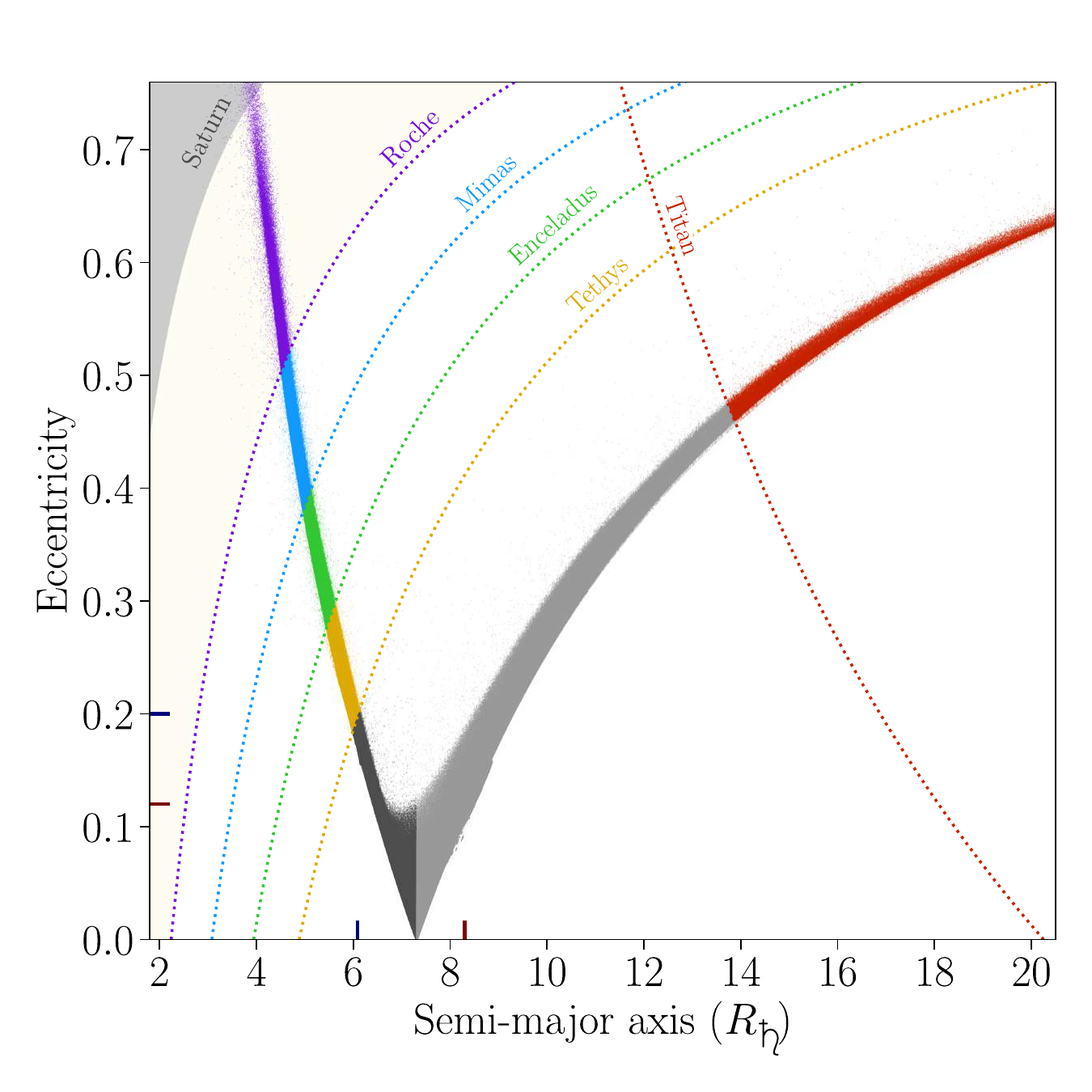}
  \\\vspace{-0.5em}
  \caption{
    The eccentricities and semi-major axes of the orbiting debris,
    for an example $\beta = 35^\circ$, $v_{\infty} = 3.0$~km~s$^{-1}$ impact,
    as illustrated in Fig.~\ref{fig:final_snaps}
    (see also Fig.~\ref{fig:e_a_B15v20}).
    Each point is one SPH particle with a mass of $\sim$$10^{14}$~kg.
    The dotted lines and corresponding particle colours
    indicate where an orbit's periapsis crosses the Roche limit
    or the orbits of other moons at their locations in the present-day system,
    or the apoapsis in the case of Titan.
    Other particles with semi-major axes inside and outside the impact point
    are coloured in grey and light grey, respectively.
    The dark-red and dark-blue pairs of axis ticks
    indicate the orbital parameters of the pre-impact moons.
	\label{fig:e_a}}
  \vspace{-1em}
\end{figure}

\begin{figure}[t]
	\centering
	\includegraphics[
    width=\columnwidth, trim={8mm 8mm 7mm 7mm}, clip]{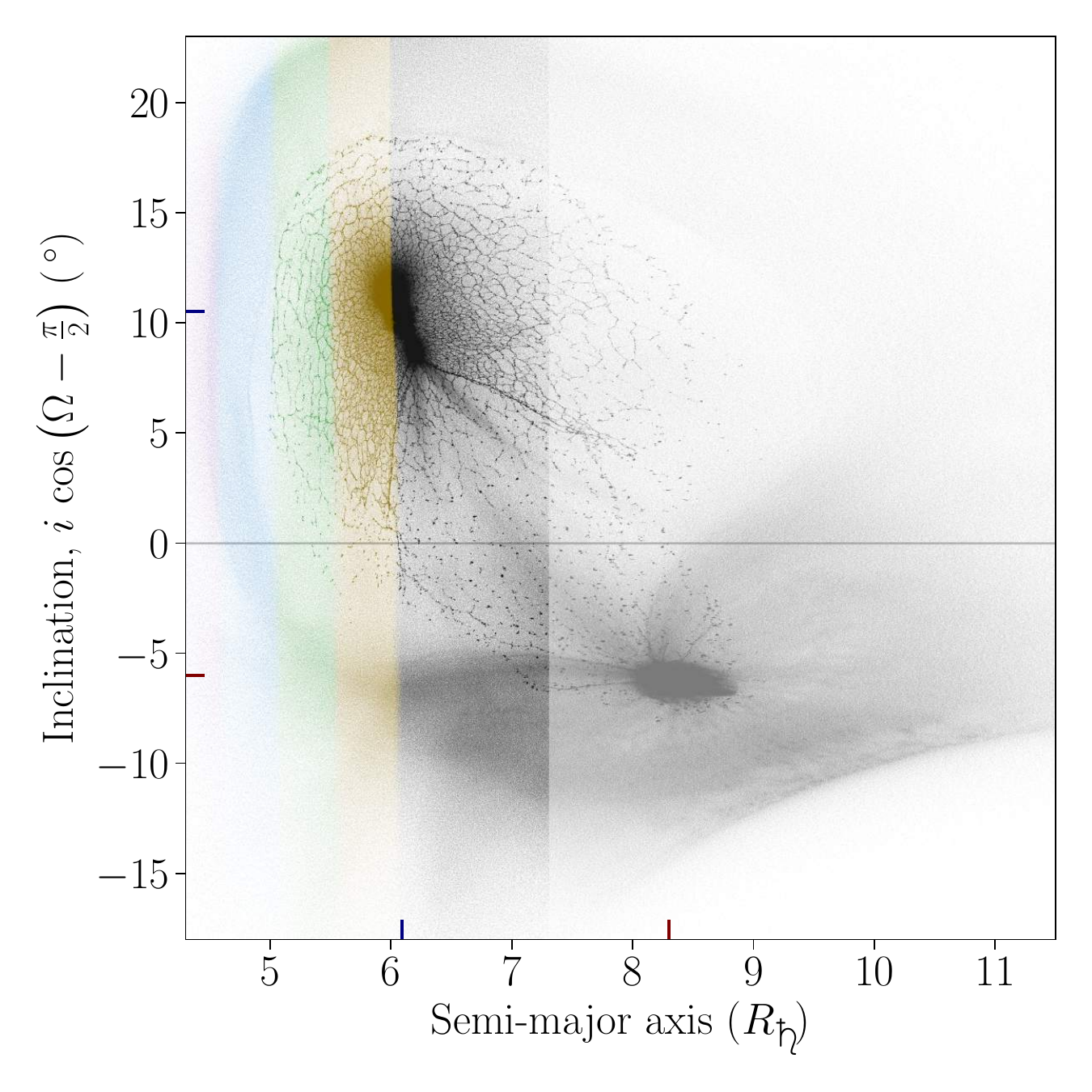}
  \\\vspace{-0.5em}
  \caption{
    The inclinations of the orbiting debris,
    with a sign set by whether the longitude of ascending node, $\Omega$,
    is at or opposite the point of impact,
    for an example $\beta = 35^\circ$, $v_{\infty} = 3.0$~km~s$^{-1}$ impact
    (see also Fig.~\ref{fig:i_a_B15v20}),
    coloured as in Fig.~\ref{fig:e_a}.
    The dark-red and dark-blue pairs of axis ticks
    indicate the orbital parameters of the pre-impact moons.
	\label{fig:i_a}}
  \vspace{-1em}
\end{figure}

\subsection{Trajectories in the Saturn system}  \label{sec:results:orbits}

We now place the SPH results in the context of the Saturn system,
following the procedure described in \S\ref{sec:methods:frame_trans}.
Material is sent far throughout the system by head-on to mid-angle impacts,
as shown in Figs.~\ref{fig:e_a} and~\ref{fig:i_a}.
Large amounts of debris and fragments are placed onto orbits
that would intersect with other satellites around the locations of
those in the present-day system,
including an extended population of debris that is directly heading
deep into the Roche limit.
This qualitative outcome is consistent across scenarios,
with the angle and speed affecting, for example,
the thickness and extent of the branches in Fig.~\ref{fig:e_a}
(see Appx.~\ref{appx:extended_results}).
We consider Saturn's Roche limit for water ice
to be at $2.27~R_{\Sa} = 136,800$~km,
which corresponds to the outer edge of the A ring.
This makes it a somewhat conservative value as,
depending on the porosity and density of the debris,
the effective limit could be higher \citep{Tiscareno+2013}.

As discussed in \S\ref{sec:introduction},
a precursor system can reasonably be expected to have multiple mid-sized satellites
in a broadly similar architecture to the system today.
The eccentric, high-velocity debris and fragments produced here could
significantly erode other moons in a collisional cascade
to distribute even more material throughout the system and into the Roche limit \pCuk,
although this will have to be explored in detail with future modelling
beyond the scope of this initial study.
Head-on to mid-angle impacts send a few to a few tens of objects
with masses above $10^{17}$~kg onto Enceladus-crossing orbits, for example,
and an order of magnitude more for Tethys.
Furthermore, a total mass of debris ranging from around that of Mimas
to over that of Enceladus (order $10^{19}$--$10^{20}$~kg)
reaches the orbits of the three inner moons.

\begin{figure}[t]
	\centering
	\includegraphics[
    width=\columnwidth, trim={2mm 5mm 22mm 25mm}, clip]{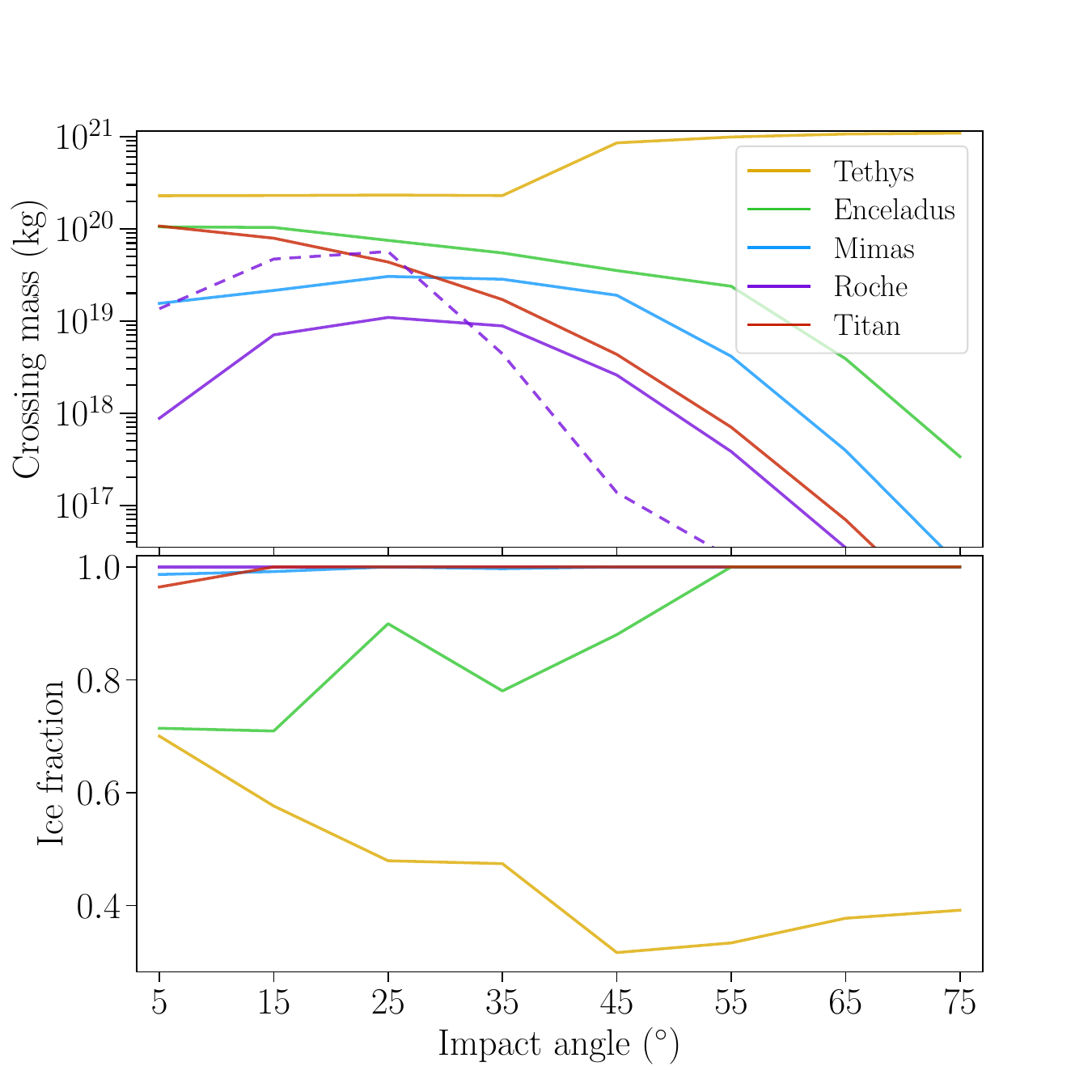}
  \\\vspace{-0.5em}
  \caption{
    The mass and composition of post-impact debris on orbits crossing
    the Roche limit or other moons,
    as a function of impact angle for $v_{\infty} = 3.0$~km~s$^{-1}$ scenarios.
    The dashed line shows the Roche-crossing mass for
    the same impact-angle scenarios with orientation angle $\phi = 90^\circ$.
    The results for all simulations are provided in Table~\ref{tab:results}.
    The plotted values are exclusive,
    so e.g. the mass shown for Enceladus's orbit
    does not include the material that also reaches Mimas or the Roche limit.
    \label{fig:m_orb_cross}}
  \vspace{-1em}
\end{figure}

The mass that crosses each location depends on the impact angle and speed,
as shown in Fig.~\ref{fig:m_orb_cross}.
Table~\ref{tab:results} contains the results for all simulations.
The distributions of debris trajectories are highly anisotropic,
so the orientation of the impact point in Saturn's frame
($\phi$, \S\ref{sec:methods:impact_scenarios}) is also important
and can significantly increase or decrease the crossing masses,
depending on the impact angle.
For these simulations, more than twice the present-day ring mass
can be immediately sent inside the Roche limit,
where it could begin to tidally and collisionally evolve
towards forming rings \citep{Dones+2007},
as discussed further in \S\ref{sec:results:implications}
-- even before considering further cascades
and other material redistribution,
or the wider parameter space of plausible impact scenarios
beyond the few examples considered here.

The Roche-entering material contains negligible rock,
matching the present-day rings' nearly pure-ice composition
\citep{Zhang+2017b,Zhang+2017a},
while the material that could encounter Tethys and Enceladus
can be several tens of percent rock by mass, depending on the scenario
(Fig.~\ref{fig:m_orb_cross}).
The results for 2~km~s$^{-1}$ scenarios are qualitatively similar,
with the crossing masses reduced by factors of a few,
approaching an order of magnitude at greater distances from the impact point,
as detailed in Table~\ref{tab:results}.

In addition to the range of possible impact speeds,
the applicability of a resonant destabilisation like that demonstrated by \tCuk
to any pair of loosely Dione- and Rhea-like analogues
reminds us that more mass could readily be delivered to the Roche limit
by, for example, a more massive pair of colliding satellites
(as examined in \S\ref{sec:results:comparisons}),
or an impact at a distance closer to Saturn.
The latter could arise from
a different orbital evolution of the outer two satellites,
or perhaps from a collision between p-Dione and a p-Tethys analogue
-- or even from an outer moon destabilised as in \citet{Wisdom+2022}.
Correspondingly, less material would likely be delivered by
a more-distant collision or a lower-mass pair,
although this is less likely than higher precursor masses
given that the present-day moons must accrete from the remnants.

A comparable $\sim$$10^{19}$--$10^{20}$~kg mass of debris
is also sent outwards by disruptive impacts to encounter Titan.
Like the material reaching the Roche limit, this is typically also pure ice,
although some rock can be delivered to Titan by close to head-on collisions.

\begin{figure}[t]
	\centering
	\includegraphics[
    width=\columnwidth, trim={8mm 7mm 5mm 4mm}, clip]{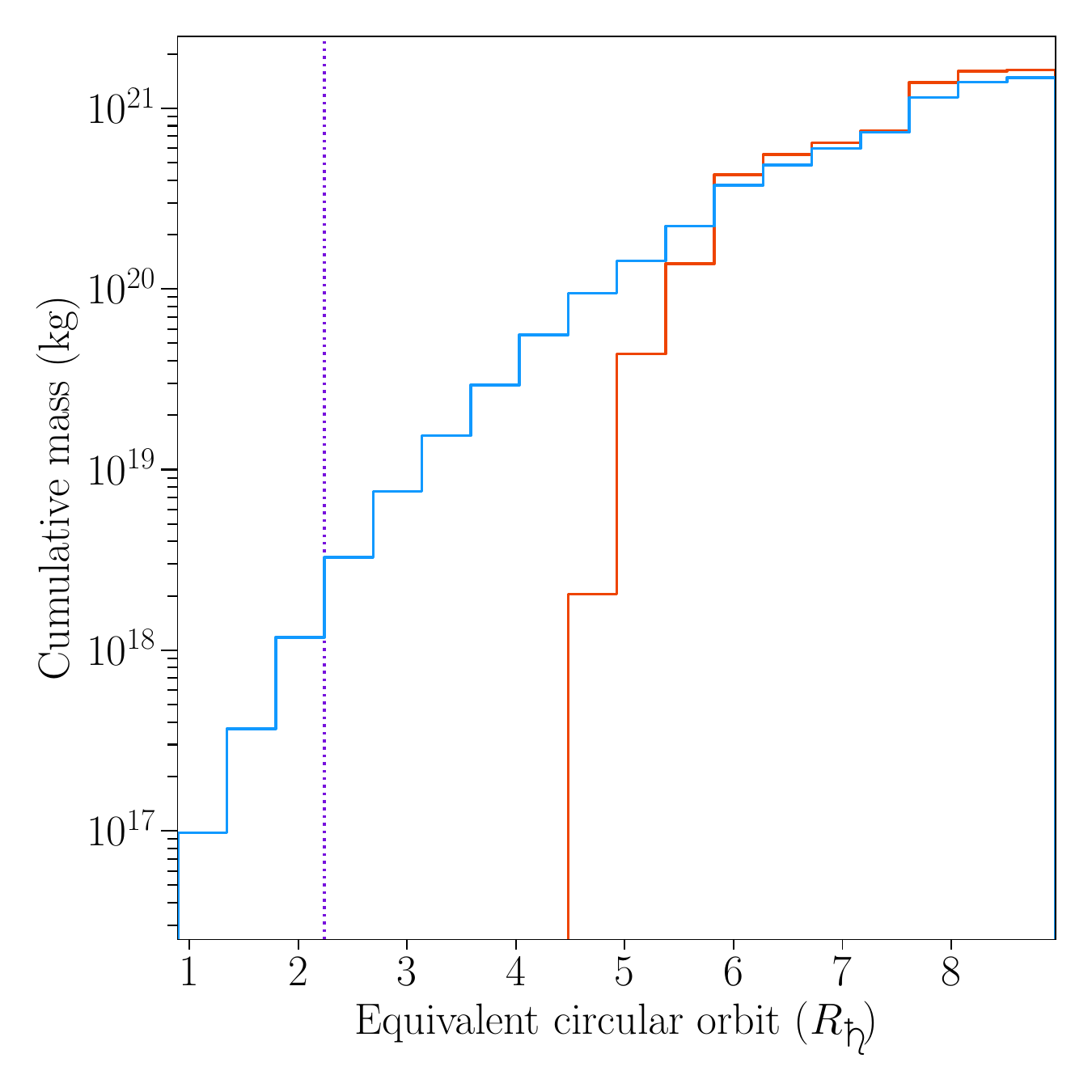}
  \\\vspace{-0.5em}
  \caption{
    The cumulative mass of debris,
    if it evolved by collisional energy dissipation
    to circularise onto an equivalent orbit
    with the same angular momentum,
    as a function of radial distance from Saturn,
    for an example $\beta = 25^\circ$, $v_{\infty} = 3.0$~km~s$^{-1}$ impact.
    The orange and blue lines show rock and ice material, respectively.
    The dotted line indicates the outer edge of the A ring at $2.27~R_{\Sa}$.
    \label{fig:m_a_eq}}
  \vspace{-1em}
\end{figure}

These initial trajectories for the debris
are only the starting place for significant further evolution
that must be examined and modelled in detail in future work.
A common approach to estimating the evolution of collisional material
is to assume that each piece of debris will evolve onto a circular orbit
with the same angular momentum.
However, disruptive collisions produce debris
with a range of angular momenta and orbital energies,
and detailed modelling beyond the scope of this paper will be required
to constrain how much additional fragmentation and spreading of material occurs.
If the system were isolated
-- though that is not the expectation here in a precursor satellite system --
then the mass of material with periapses within the Roche limit
and the mass of material with equivalent circular orbits within the Roche limit
could be interpreted as crude upper- and lower-bound estimates
for the mass likely to evolve there.

With these caveats in mind, Fig.~\ref{fig:m_a_eq}
shows the mass distribution of material with equivalent circular radius,
$a_{\rm eq} = a (1 - e^2) \cos(i)^2$.
A significant mass of pure-ice material can circularise inside the Roche limit
even without subsequent disruption,
though not immediately as much mass as that of the present-day rings
for this example collision.
The most dispersive scenarios tested here
can yield up to $\sim$$10^{19}$~kg of ice with $a_{\rm eq} < R_{\rm Roche}$,
compared with $\sim$$5 \times 10^{19}$~kg with periapses $< R_{\rm Roche}$.
Therefore, under the conservative assumption
of perfectly angular-momentum-retaining collisions
and for the specific scenarios simulated here,
some further redistribution of material via cascade collisions or other processes
could be required to form the rings.
Such evolution could be supported or even made unnecessary
by a more dispersive base impact scenario,
such as with larger or faster colliding moons,
as discussed in \S\ref{sec:results:comparisons}.

We also note that a comparable but perhaps greater challenge
could be faced by hypotheses of ring formation
from the tidal disruption of an external body or a destabilised moon
\citep{Dones1991,Hyodo+2017c,Wisdom+2022},
with significantly greater initial semi-major axes and eccentricities.
Such scenarios could still lead to the creation of rings
but lack the range of highly reduced orbital energies and angular momenta
of the initial debris that is produced
by the dissipative collisions considered here.

Fig.~\ref{fig:m_a_eq} also further demonstrates
the lack of non-ice pollution in the initial ring-forming material,
which for the other hypotheses could require a fortuitously ice-rich progenitor
to match the observations \citep{Dones1991,Wisdom+2022,Estrada+Durisen2023}.
Here, no rocky material has an equivalent circular orbit
within the Roche limit or even inside of Enceladus's orbit in this example,
and none inside of Mimas's orbit for any of our impact simulations.

\subsection{Convergence and comparisons}  \label{sec:results:comparisons}

\begin{figure}[t]
	\centering
	\includegraphics[
    width=\columnwidth, trim={4mm 5mm 22mm 25mm}, clip]{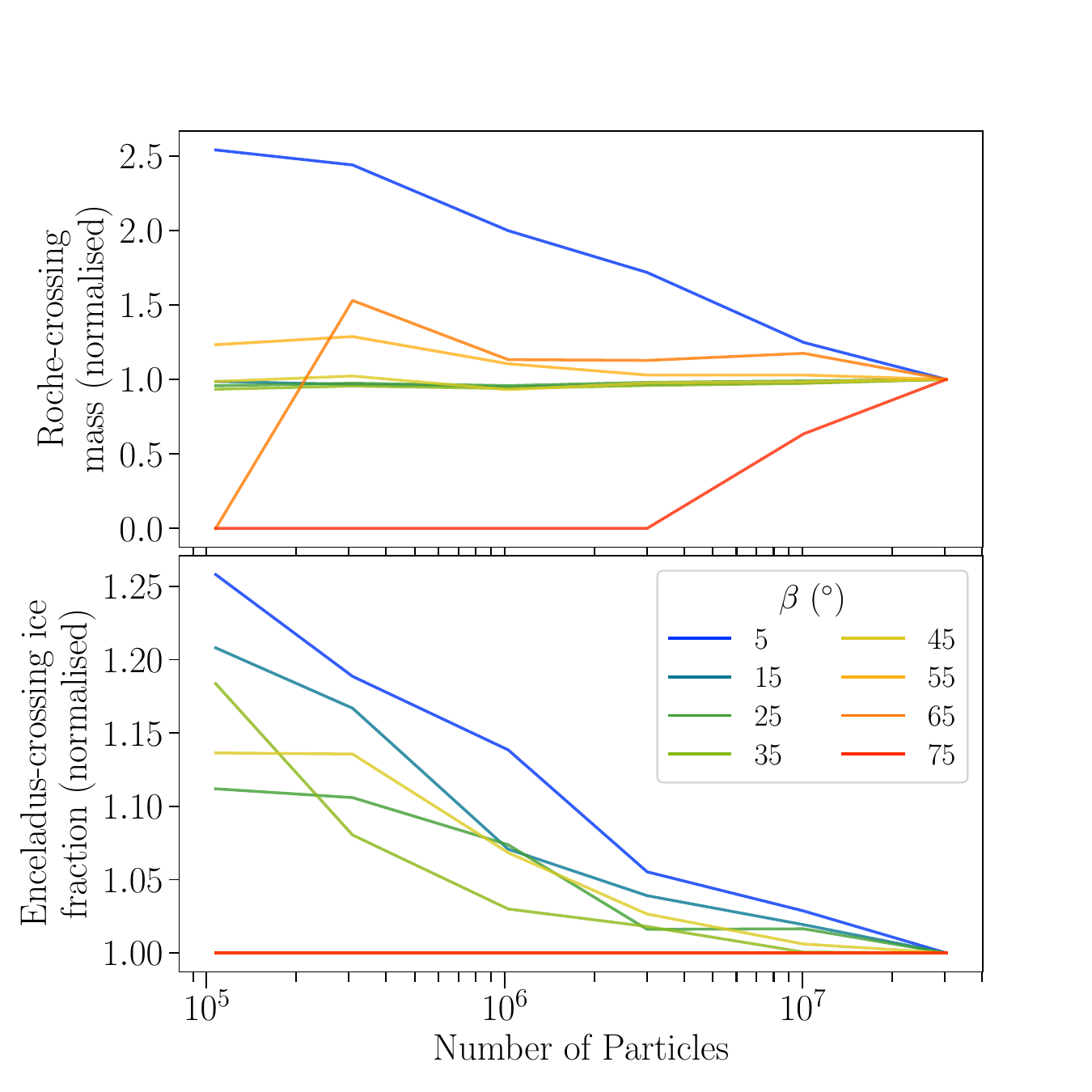}
  \\\vspace{-0.5em}
  \caption{
    The mass of post-impact debris on orbits crossing the Roche limit
    and the composition of debris crossing Enceladus's orbit
    (as Roche-crossing material is pure ice at all resolutions),
    as functions of the numerical resolution,
    for $v_{\infty} = 3.0$~km~s$^{-1}$ scenarios.
    The results for each impact angle are normalised
    to that of the highest resolution to examine convergence.
    \label{fig:m_orb_cross_res}}
  \vspace{-1em}
\end{figure}

The total mass of debris that is placed onto orbits reaching the Roche limit
and crossing the orbits of other moons numerically converges
to within $\sim$1\% by $10^7$ particles for disruptive scenarios,
up to a few tens of percent at other impact angles,
as shown in Fig.~\ref{fig:m_orb_cross_res} for 3~km~s$^{-1}$ impacts,
with complete results in Table~\ref{tab:results}.
Low resolutions below $\sim$$10^6$ particles yield
unconverged masses that differ by many tens of percent for various impact angles.
The masses for low-angle impacts converge at lower resolutions,
apart from very close to head-on collisions,
for which the mass crossing the Roche limit is lower
and thus the resolution requirements for reliability are higher.
This aligns with previous studies
where the primary outcomes of high-speed, disruptive impacts
converge relatively quickly,
while impacts closer to the mutual escape speed
can require much higher resolutions,
as do extracted results of lower-mass features \citep{Kegerreis+2020}.
For 2~km~s$^{-1}$ impacts,
the mass crossing the Roche limit is lower
and thus convergence requires higher resolution,
with most angles yielding differences of several percent
between $10^7$ and $10^{7.5}$ particles.
In all cases, the ice fraction of the orbit-crossing debris
converges to within a few percent by $10^{6.5}$--$10^{7}$ particles
(Fig.~\ref{fig:m_orb_cross_res}).
Lower resolutions systematically underestimate the mass of rock delivered.

For the mass function of debris,
higher resolutions are important both
for converging on the distribution of large objects
and for enabling the inspection of smaller ones.
We find that the overall shape of the high-mass end of the mass function
is typically well converged by $10^{7}$ particles.
The distribution of smaller objects with masses below $\sim$$10^{18}$~kg
is still broadly similar for $\geq$$10^{7}$ particles,
but matches less reliably.
The highest resolution used here enables the probing of objects
down to masses of about $10^{16}$~kg,
corresponding to $\sim$100 SPH particles,
with potential numerical concerns for the analysis of smaller objects.

We also tested repeat simulations with reoriented initial conditions
(\S\ref{sec:methods:impact_scenarios}),
which with infinite resolution should give identical outcomes,
but can produce non-negligibly different results
even at high resolutions \citep{Kegerreis+2022}.
Fortunately, in these high-speed scenarios the differences are minimal
for the orbit-crossing masses (standard deviations $<$$1$\%),
and the variation in the debris mass functions is of a similar magnitude
to that between different linking lengths shown in Fig.~\ref{fig:mass_function}.

A potentially more significant limitation
is the isolation of the SPH simulations from Saturn,
as discussed in \S\ref{sec:methods:frame_trans}.
Tidal effects would likely affect
the separation and merging behaviour of debris fragments,
and the orbital estimates become less accurate.
However, for the first several hours after the impact,
the mass function is still rapidly evolving.
By $\sim$8--9~h (or sooner in highly grazing scenarios),
the results remain broadly consistent with time.
Some fragments continue to merge, collide, and separate,
but this yields far less variation than found between different impact scenarios.

Replacing the impacting p-Dione with a larger p-Rhea
produces similar results to the scenarios examined thus far.
The mass functions of fragments follow the same trends
shown in Fig.~\ref{fig:mass_function} (and Fig.~\ref{fig:mass_function_2kms}),
though as a symmetric collision there are two largest remnants rather than one,
and at middle angles not quite as many large, secondary fragments are produced.
The total mass that crosses the Roche limit
is within a factor of two of the p-Dione--p-Rhea results in most cases,
but is over an order of magnitude higher for close to head-on impacts
(see Table~\ref{tab:results}).
The mass crossing the orbits of other present-day moons
is similarly comparable to the p-Dione--p-Rhea cases,
but is typically somewhat higher for head-on to mid-angle impacts
and lower at more grazing angles.

The above differences for p-Rhea--p-Rhea collisions
are likely driven largely by the impact geometry,
with some effects also expected from the slightly greater self gravity
of the moons and their post-impact remnants.
At the same input impact angle,
the centres of the two moons are more distant at the time of contact,
so a smaller portion of the p-Rhea impactor intersects with the p-Rhea target
than in the case of the smaller p-Dione impactor.
This is compounded by Rhea's smaller rock fraction and lower density,
such that grazing collisions can eject significant outer ice material
but the rocky cores are less readily disrupted.
This is reflected in the ice fractions of the orbit-crossing debris,
which contain rock out to less grazing angles
than in the p-Dione--p-Rhea case, as detailed in Table~\ref{tab:results}.

We also tested a few scenarios with
systematically varying masses of the precursor moons,
with the p-Rhea mass simplified to double that of the p-Dione.
The overall outcomes are highly similar to the
fiducial simulations at the same impact angle.
The mass crossing the Roche limit increases with the colliding mass
almost proportionally,
and roughly follows a power law with exponent $\sim$$0.85$--$0.95$
depending on the impact angle (Table~\ref{tab:results}).
This substantiates the expectation that significantly more mass could
be distributed throughout the system if the precursor moons were more massive
than the present-day analogues considered for the primary simulations here.

In contrast to this general consistency in outcomes
that we find across scenarios,
our results show some differences from those of
\citet[hereafter \citetalias{Hyodo+Charnoz2017}]{Hyodo+Charnoz2017},
who ran SPH impact simulations of two Rhea-mass bodies
and found less spread of debris through the Saturn system,
with a $45^\circ$ impact delivering no material
directly to the Roche limit, for example.
They then took 10\% of the particles from
the $45^\circ$ SPH simulation as inputs for five $N$-body integrations,
with a different random subset of particles in each run.
Further fragmentation was not included,
but particles could merge, following a hard-sphere model,
which they found leads most of the debris to re-accrete into a single body
on a timescale of a few kyr.

Our SPH simulations differ in a few respects.
One numerical difference is the SPH resolution used, as discussed above,
which is over two orders of magnitude higher here.
However, even with our test simulations at a comparable resolution to theirs,
we still find more material on orbit-crossing trajectories
for the same impact angle.
Minor variations in the initial conditions preclude
a direct comparison of the similar numerical techniques:
First, for the impact point and the centre-of-mass's orbit,
\citetalias{Hyodo+Charnoz2017} set a circular, equatorial orbit
with a 3~km~s$^{-1}$ relative impact velocity.
This places the impact point $\sim$13\% farther away from Saturn than here.
The predicted orbits of the debris are highly sensitive to the
transformation of the SPH results into the Saturn system
(see \S\ref{sec:methods:frame_trans},
underspecified in \citetalias{Hyodo+Charnoz2017}
with respect to the orientation of the moons),
and a more distant impact point reduces the inward reach of debris.

Second, \citetalias{Hyodo+Charnoz2017} collided Rhea-mass bodies
with core mass fractions of 40\% and 60\%,
compared with only 40\% in both p-Rhea bodies here.
This greater core fraction might also have reduced the amount of widespread ejecta.
The interior structures (and masses) of the moons in a precursor system are unknown,
so future work to systematically examine the outcomes
from collisions of different initial bodies
would help to constrain this uncertainty.
Similarly, simulations that include Saturn
and model impacts at various distances from the planet
would reveal the importance both of the planet's gravity
and of the location of the point of impact.

The conclusions that we draw from the SPH results
in the context of the wider Saturn system
also differ from \citetalias{Hyodo+Charnoz2017},
as we speculate that collisions with other inner precursor moons
could further support the delivery of material into the Roche limit,
even with significantly less initial spread of debris than we find here.
However, detailed $N$-body simulations that include both fragmentation
and a full precursor satellite system will be required
to make more robust predictions.
It also may be that a more-disruptive, less-grazing impact
than the 45$^\circ$ one examined by \citetalias{Hyodo+Charnoz2017}
will be required for successful long-term distribution of material,
even when fragmentation is accounted for.

\citetalias{Hyodo+Charnoz2017} also estimated analytically that
the $\sim$kyr accretion timescale of their disk, neglecting fragmentation,
would be shorter than the timescale of viscous spreading,
which \tCuk had speculated could promote the delivery of material
to inside the Roche limit.
However, the immediate production of
Roche-bound and Roche-circularising debris that we find here
and the possibilities of collisional cascades with other moons
could avoid the requirement for a ring-forming disk to spread in this way.
Furthermore, as also noted by \tCuk,
the first set of satellites that re-accrete from such a disk
would likely be numerous and unstable.
A longer timescale could be required
for these satellites to merge into the relatively stable system we see today,
during which the likely chaotic destabilisation and collisions
could further distribute debris.

Both our primary simulations and those of \citetalias{Hyodo+Charnoz2017}
used \citet{Tillotson1962} equations of state
(and neglected material strength),
which do not include phase boundaries
and can prompt unrealistic behaviour for highly shocked ejecta \citep{Stewart+2020}.
A full exploration of the importance and effects of these systematic uncertainties
is beyond the scope of this project,
but we make an initial test of the sensitivity of our results
using a repeated subset of the p-Rhea--p-Rhea impacts
simulated with the more sophisticated ANEOS and AQUA EoS
\citep{Stewart+2020,Haldemann+2020}.
The overall behaviour is similar to the Tillotson results,
although in general the ice ejecta appears to be less diffusely dispersed.
A greater proportion of ice also re-accretes into objects
with masses around $10^{17}$--$10^{18}$~kg,
yielding a somewhat shallower size distribution
and typical ice fractions in fragments ranging from $\sim$20--100\%,
compared with a previous $\sim$5--40\% for these p-Rhea--p-Rhea impacts.
Most of the total masses of orbit-crossing debris
are within a few tens of percent of the primary results,
with some greater differences for the slower 2~km~s$^{-1}$ collisions
(Table~\ref{tab:results}).
The variations due to the impact angle and speed
continue to be more important than the choices of equation of state,
moon masses, compositions, and numerical resolution,
though the differences may become more significant for
future work on the detailed subsequent evolution of the debris.

\subsection{Implications and future work}  \label{sec:results:implications}

The outputs of simulations like these provide the inputs for
the next key stage of modelling the evolution
of the initially eccentric and inclined fragments and debris.
Combinations of $N$-body integrations and further impact simulations
can then begin to determine the consequences for the other mid-sized moons
and the possible growth or rejuvenation of circularised rings
within Saturn's Roche limit.

Returning first to a brief discussion of the initial conditions for this scenario,
the original hypothesis of recent re-accretion of the inner moons by \tCuk
was based on the assumption of equilibrium tides.
However, as discussed in \S\ref{sec:introduction},
more sophisticated tidal models
and the possibility of resonance locking \citep{Lainey+2020}
do not avoid the recent encounter that an ancient Rhea
would have had with the evection resonance.
It is unlikely that this crossing could be reconciled with
the relatively low-$e$, low-$i$ orbit of Rhea
without significant collisional damping, possibly even re-accretion (cf. \tCuk),
so an ancient age for Rhea remains highly unlikely.
Additionally, a resonant tidal response offers
many more possibilities for the system to be disrupted,
as the energy that can be put into exciting satellite orbits by resonant tides
is orders of magnitude larger than that provided by equilibrium tides.
In future work, we hope to revisit the potential destabilisation of the system
and expand on the input models that could lead
to this general class of recent-collision scenarios,
whether from evection excitation or, for example,
the ejected outer satellite proposed by \citet{Wisdom+2022}.

There are also some limitations of this work that merit future study.
While we explored a range of impact angles for two plausible speeds
and brief tests of other moon masses,
this scenario allows for a diverse variety of impact conditions,
including different sizes and compositions of the colliding moons
in addition to the angle, speed, and distance from Saturn of the impact.
We also neglected material strength, which is unlikely to dominate
in such high-speed impacts but could have uncertain
effects on the details and thermodynamics of the ejecta
-- as would the use of more sophisticated equations of state
than in the primary suite of simulations.
Furthermore, the SPH simulations were performed in an isolated box
before being placed around Saturn.
The presence of Saturn throughout the hours of the simulation
could promote the separation of fragments via tidal forces, for example,
and its absence adds uncertainty to the values of the distributed mass.
Therefore, while our primary conclusion that a significant amount of debris
can be sent throughout the Saturn system to cross other moons and the Roche limit
should not be sensitive to these limitations,
the precise details of the debris
and of the parameter space of successful specific scenarios
may be expected to change as these simplifications are addressed.

\begin{figure}[t]
	\centering
	\includegraphics[
    width=\columnwidth, trim={7mm 7mm 7mm 7mm}, clip]{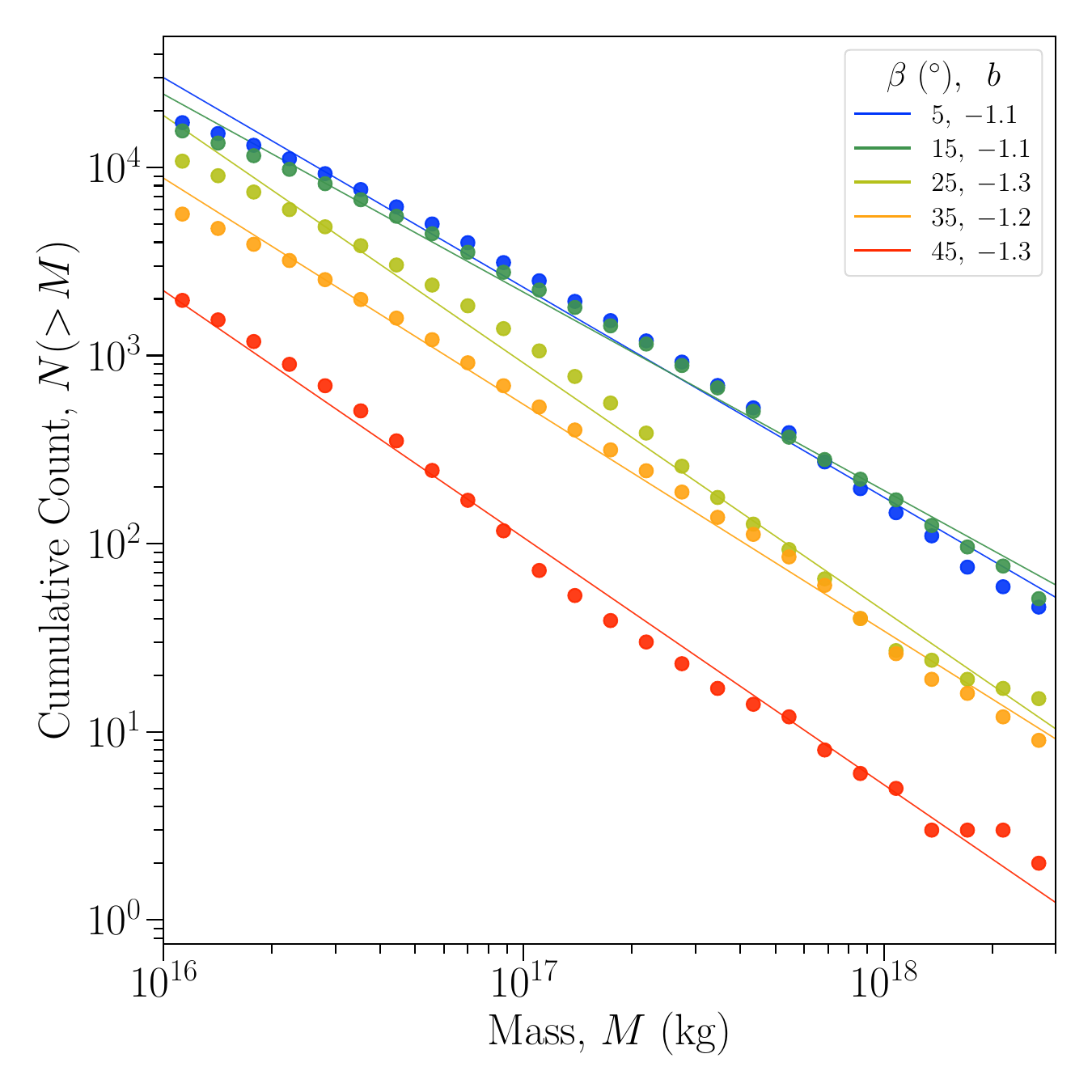}
  \\\vspace{-0.5em}
  \caption{
    The cumulative number distributions of debris objects
    produced by head-on to mid-angle 3~km~s$^{-1}$ impacts,
    as also characterised by Fig.~\ref{fig:mass_function}.
    The lines show fitted power laws $N \propto M^b$,
    with exponents $b$ given in the legend,
    for the mass range $10^{16}$--$2 \times 10^{18}$~kg.
    The fits do not include the many smaller objects
    that are less robustly resolved,
    nor the lower numbers of even larger objects
    that are less well represented by a single power law in some cases.
    \label{fig:count_dist}}
  \vspace{-1em}
\end{figure}

The ejected fragments and debris will continue to interact with each other,
with the other moons and any existing rings, and with Saturn's tidal field,
before the system evolves towards its current state.
However, once the present-day moons are mostly in place,
any debris that is still in orbit
-- including secondary and further cascade ejecta --
could imprint itself in the crater populations seen today.
The craters on Saturn's moons indicate a population
of at least some planeto- rather than helio-centric impactors \citep{Ferguson+2022},
and do not match the distributions on the other outer planets' satellites.
It is likely premature to consider
the immediate fragment distributions from this study
as a crater-forming population,
since they can be expected to evolve significantly over time,
but they can represent an initial approach
towards future comparisons and constraints.

With this caveat in mind, Fig.~\ref{fig:count_dist} shows
the cumulative number distributions of debris objects
from low-to-mid impact angles at an impact velocity of 3~km~s$^{-1}$
(see also Fig. \ref{fig:mass_function}).
The distributions can be fitted crudely with a $\sim 1/M$ power law,
with a slight suggestion of a trend to steeper slopes for higher impact angles,
as comparatively fewer medium-to-large objects are produced
by more grazing collisions.
One can then begin to speculate how this population of impactors
could affect the re-accreting moons,
and what the impact rate on these bodies might be
-- again neglecting the fact that on these elliptical orbits
significant erosion and fragmentation could first be expected.
For example, a dense, planeto-centric population may provide
a sufficient ``fluence'' of objects to saturate satellite surfaces
in a relatively short period of time, perhaps $\sim$kyr,
so that they may appear as cratered as ancient, Gyr-old terrain
might be from a heliocentric population \citep{Lissauer+1988,Zahnle+2003}.
Furthermore, the range of impact speeds
will also be lower for planeto-centric orbits,
so different scaling laws for impactor size to crater diameter may be required.
The thermal state of the evolving moons should also be considered,
in terms of the cooling time that might be required before
craters can be preserved without significant relaxation \citep{White+2017}.
However, the low gravitational binding energy of the small moons
means that large impactors may cause more erosion than heating.

In addition to the disruption, accretion,
and eventual surfacing implications for Saturn's inner mid-sized moons,
approximately an Enceladus mass of debris
can be placed onto orbits that intersect with Titan.
Titan's unique atmosphere and surface present many unanswered questions
regarding their evolution and history \citep{Nixon+2018}.
This includes a possible young age for the current methane system
of a few tens to hundreds of Myr
associated with the rapid conversion of methane into heavier hydrocarbons.
The effects of this high-energy delivery of a large mass of icy debris
are beyond the focus of this study,
but could open up new options for models of Titan's evolution.
Furthermore, \tCuk also proposed that resonant interactions of Titan
with potentially long-lived outer parts of the debris disk
and a transient satellite that accretes there
could excite Titan's eccentricity enough to explain
its significant value ($\sim$0.03) today.

Finally, we could consider this general scenario
for the destabilisation and disruption of a satellite system
as a potential evolutionary pathway for other planets' moons and rings.
However, amongst the other giant planet satellite systems in our own solar system,
a similar evection-resonance destabilisation
would likely not have occurred recently, if at all.
In the case of Jupiter, where the evection resonance lies between Io and Europa%
\footnote{
    This information for these and the other giant and ice giant planet systems
    can be obtained from the JPL Planetary Satellite Mean Elements table
    at \href{https://ssd.jpl.nasa.gov}{ssd.jpl.nasa.gov}.
},
the resonant configuration of the inner three Galilean moons
has long been known to be stable \citep{Laplace+1829} and is likely ancient,
perhaps as old as the Solar System \citep{Goldreich+Soter1966}.
At Uranus, evection lies between Ariel and Umbriel,
neither of which are likely to have crossed
owing to the planet's large tidal $Q$ \citep{Cuk+2020}.
Lastly, whether such a destabilisation could have happened at Neptune is unknown,
since any pre-existing satellite system can be expected to have been disrupted
owing to to the capture of Triton \citep{Goldreich+1989,Agnor+Hamilton2006}.

\section{Conclusions} \label{sec:conclusions}

Dynamical instabilities in a precursor system of Saturnian mid-sized icy moons
can lead to high-velocity collisions
that scatter fragments and debris throughout the system \citep{Cuk+2016b}.
Using SPH simulations with over two orders of magnitude
higher resolution than previous studies,
we find that a range of plausible impacts
can deliver significant mass directly inside the Roche limit,
with a ring-like composition of pure ice.
Furthermore, more than a Mimas mass of material
-- and even more than an Enceladus mass in some cases --
is placed onto crossing orbits with present-day
Mimas, Enceladus, and Tethys (and Titan),
facilitating the possibility of a collisional cascade
to further distribute material across the system.

Head-on to mid-angle collisions produce large numbers of cohesive fragments
and re-accreted objects with masses above $10^{18}$--$10^{19}$~kg
($\sim$0.1--1 Mimas masses).
Most of these are -- at 9~hours after impact --
primarily composed of rock in a more diffuse field of ejected ice.
Grazing impacts produce little debris
but the colliding moons' orbits are also negligibly changed,
so the scenario retains the opportunity of a more disruptive collision
on a future orbit as the satellites continue to cross paths.
Slower collisions are less effective at distributing material,
but they can still send small amounts of debris across the Roche limit,
and send significantly more that would intersect with other precursor moons
to potentially cause further disruption.
Impacts between two equal-sized moons yield similar results and trends.

Even before considering subsequent collisions
and future modelling of the further creation and redistribution of debris,
we find that the example impacts considered here
can place over $5 \times 10^{19}$~kg of ice onto Roche-crossing orbits (and no rock),
and $\sim$$10^{18}$--$10^{19}$~kg of pure ice with
equivalent circular orbits inside the Roche limit.
However, that these masses are factors of a few
above and below the mass of the present-day rings
is less important for this proof-of-concept study
than the qualitative placement of significant masses onto Roche- and moon-crossing orbits
across a range of impact scenarios
-- to which the production of immediately Roche-circularising material
is an encouraging bonus.
Furthermore, the mass on Roche-crossing orbits scales close to proportionally
with the mass of the colliding moons,
so the larger moons that might be expected in a precursor system
could more readily produce massive rings.

The mass of debris entering the Roche limit
converges for moderately high numerical resolutions in most cases,
while the mass distribution requires at least $\sim$$10^7$ SPH particles
to converge reliably on the population of mid-to-large fragments,
let alone smaller objects.

We conclude that the impact of two destabilised icy moons is a promising scenario
for the recent formation or rejuvenation of Saturn's rings
and re-accretion of mid-sized moons.
Future work on the long-term evolution of the orbit-crossing debris,
combined with further and more detailed modelling
of collisions between both icy moons and smaller fragments,
will help to constrain the implications of this scenario
for Saturn's rings, its moons, their craters,
and other surface environments.

\vfill

\acknowledgments

J.A.K. acknowledges support from a NASA Postdoctoral Program Fellowship,
administered by Oak Ridge Associated Universities.
J.N.C. was supported for part of this work by the Cassini project,
and partly by a NASA Cassini Data Analysis Program grant to P.R.E.
We thank K. Zahnle for valuable input and discussion.
We also thank A. Rhoden for an early look at a pending chapter
on cratering evidence for icy moon evolution.
L.F.A.T. would like to dedicate this article to the memory of Michael S. Warren.
We thank the anonymous reviewer for their helpful comments.
The research in this paper made use of the \swift open-source simulation code
\citep{Schaller+2018,Schaller+2023}, version 0.9.0.
This work was supported by Science and Technology Facilities Council (STFC)
grants ST/P000541/1, ST/T000244/1, and ST/X001075/1,
and used the EPSRC funded ARCHIE-WeSt High Performance Computer
(www.archie-west.ac.uk), EPSRC grant EP/K000586/1,
and the DiRAC@Durham facility managed by the
Institute for Computational Cosmology
on behalf of the STFC DiRAC HPC Facility (www.dirac.ac.uk).
This equipment was funded by BEIS via STFC capital grants
ST/K00042X/1, ST/P002293/1, ST/R002371/1 and ST/S002502/1,
Durham University and STFC operations grant ST/R000832/1.
DiRAC is part of the National e-Infrastructure.

%



\software{
  \swift (\href{www.swiftsim.com}{www.swiftsim.com},
  \citealt{Schaller+2018,Schaller+2023}, version 0.9.0);
  \woma (\href{https://pypi.org/project/woma/}{pypi.org/project/woma/},
  \citealt{RuizBonilla+2021}).
}
\vfill




\begin{figure*}[t]
  \centering
  \includegraphics[
  width=0.495\textwidth, trim={18mm 8mm 25mm 7mm}, clip]{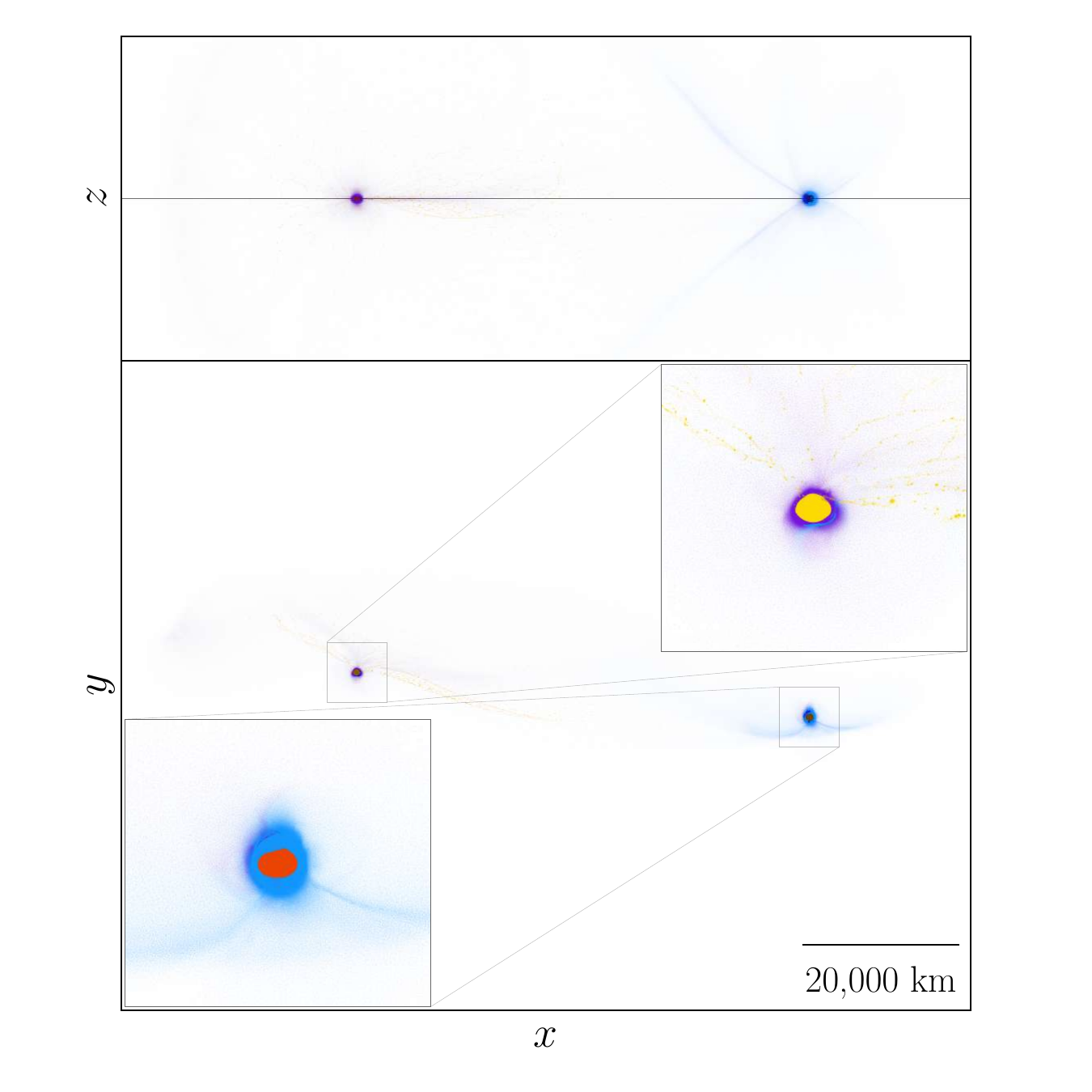}
  \includegraphics[
  width=0.495\textwidth, trim={18mm 8mm 25mm 7mm}, clip]{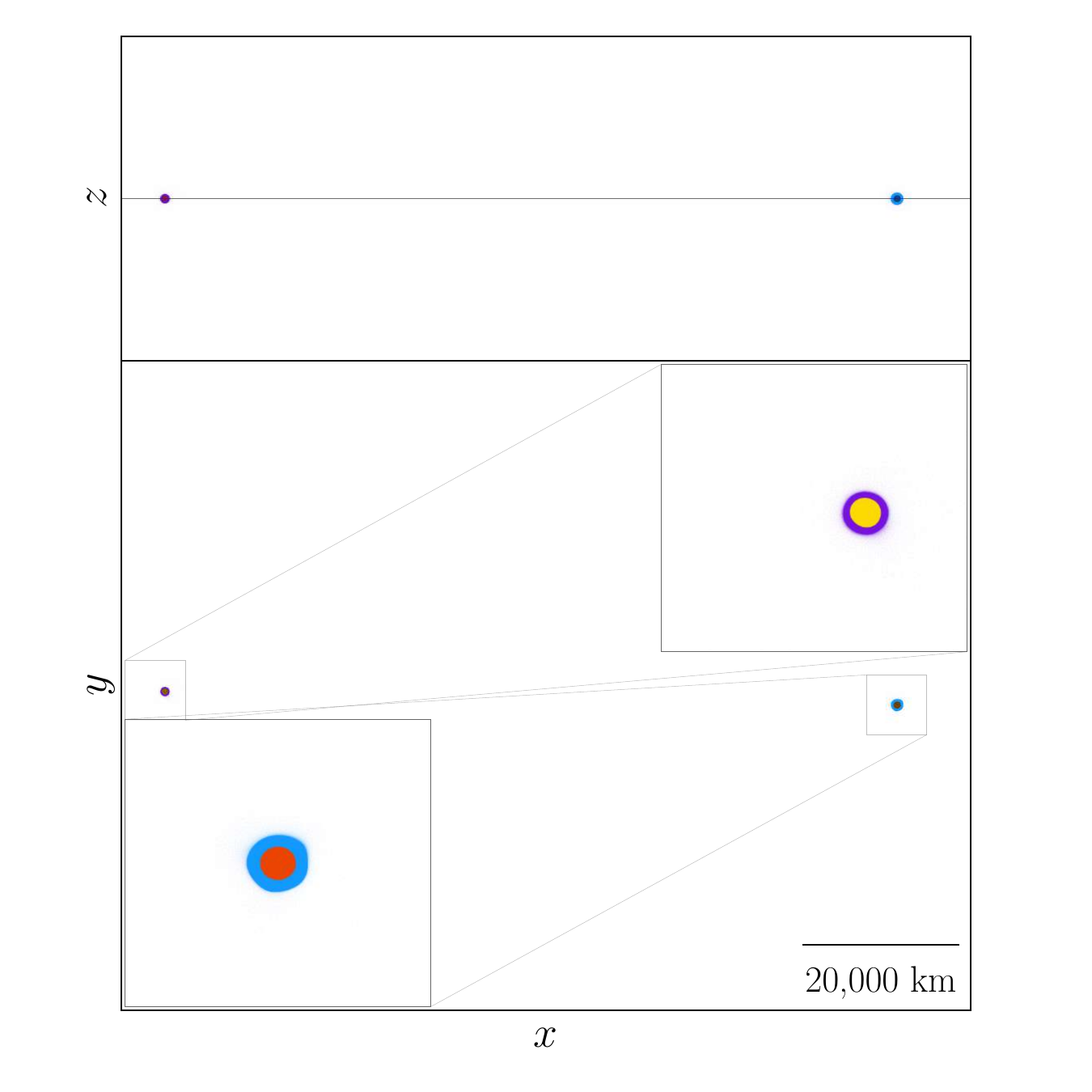}
  \\\vspace{-0.5em}
  \caption{
  Illustrative final snapshots from simulations with $10^{7.5}$ particles
  for impact angles and speeds at infinity of
  (left) $\beta = 45^\circ$, $v_{\infty} = 2.0$~km~s$^{-1}$ and
  (right) $\beta = 75^\circ$, $v_{\infty} = 3.0$~km~s$^{-1}$,
  shown as in Fig.~\ref{fig:final_snaps}.
  \label{fig:final_snaps_graze}}
  \vspace{0.5em}
\end{figure*}

\appendix

\section{Extended results for other impact scenarios}  \label{appx:extended_results}

\begin{figure}[h]
	\centering
	\includegraphics[
    width=\columnwidth, trim={7mm 7mm 6mm 6mm}, clip]{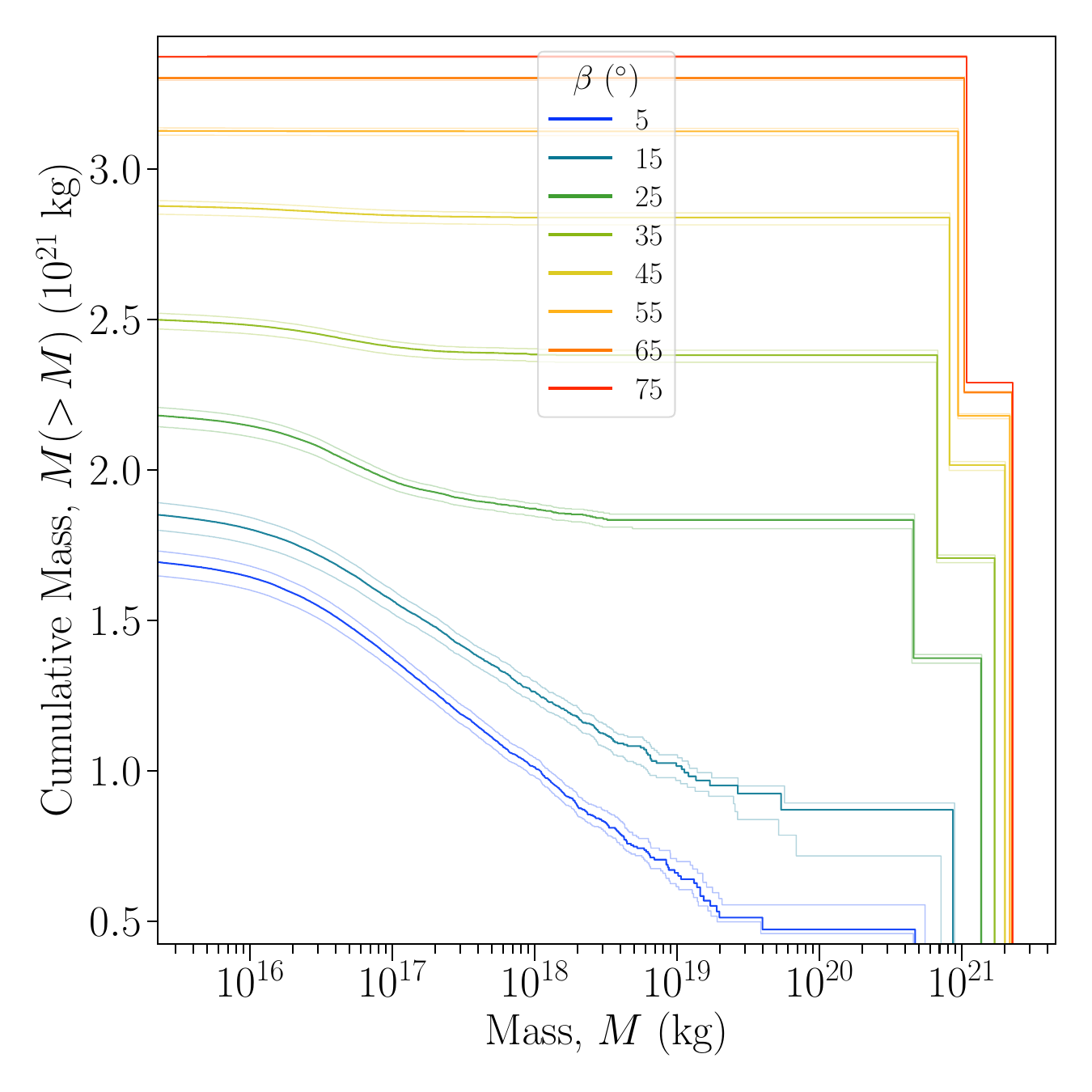}
  \\\vspace{-0.5em}
  \caption{
    The cumulative mass distributions of objects
    produced by impacts with $v_{\infty} = 2$~km~s$^{-1}$,
    shown as in Fig.~\ref{fig:mass_function}.
    \label{fig:mass_function_2kms}}
  \vspace{-1em}
\end{figure}

For comparison with the head-on to mid-angle impacts in Fig.~\ref{fig:final_snaps},
example outcomes for more grazing collisions
are illustrated in Fig.~\ref{fig:final_snaps_graze}.
Less debris is ejected, and the mostly intact remnants of the two moons
continue on little-changed trajectories
so are likely to re-collide on a future orbit,
as discussed in the main text.

The mass distributions from slower impacts
are shown in Fig.~\ref{fig:mass_function_2kms},
for comparison with the faster ones in Fig.~\ref{fig:mass_function}.
The general outcomes and trends are similar,
but the debris become dominated by the two remnants of the initial moons
at a less grazing impact angle.

\begin{figure}[h]
	\centering
	\includegraphics[
    width=\columnwidth, trim={7.5mm 8mm 7mm 17mm}, clip]{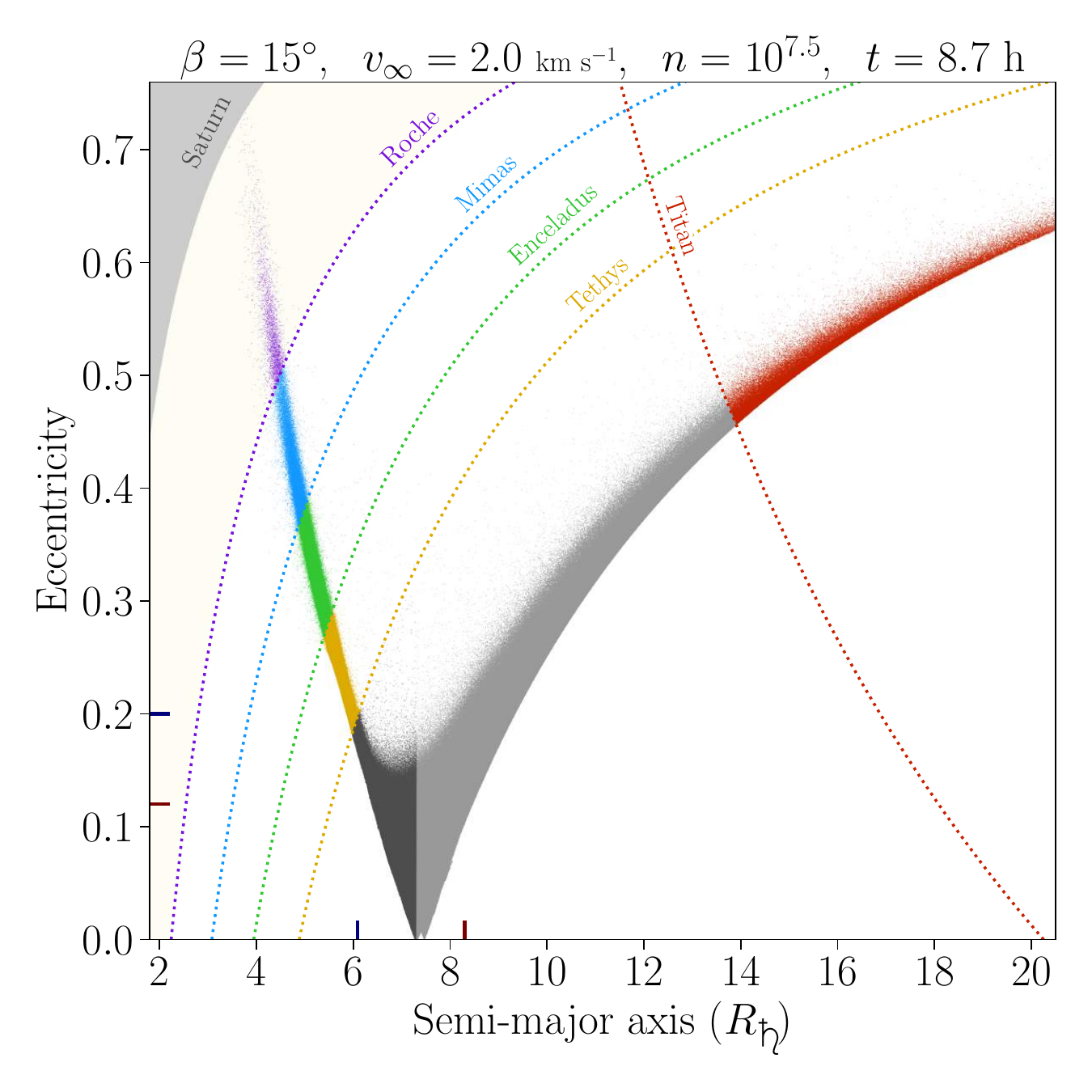}
  \\\vspace{-0.5em}
  \caption{
    The eccentricities and semi-major axes of the orbiting debris,
    in the final snapshot of an example
    $\beta = 15^\circ$, $v_{\infty} = 2.0$~km~s$^{-1}$ impact,
    as illustrated in Fig.~\ref{fig:final_snaps},
    shown as in Fig.~\ref{fig:e_a}.
    \label{fig:e_a_B15v20}}
  \vspace{-1em}
\end{figure}

\begin{figure}[h]
	\centering
	\includegraphics[
    width=\columnwidth, trim={8mm 8mm 7mm 7mm}, clip]{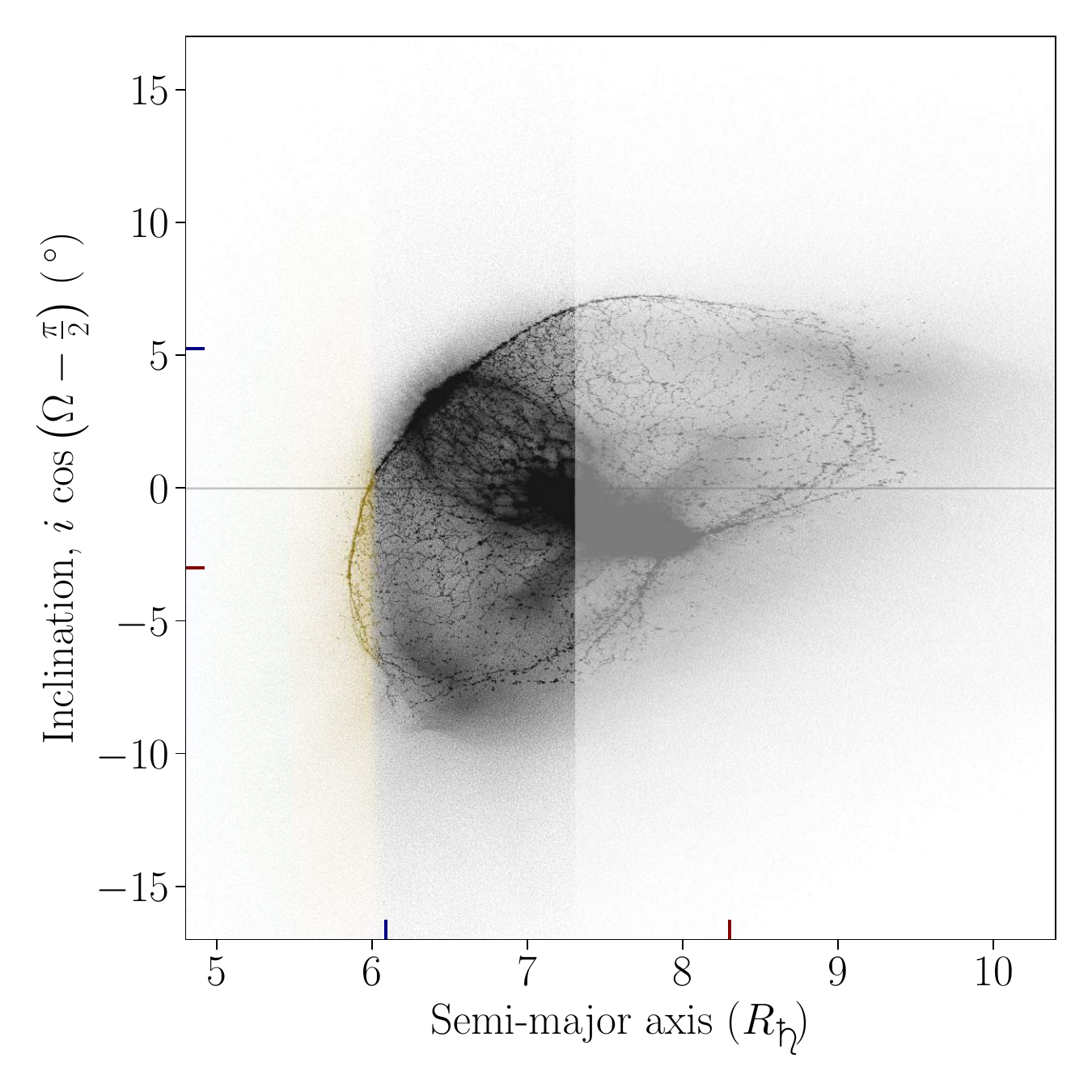}
  \\\vspace{-0.5em}
  \caption{
    The inclinations of the orbiting debris
    in the final snapshot of an example
    $\beta = 15^\circ$, $v_{\infty} = 2.0$~km~s$^{-1}$ impact,
    as illustrated in Fig.~\ref{fig:final_snaps},
    shown as in Fig.~\ref{fig:i_a}.
    \label{fig:i_a_B15v20}}
  \vspace{-1em}
\end{figure}

The resulting Saturn-frame orbits of the debris
from the slower, more head-on scenario
illustrated in Fig.~\ref{fig:final_snaps}(left)
are shown in Figs.~\ref{fig:e_a_B15v20} and~\ref{fig:i_a_B15v20}.
The overall results are similar to the faster, higher-angle example
of Figs.~\ref{fig:e_a} and~\ref{fig:i_a}.
The debris is more spread out around the impact point,
which is representative for more head-on collisions in general,
with somewhat less material reaching the Roche limit,
which is representative for lower speeds in general.
The inclinations are also significantly less high,
because the initial moons' material has been more thoroughly mixed,
and the debris therefore lies on orbits closer to the near-equatorial one
of the centre of mass.


\section{Orbital elements}  \label{appx:orbits}

After the SPH particle data are transformed into Saturn's frame,
as described in \S\ref{sec:methods:frame_trans},
we estimate the orbit of each particle or fragment
from its position, $\vec{r}$, and velocity, $\vec{v}$.
The vis-viva equation yields the semi-major axis, $a$:
\begin{equation}
  a = \left(\dfrac{2}{r} - \dfrac{v^2}{\mu}\right)^{-1} \;,
\end{equation}
where $r = |\vec{r}|$, $v = |\vec{v}|$, and $\mu = G (M_\Sa + m)$,
with $m$ the particle or fragment's mass
and $G$ the gravitational constant.

The eccentricity, $e$, and inclination, $i$,
are then derived from the specific angular momentum,
$\vec{h} = \vec{r} \times \vec{v}$:
\begin{align}
  \vec{e} &= \dfrac{\vec{v} \times \vec{h}}{\mu} - \dfrac{\vec{r}}{r} \;, \nonumber \\
  i &= \cos^{-1}\left( \dfrac{h_z}{h} \right) \;.
\end{align}
The longitude of ascending node, $\Omega$, is then
\begin{equation}
  \Omega = \cos^{-1}\left( \dfrac{-h_x}{\sqrt{h_x^2 + h_y^2}} \right) \;,
\end{equation}
where the sign of $\Omega$ is flipped if $h_x < 0$.

\vspace{2em}

\section{Results tables}  \label{appx:tables}

The primary results from all simulations are listed in Table~\ref{tab:results},
arranged in groups by subset, impact angle, and speed.
The SPH simulation data will be made available on reasonable request.
Table~\ref{tab:results} is available in the online version
in a machine-readable format.


\onecolumngrid
\begin{center}
  {
  \small
  \begin{longtable}{crc | ccccc rrrrr}
    \caption{
      Outcomes of icy-moon impact simulations.
      Columns list the impact angle, $\beta$;
      speed at infinity, $v_\infty$;
      the mass of debris on orbits crossing the Roche limit and present-day moons;
      and the fraction of which is ice.
      For the base suite of simulations,
      each entry is followed by three more lines
      with the SPH frame rotated about the $x$ axis
      by $\phi = 90^\circ, 180^\circ, 270^\circ$ (see Fig.\ref{fig:frames_diagram}).
      The `type' notes other parameters that are changed from the base scenario
      for the following subsets of simulations,
      as described in \S\ref{sec:methods:impact_scenarios},
      within which the angle and speed are then varied, namely:
      the number of particles, $N$ (base $10^{7.5}$);
      the body colliding with p-Rhea (base p-Dione);
      the EoS (base Tillotson),
      and the mass of p-Dione for the separate tests of moon masses,
      $M_{\rm pD}$ in units of $10^{21}$~kg,
      with corresponding p-Rhea mass $M_{\rm pR} = 2 \, M_{\rm pD}$.
      Note: the orbit-crossing masses
      include only the mass that reaches that orbit,
      and exclude material that also reaches further ones
      (e.g.\ the Enceladus values do not include Mimas- or Roche-crossing material).
    } \label{tab:results} \\

    \hline
    \hline
     & $\beta\;$ & $v_\infty$ %
    & \multicolumn{5}{c}{Mass (kg)}
    & \multicolumn{5}{c}{Ice fraction (\%)}
    \\
    Type & $(^\circ)$ & (km s$^{-1}$) %
    & Te. & En. & Mi. & Ro. & Ti.
    & Te. & En. & Mi. & Ro. & Ti.
    \\
    \hline\\[-1em]  
    \endfirsthead

    \multicolumn{4}{l}{\textbf{\tablename\ \thetable{}} -- continued} \\
    \hline
     & $\beta\;$ & $v_\infty$ %
    & \multicolumn{5}{c}{Mass (kg)}
    & \multicolumn{5}{c}{Ice fraction (\%)}
    \\
    Type & $(^\circ)$ & (km s$^{-1}$) %
    & Te. & En. & Mi. & Ro. & Ti.
    & Te. & En. & Mi. & Ro. & Ti.
    \\
    \hline\\[-1em]  
    \endhead

    \hline
    \endfoot

    \hline
    \hline
    \endlastfoot

    \multirow{2}{*}{base} & $5$ & $2.0$ & $2.1 \!\times\! 10^{19}$ & $7.4 \!\times\! 10^{17}$ & $7.5 \!\times\! 10^{16}$ & $1.3 \!\times\! 10^{16}$ & $1.5 \!\times\! 10^{19}$ & $68$ & $100$ & $100$ & $100$ & $100$  \\
       & & & $1.1 \!\times\! 10^{20}$ & $1.2 \!\times\! 10^{19}$ & $1.9 \!\times\! 10^{18}$ & $1.8 \!\times\! 10^{17}$ & $2.2 \!\times\! 10^{19}$ & $52$ & $100$ & $100$ & $100$ & $100$  \\
       & & & $5.0 \!\times\! 10^{19}$ & $3.6 \!\times\! 10^{18}$ & $3.7 \!\times\! 10^{17}$ & $4.4 \!\times\! 10^{16}$ & $1.8 \!\times\! 10^{19}$ & $62$ & $100$ & $100$ & $100$ & $100$  \\
       & & & $1.8 \!\times\! 10^{17}$ & $2.9 \!\times\! 10^{16}$ & $1.2 \!\times\! 10^{16}$ & $1.1 \!\times\! 10^{15}$ & $8.4 \!\times\! 10^{18}$ & $100$ & $100$ & $100$ & $100$ & $100$  \\
       & $15$ & $2.0$ & $5.5 \!\times\! 10^{19}$ & $9.9 \!\times\! 10^{18}$ & $2.6 \!\times\! 10^{18}$ & $3.3 \!\times\! 10^{17}$ & $1.1 \!\times\! 10^{19}$ & $65$ & $100$ & $100$ & $100$ & $100$  \\
       & & & $3.5 \!\times\! 10^{20}$ & $3.6 \!\times\! 10^{19}$ & $1.9 \!\times\! 10^{19}$ & $6.4 \!\times\! 10^{18}$ & $1.7 \!\times\! 10^{19}$ & $44$ & $100$ & $100$ & $100$ & $100$  \\
       & & & $1.0 \!\times\! 10^{20}$ & $1.8 \!\times\! 10^{19}$ & $6.7 \!\times\! 10^{18}$ & $1.2 \!\times\! 10^{18}$ & $1.6 \!\times\! 10^{19}$ & $69$ & $100$ & $100$ & $100$ & $100$  \\
       & & & $8.6 \!\times\! 10^{16}$ & $1.4 \!\times\! 10^{16}$ & $1.0 \!\times\! 10^{15}$ & $0.0$ & $5.1 \!\times\! 10^{17}$ & $100$ & $100$ & $100$ & $0$ & $100$  \\
       & $25$ & $2.0$ & $7.8 \!\times\! 10^{19}$ & $2.3 \!\times\! 10^{19}$ & $8.5 \!\times\! 10^{18}$ & $1.1 \!\times\! 10^{18}$ & $5.7 \!\times\! 10^{18}$ & $92$ & $100$ & $100$ & $100$ & $100$  \\
       & & & $4.8 \!\times\! 10^{20}$ & $6.3 \!\times\! 10^{19}$ & $4.1 \!\times\! 10^{19}$ & $6.3 \!\times\! 10^{18}$ & $1.7 \!\times\! 10^{18}$ & $40$ & $88$ & $100$ & $100$ & $100$  \\
       & & & $1.1 \!\times\! 10^{20}$ & $3.7 \!\times\! 10^{19}$ & $1.5 \!\times\! 10^{19}$ & $2.4 \!\times\! 10^{18}$ & $6.9 \!\times\! 10^{18}$ & $76$ & $100$ & $100$ & $100$ & $100$  \\
       & & & $2.7 \!\times\! 10^{18}$ & $2.8 \!\times\! 10^{17}$ & $2.5 \!\times\! 10^{16}$ & $0.0$ & $3.7 \!\times\! 10^{17}$ & $100$ & $100$ & $100$ & $0$ & $100$  \\
       & $35$ & $2.0$ & $1.3 \!\times\! 10^{20}$ & $3.7 \!\times\! 10^{19}$ & $1.0 \!\times\! 10^{19}$ & $7.6 \!\times\! 10^{17}$ & $2.0 \!\times\! 10^{18}$ & $72$ & $99$ & $100$ & $100$ & $100$  \\
       & & & $8.5 \!\times\! 10^{20}$ & $6.4 \!\times\! 10^{19}$ & $3.3 \!\times\! 10^{19}$ & $4.6 \!\times\! 10^{17}$ & $3.8 \!\times\! 10^{17}$ & $34$ & $81$ & $100$ & $100$ & $100$  \\
       & & & $1.7 \!\times\! 10^{20}$ & $4.1 \!\times\! 10^{19}$ & $1.8 \!\times\! 10^{19}$ & $1.2 \!\times\! 10^{18}$ & $1.8 \!\times\! 10^{18}$ & $57$ & $93$ & $100$ & $100$ & $100$  \\
       & & & $3.5 \!\times\! 10^{19}$ & $2.6 \!\times\! 10^{18}$ & $1.4 \!\times\! 10^{17}$ & $4.1 \!\times\! 10^{15}$ & $2.2 \!\times\! 10^{17}$ & $100$ & $100$ & $100$ & $100$ & $100$  \\
       & $45$ & $2.0$ & $9.3 \!\times\! 10^{20}$ & $3.3 \!\times\! 10^{19}$ & $3.9 \!\times\! 10^{18}$ & $2.1 \!\times\! 10^{17}$ & $4.9 \!\times\! 10^{17}$ & $34$ & $100$ & $100$ & $100$ & $100$  \\
       & & & $9.5 \!\times\! 10^{20}$ & $5.0 \!\times\! 10^{19}$ & $4.7 \!\times\! 10^{17}$ & $1.2 \!\times\! 10^{17}$ & $9.2 \!\times\! 10^{16}$ & $34$ & $97$ & $100$ & $100$ & $100$  \\
       & & & $9.3 \!\times\! 10^{20}$ & $3.6 \!\times\! 10^{19}$ & $4.4 \!\times\! 10^{18}$ & $2.4 \!\times\! 10^{17}$ & $4.1 \!\times\! 10^{17}$ & $33$ & $98$ & $100$ & $100$ & $100$  \\
       & & & $6.0 \!\times\! 10^{20}$ & $8.6 \!\times\! 10^{18}$ & $1.8 \!\times\! 10^{17}$ & $7.4 \!\times\! 10^{15}$ & $9.2 \!\times\! 10^{16}$ & $45$ & $100$ & $100$ & $100$ & $100$  \\
       & $55$ & $2.0$ & $2.4 \!\times\! 10^{19}$ & $3.5 \!\times\! 10^{18}$ & $3.8 \!\times\! 10^{17}$ & $4.6 \!\times\! 10^{16}$ & $1.3 \!\times\! 10^{17}$ & $100$ & $100$ & $100$ & $100$ & $100$  \\
       & & & $1.0 \!\times\! 10^{21}$ & $7.6 \!\times\! 10^{17}$ & $8.1 \!\times\! 10^{16}$ & $2.6 \!\times\! 10^{16}$ & $2.4 \!\times\! 10^{16}$ & $37$ & $100$ & $100$ & $100$ & $100$  \\
       & & & $1.0 \!\times\! 10^{21}$ & $9.9 \!\times\! 10^{18}$ & $4.8 \!\times\! 10^{17}$ & $4.8 \!\times\! 10^{16}$ & $7.1 \!\times\! 10^{16}$ & $36$ & $100$ & $100$ & $100$ & $100$  \\
       & & & $1.0 \!\times\! 10^{21}$ & $3.8 \!\times\! 10^{18}$ & $1.0 \!\times\! 10^{17}$ & $2.9 \!\times\! 10^{15}$ & $2.8 \!\times\! 10^{16}$ & $36$ & $100$ & $100$ & $100$ & $100$  \\
       & $65$ & $2.0$ & $1.1 \!\times\! 10^{21}$ & $1.1 \!\times\! 10^{18}$ & $5.8 \!\times\! 10^{16}$ & $9.1 \!\times\! 10^{15}$ & $1.4 \!\times\! 10^{16}$ & $38$ & $100$ & $100$ & $100$ & $100$  \\
       & & & $1.1 \!\times\! 10^{21}$ & $7.8 \!\times\! 10^{16}$ & $1.2 \!\times\! 10^{16}$ & $4.4 \!\times\! 10^{15}$ & $2.6 \!\times\! 10^{15}$ & $38$ & $100$ & $100$ & $100$ & $100$  \\
       & & & $1.1 \!\times\! 10^{21}$ & $8.8 \!\times\! 10^{17}$ & $4.0 \!\times\! 10^{16}$ & $1.1 \!\times\! 10^{16}$ & $1.2 \!\times\! 10^{16}$ & $38$ & $100$ & $100$ & $100$ & $100$  \\
       & & & $1.1 \!\times\! 10^{21}$ & $6.4 \!\times\! 10^{17}$ & $2.9 \!\times\! 10^{16}$ & $2.2 \!\times\! 10^{14}$ & $4.6 \!\times\! 10^{15}$ & $38$ & $100$ & $100$ & $100$ & $100$  \\
       & $75$ & $2.0$ & $1.1 \!\times\! 10^{21}$ & $9.2 \!\times\! 10^{16}$ & $5.3 \!\times\! 10^{15}$ & $1.1 \!\times\! 10^{14}$ & $5.6 \!\times\! 10^{14}$ & $39$ & $100$ & $100$ & $100$ & $100$  \\
       & & & $1.1 \!\times\! 10^{21}$ & $4.4 \!\times\! 10^{15}$ & $0.0$ & $0.0$ & $0.0$ & $39$ & $100$ & $0$ & $0$ & $0$  \\
       & & & $1.1 \!\times\! 10^{21}$ & $5.6 \!\times\! 10^{16}$ & $4.2 \!\times\! 10^{15}$ & $1.1 \!\times\! 10^{14}$ & $1.1 \!\times\! 10^{14}$ & $39$ & $100$ & $100$ & $100$ & $100$  \\
       & & & $1.1 \!\times\! 10^{21}$ & $1.3 \!\times\! 10^{17}$ & $3.6 \!\times\! 10^{15}$ & $0.0$ & $0.0$ & $39$ & $100$ & $100$ & $0$ & $0$  \\
       & $5$ & $3.0$ & $2.3 \!\times\! 10^{20}$ & $1.0 \!\times\! 10^{20}$ & $1.6 \!\times\! 10^{19}$ & $8.8 \!\times\! 10^{17}$ & $1.1 \!\times\! 10^{20}$ & $70$ & $71$ & $99$ & $100$ & $96$  \\
       & & & $2.6 \!\times\! 10^{20}$ & $1.6 \!\times\! 10^{20}$ & $6.9 \!\times\! 10^{19}$ & $1.4 \!\times\! 10^{19}$ & $1.2 \!\times\! 10^{20}$ & $59$ & $63$ & $67$ & $100$ & $95$  \\
       & & & $2.4 \!\times\! 10^{20}$ & $1.3 \!\times\! 10^{20}$ & $2.5 \!\times\! 10^{19}$ & $2.9 \!\times\! 10^{18}$ & $1.1 \!\times\! 10^{20}$ & $66$ & $67$ & $96$ & $100$ & $96$  \\
       & & & $1.6 \!\times\! 10^{20}$ & $2.2 \!\times\! 10^{19}$ & $8.1 \!\times\! 10^{16}$ & $4.9 \!\times\! 10^{16}$ & $9.1 \!\times\! 10^{19}$ & $84$ & $99$ & $100$ & $100$ & $100$  \\
       & $15$ & $3.0$ & $2.3 \!\times\! 10^{20}$ & $1.0 \!\times\! 10^{20}$ & $2.1 \!\times\! 10^{19}$ & $7.1 \!\times\! 10^{18}$ & $7.9 \!\times\! 10^{19}$ & $58$ & $71$ & $99$ & $100$ & $100$  \\
       & & & $3.7 \!\times\! 10^{20}$ & $2.6 \!\times\! 10^{20}$ & $7.9 \!\times\! 10^{19}$ & $4.7 \!\times\! 10^{19}$ & $1.0 \!\times\! 10^{20}$ & $40$ & $42$ & $98$ & $100$ & $100$  \\
       & & & $2.7 \!\times\! 10^{20}$ & $1.3 \!\times\! 10^{20}$ & $3.6 \!\times\! 10^{19}$ & $1.3 \!\times\! 10^{19}$ & $9.0 \!\times\! 10^{19}$ & $50$ & $68$ & $100$ & $100$ & $100$  \\
       & & & $5.8 \!\times\! 10^{18}$ & $5.2 \!\times\! 10^{16}$ & $2.8 \!\times\! 10^{16}$ & $1.1 \!\times\! 10^{16}$ & $4.5 \!\times\! 10^{19}$ & $100$ & $100$ & $100$ & $100$ & $100$  \\
       & $25$ & $3.0$ & $2.3 \!\times\! 10^{20}$ & $7.5 \!\times\! 10^{19}$ & $3.0 \!\times\! 10^{19}$ & $1.1 \!\times\! 10^{19}$ & $4.4 \!\times\! 10^{19}$ & $48$ & $90$ & $100$ & $100$ & $100$  \\
       & & & $6.3 \!\times\! 10^{20}$ & $1.6 \!\times\! 10^{20}$ & $7.7 \!\times\! 10^{19}$ & $5.7 \!\times\! 10^{19}$ & $2.5 \!\times\! 10^{19}$ & $31$ & $58$ & $78$ & $98$ & $100$  \\
       & & & $3.4 \!\times\! 10^{20}$ & $9.4 \!\times\! 10^{19}$ & $4.5 \!\times\! 10^{19}$ & $1.8 \!\times\! 10^{19}$ & $4.9 \!\times\! 10^{19}$ & $41$ & $75$ & $100$ & $100$ & $100$  \\
       & & & $1.9 \!\times\! 10^{18}$ & $1.0 \!\times\! 10^{17}$ & $1.3 \!\times\! 10^{16}$ & $2.2 \!\times\! 10^{14}$ & $1.9 \!\times\! 10^{18}$ & $100$ & $100$ & $100$ & $100$ & $100$  \\
       & $35$ & $3.0$ & $2.3 \!\times\! 10^{20}$ & $5.5 \!\times\! 10^{19}$ & $2.8 \!\times\! 10^{19}$ & $8.8 \!\times\! 10^{18}$ & $1.7 \!\times\! 10^{19}$ & $47$ & $78$ & $100$ & $100$ & $100$  \\
       & & & $7.4 \!\times\! 10^{20}$ & $1.8 \!\times\! 10^{20}$ & $6.6 \!\times\! 10^{19}$ & $4.4 \!\times\! 10^{18}$ & $6.6 \!\times\! 10^{17}$ & $32$ & $57$ & $88$ & $100$ & $100$  \\
       & & & $5.2 \!\times\! 10^{20}$ & $6.6 \!\times\! 10^{19}$ & $3.3 \!\times\! 10^{19}$ & $1.4 \!\times\! 10^{19}$ & $1.6 \!\times\! 10^{19}$ & $36$ & $72$ & $96$ & $100$ & $100$  \\
       & & & $1.2 \!\times\! 10^{19}$ & $7.6 \!\times\! 10^{17}$ & $1.1 \!\times\! 10^{17}$ & $4.4 \!\times\! 10^{15}$ & $8.2 \!\times\! 10^{17}$ & $100$ & $100$ & $100$ & $100$ & $100$  \\
       & $45$ & $3.0$ & $8.6 \!\times\! 10^{20}$ & $3.5 \!\times\! 10^{19}$ & $1.9 \!\times\! 10^{19}$ & $2.6 \!\times\! 10^{18}$ & $4.3 \!\times\! 10^{18}$ & $32$ & $88$ & $100$ & $100$ & $100$  \\
       & & & $9.3 \!\times\! 10^{20}$ & $8.2 \!\times\! 10^{19}$ & $5.6 \!\times\! 10^{17}$ & $1.4 \!\times\! 10^{17}$ & $1.6 \!\times\! 10^{17}$ & $33$ & $89$ & $100$ & $100$ & $100$  \\
       & & & $8.8 \!\times\! 10^{20}$ & $4.0 \!\times\! 10^{19}$ & $2.2 \!\times\! 10^{19}$ & $2.8 \!\times\! 10^{18}$ & $3.5 \!\times\! 10^{18}$ & $32$ & $85$ & $100$ & $100$ & $100$  \\
       & & & $3.6 \!\times\! 10^{19}$ & $3.4 \!\times\! 10^{18}$ & $2.9 \!\times\! 10^{17}$ & $1.7 \!\times\! 10^{16}$ & $3.6 \!\times\! 10^{17}$ & $99$ & $100$ & $100$ & $100$ & $100$  \\
       & $55$ & $3.0$ & $9.9 \!\times\! 10^{20}$ & $2.4 \!\times\! 10^{19}$ & $4.1 \!\times\! 10^{18}$ & $3.8 \!\times\! 10^{17}$ & $7.0 \!\times\! 10^{17}$ & $33$ & $100$ & $100$ & $100$ & $100$  \\
       & & & $1.0 \!\times\! 10^{21}$ & $5.6 \!\times\! 10^{17}$ & $7.1 \!\times\! 10^{16}$ & $2.7 \!\times\! 10^{16}$ & $2.9 \!\times\! 10^{16}$ & $37$ & $100$ & $100$ & $100$ & $100$  \\
       & & & $9.9 \!\times\! 10^{20}$ & $2.5 \!\times\! 10^{19}$ & $4.1 \!\times\! 10^{18}$ & $3.3 \!\times\! 10^{17}$ & $4.7 \!\times\! 10^{17}$ & $33$ & $100$ & $100$ & $100$ & $100$  \\
       & & & $9.8 \!\times\! 10^{20}$ & $3.9 \!\times\! 10^{18}$ & $2.9 \!\times\! 10^{17}$ & $1.6 \!\times\! 10^{16}$ & $1.0 \!\times\! 10^{17}$ & $33$ & $100$ & $100$ & $100$ & $100$  \\
       & $65$ & $3.0$ & $1.1 \!\times\! 10^{21}$ & $3.9 \!\times\! 10^{18}$ & $3.9 \!\times\! 10^{17}$ & $3.5 \!\times\! 10^{16}$ & $7.0 \!\times\! 10^{16}$ & $38$ & $100$ & $100$ & $100$ & $100$  \\
       & & & $1.1 \!\times\! 10^{21}$ & $4.7 \!\times\! 10^{16}$ & $1.1 \!\times\! 10^{16}$ & $8.9 \!\times\! 10^{14}$ & $3.0 \!\times\! 10^{15}$ & $38$ & $100$ & $100$ & $100$ & $100$  \\
       & & & $1.1 \!\times\! 10^{21}$ & $3.7 \!\times\! 10^{18}$ & $3.1 \!\times\! 10^{17}$ & $3.0 \!\times\! 10^{16}$ & $4.3 \!\times\! 10^{16}$ & $38$ & $100$ & $100$ & $100$ & $100$  \\
       & & & $1.1 \!\times\! 10^{21}$ & $1.2 \!\times\! 10^{18}$ & $1.6 \!\times\! 10^{17}$ & $6.0 \!\times\! 10^{15}$ & $2.5 \!\times\! 10^{16}$ & $38$ & $100$ & $100$ & $100$ & $100$  \\
       & $75$ & $3.0$ & $1.1 \!\times\! 10^{21}$ & $3.4 \!\times\! 10^{17}$ & $2.2 \!\times\! 10^{16}$ & $2.1 \!\times\! 10^{15}$ & $4.6 \!\times\! 10^{15}$ & $39$ & $100$ & $100$ & $100$ & $100$  \\
       & & & $1.1 \!\times\! 10^{21}$ & $2.9 \!\times\! 10^{15}$ & $0.0$ & $0.0$ & $0.0$ & $39$ & $100$ & $0$ & $0$ & $0$  \\
       & & & $1.1 \!\times\! 10^{21}$ & $2.9 \!\times\! 10^{17}$ & $1.4 \!\times\! 10^{16}$ & $2.1 \!\times\! 10^{15}$ & $1.9 \!\times\! 10^{15}$ & $39$ & $100$ & $100$ & $100$ & $100$  \\
       & & & $1.1 \!\times\! 10^{21}$ & $3.7 \!\times\! 10^{17}$ & $4.7 \!\times\! 10^{16}$ & $2.2 \!\times\! 10^{14}$ & $2.4 \!\times\! 10^{15}$ & $39$ & $100$ & $100$ & $100$ & $100$  \\
      \cline{0-0} \multirow{2}{*}{$N=10^{7}$} & $5$ & $2.0$ & $1.7 \!\times\! 10^{19}$ & $9.2 \!\times\! 10^{17}$ & $1.0 \!\times\! 10^{17}$ & $1.1 \!\times\! 10^{16}$ & $1.4 \!\times\! 10^{19}$ & $80$ & $100$ & $100$ & $100$ & $100$  \\
       & $15$ & $2.0$ & $5.4 \!\times\! 10^{19}$ & $9.8 \!\times\! 10^{18}$ & $2.4 \!\times\! 10^{18}$ & $3.0 \!\times\! 10^{17}$ & $1.0 \!\times\! 10^{19}$ & $67$ & $100$ & $100$ & $100$ & $100$  \\
       & $25$ & $2.0$ & $7.8 \!\times\! 10^{19}$ & $2.3 \!\times\! 10^{19}$ & $8.1 \!\times\! 10^{18}$ & $1.0 \!\times\! 10^{18}$ & $5.5 \!\times\! 10^{18}$ & $91$ & $100$ & $100$ & $100$ & $100$  \\
       & $35$ & $2.0$ & $1.3 \!\times\! 10^{20}$ & $3.6 \!\times\! 10^{19}$ & $9.9 \!\times\! 10^{18}$ & $7.3 \!\times\! 10^{17}$ & $1.9 \!\times\! 10^{18}$ & $71$ & $100$ & $100$ & $100$ & $100$  \\
       & $45$ & $2.0$ & $9.3 \!\times\! 10^{20}$ & $3.3 \!\times\! 10^{19}$ & $3.8 \!\times\! 10^{18}$ & $2.1 \!\times\! 10^{17}$ & $5.0 \!\times\! 10^{17}$ & $34$ & $100$ & $100$ & $100$ & $100$  \\
       & $55$ & $2.0$ & $1.0 \!\times\! 10^{21}$ & $9.9 \!\times\! 10^{18}$ & $5.9 \!\times\! 10^{17}$ & $5.0 \!\times\! 10^{16}$ & $9.0 \!\times\! 10^{16}$ & $36$ & $100$ & $100$ & $100$ & $100$  \\
       & $65$ & $2.0$ & $1.1 \!\times\! 10^{21}$ & $1.0 \!\times\! 10^{18}$ & $6.1 \!\times\! 10^{16}$ & $9.4 \!\times\! 10^{15}$ & $1.1 \!\times\! 10^{16}$ & $38$ & $100$ & $100$ & $100$ & $100$  \\
       & $75$ & $2.0$ & $1.1 \!\times\! 10^{21}$ & $9.1 \!\times\! 10^{16}$ & $3.7 \!\times\! 10^{15}$ & $0.0$ & $0.0$ & $39$ & $100$ & $100$ & $0$ & $0$  \\
       & $5$ & $3.0$ & $2.3 \!\times\! 10^{20}$ & $1.0 \!\times\! 10^{20}$ & $1.5 \!\times\! 10^{19}$ & $1.1 \!\times\! 10^{18}$ & $1.1 \!\times\! 10^{20}$ & $70$ & $73$ & $100$ & $100$ & $98$  \\
       & $15$ & $3.0$ & $2.3 \!\times\! 10^{20}$ & $1.0 \!\times\! 10^{20}$ & $2.1 \!\times\! 10^{19}$ & $6.9 \!\times\! 10^{18}$ & $7.8 \!\times\! 10^{19}$ & $59$ & $72$ & $100$ & $100$ & $100$  \\
       & $25$ & $3.0$ & $2.4 \!\times\! 10^{20}$ & $7.4 \!\times\! 10^{19}$ & $3.0 \!\times\! 10^{19}$ & $1.1 \!\times\! 10^{19}$ & $4.3 \!\times\! 10^{19}$ & $47$ & $91$ & $100$ & $100$ & $100$  \\
       & $35$ & $3.0$ & $2.2 \!\times\! 10^{20}$ & $5.4 \!\times\! 10^{19}$ & $2.8 \!\times\! 10^{19}$ & $8.6 \!\times\! 10^{18}$ & $1.7 \!\times\! 10^{19}$ & $48$ & $78$ & $100$ & $100$ & $100$  \\
       & $45$ & $3.0$ & $8.7 \!\times\! 10^{20}$ & $3.5 \!\times\! 10^{19}$ & $1.9 \!\times\! 10^{19}$ & $2.5 \!\times\! 10^{18}$ & $4.3 \!\times\! 10^{18}$ & $32$ & $89$ & $100$ & $100$ & $100$  \\
       & $55$ & $3.0$ & $9.9 \!\times\! 10^{20}$ & $2.3 \!\times\! 10^{19}$ & $4.2 \!\times\! 10^{18}$ & $3.9 \!\times\! 10^{17}$ & $7.1 \!\times\! 10^{17}$ & $33$ & $100$ & $100$ & $100$ & $100$  \\
       & $65$ & $3.0$ & $1.1 \!\times\! 10^{21}$ & $3.8 \!\times\! 10^{18}$ & $4.2 \!\times\! 10^{17}$ & $4.1 \!\times\! 10^{16}$ & $6.2 \!\times\! 10^{16}$ & $38$ & $100$ & $100$ & $100$ & $100$  \\
       & $75$ & $3.0$ & $1.1 \!\times\! 10^{21}$ & $3.6 \!\times\! 10^{17}$ & $2.8 \!\times\! 10^{16}$ & $1.3 \!\times\! 10^{15}$ & $1.3 \!\times\! 10^{15}$ & $39$ & $100$ & $100$ & $100$ & $100$  \\
      \cline{0-0} \multirow{2}{*}{$N=10^{6.5}$} & $5$ & $2.0$ & $1.6 \!\times\! 10^{19}$ & $1.3 \!\times\! 10^{18}$ & $1.2 \!\times\! 10^{17}$ & $7.8 \!\times\! 10^{15}$ & $1.4 \!\times\! 10^{19}$ & $99$ & $100$ & $100$ & $100$ & $100$  \\
       & $15$ & $2.0$ & $5.2 \!\times\! 10^{19}$ & $9.6 \!\times\! 10^{18}$ & $2.4 \!\times\! 10^{18}$ & $2.8 \!\times\! 10^{17}$ & $9.9 \!\times\! 10^{18}$ & $72$ & $100$ & $100$ & $100$ & $100$  \\
       & $25$ & $2.0$ & $7.7 \!\times\! 10^{19}$ & $2.4 \!\times\! 10^{19}$ & $7.6 \!\times\! 10^{18}$ & $9.7 \!\times\! 10^{17}$ & $5.2 \!\times\! 10^{18}$ & $92$ & $100$ & $100$ & $100$ & $100$  \\
       & $35$ & $2.0$ & $1.2 \!\times\! 10^{20}$ & $3.7 \!\times\! 10^{19}$ & $9.4 \!\times\! 10^{18}$ & $7.1 \!\times\! 10^{17}$ & $1.9 \!\times\! 10^{18}$ & $68$ & $100$ & $100$ & $100$ & $100$  \\
       & $45$ & $2.0$ & $9.4 \!\times\! 10^{20}$ & $3.3 \!\times\! 10^{19}$ & $3.7 \!\times\! 10^{18}$ & $2.1 \!\times\! 10^{17}$ & $5.1 \!\times\! 10^{17}$ & $34$ & $100$ & $100$ & $100$ & $100$  \\
       & $55$ & $2.0$ & $1.0 \!\times\! 10^{21}$ & $9.9 \!\times\! 10^{18}$ & $6.1 \!\times\! 10^{17}$ & $4.9 \!\times\! 10^{16}$ & $9.1 \!\times\! 10^{16}$ & $36$ & $100$ & $100$ & $100$ & $100$  \\
       & $65$ & $2.0$ & $1.1 \!\times\! 10^{21}$ & $1.1 \!\times\! 10^{18}$ & $6.4 \!\times\! 10^{16}$ & $5.6 \!\times\! 10^{15}$ & $1.1 \!\times\! 10^{16}$ & $38$ & $100$ & $100$ & $100$ & $100$  \\
       & $75$ & $2.0$ & $1.1 \!\times\! 10^{21}$ & $9.8 \!\times\! 10^{16}$ & $1.1 \!\times\! 10^{15}$ & $0.0$ & $0.0$ & $39$ & $100$ & $100$ & $0$ & $0$  \\
       & $5$ & $3.0$ & $2.3 \!\times\! 10^{20}$ & $1.0 \!\times\! 10^{20}$ & $1.5 \!\times\! 10^{19}$ & $1.5 \!\times\! 10^{18}$ & $1.0 \!\times\! 10^{20}$ & $69$ & $75$ & $100$ & $100$ & $100$  \\
       & $15$ & $3.0$ & $2.3 \!\times\! 10^{20}$ & $1.0 \!\times\! 10^{20}$ & $2.1 \!\times\! 10^{19}$ & $6.8 \!\times\! 10^{18}$ & $7.8 \!\times\! 10^{19}$ & $58$ & $74$ & $100$ & $100$ & $100$  \\
       & $25$ & $3.0$ & $2.4 \!\times\! 10^{20}$ & $7.2 \!\times\! 10^{19}$ & $3.1 \!\times\! 10^{19}$ & $1.1 \!\times\! 10^{19}$ & $4.3 \!\times\! 10^{19}$ & $47$ & $91$ & $100$ & $100$ & $100$  \\
       & $35$ & $3.0$ & $2.3 \!\times\! 10^{20}$ & $5.3 \!\times\! 10^{19}$ & $2.9 \!\times\! 10^{19}$ & $8.5 \!\times\! 10^{18}$ & $1.7 \!\times\! 10^{19}$ & $48$ & $79$ & $100$ & $100$ & $100$  \\
       & $45$ & $3.0$ & $8.6 \!\times\! 10^{20}$ & $3.5 \!\times\! 10^{19}$ & $1.9 \!\times\! 10^{19}$ & $2.5 \!\times\! 10^{18}$ & $4.2 \!\times\! 10^{18}$ & $32$ & $90$ & $100$ & $100$ & $100$  \\
       & $55$ & $3.0$ & $9.9 \!\times\! 10^{20}$ & $2.3 \!\times\! 10^{19}$ & $4.2 \!\times\! 10^{18}$ & $3.9 \!\times\! 10^{17}$ & $7.3 \!\times\! 10^{17}$ & $33$ & $100$ & $100$ & $100$ & $100$  \\
       & $65$ & $3.0$ & $1.1 \!\times\! 10^{21}$ & $4.3 \!\times\! 10^{18}$ & $5.2 \!\times\! 10^{17}$ & $3.9 \!\times\! 10^{16}$ & $6.1 \!\times\! 10^{16}$ & $38$ & $100$ & $100$ & $100$ & $100$  \\
       & $75$ & $3.0$ & $1.1 \!\times\! 10^{21}$ & $4.2 \!\times\! 10^{17}$ & $2.7 \!\times\! 10^{16}$ & $0.0$ & $1.1 \!\times\! 10^{15}$ & $39$ & $100$ & $100$ & $0$ & $100$  \\
      \cline{0-0} \multirow{2}{*}{$N=10^{6}$} & $5$ & $2.0$ & $1.6 \!\times\! 10^{19}$ & $1.5 \!\times\! 10^{18}$ & $2.0 \!\times\! 10^{17}$ & $6.5 \!\times\! 10^{15}$ & $1.3 \!\times\! 10^{19}$ & $100$ & $100$ & $100$ & $100$ & $100$  \\
       & $15$ & $2.0$ & $4.8 \!\times\! 10^{19}$ & $9.3 \!\times\! 10^{18}$ & $2.3 \!\times\! 10^{18}$ & $2.9 \!\times\! 10^{17}$ & $9.5 \!\times\! 10^{18}$ & $79$ & $100$ & $100$ & $100$ & $100$  \\
       & $25$ & $2.0$ & $7.6 \!\times\! 10^{19}$ & $2.4 \!\times\! 10^{19}$ & $7.1 \!\times\! 10^{18}$ & $9.6 \!\times\! 10^{17}$ & $5.2 \!\times\! 10^{18}$ & $93$ & $100$ & $100$ & $100$ & $100$  \\
       & $35$ & $2.0$ & $1.2 \!\times\! 10^{20}$ & $3.7 \!\times\! 10^{19}$ & $8.9 \!\times\! 10^{18}$ & $7.0 \!\times\! 10^{17}$ & $1.9 \!\times\! 10^{18}$ & $69$ & $100$ & $100$ & $100$ & $100$  \\
       & $45$ & $2.0$ & $9.4 \!\times\! 10^{20}$ & $3.2 \!\times\! 10^{19}$ & $3.4 \!\times\! 10^{18}$ & $2.4 \!\times\! 10^{17}$ & $5.2 \!\times\! 10^{17}$ & $34$ & $100$ & $100$ & $100$ & $100$  \\
       & $55$ & $2.0$ & $1.0 \!\times\! 10^{21}$ & $9.9 \!\times\! 10^{18}$ & $6.1 \!\times\! 10^{17}$ & $5.2 \!\times\! 10^{16}$ & $9.9 \!\times\! 10^{16}$ & $35$ & $100$ & $100$ & $100$ & $100$  \\
       & $65$ & $2.0$ & $1.1 \!\times\! 10^{21}$ & $1.2 \!\times\! 10^{18}$ & $6.2 \!\times\! 10^{16}$ & $3.3 \!\times\! 10^{15}$ & $3.3 \!\times\! 10^{15}$ & $38$ & $100$ & $100$ & $100$ & $100$  \\
       & $75$ & $2.0$ & $1.1 \!\times\! 10^{21}$ & $7.5 \!\times\! 10^{16}$ & $0.0$ & $0.0$ & $0.0$ & $39$ & $100$ & $0$ & $0$ & $0$  \\
       & $5$ & $3.0$ & $2.3 \!\times\! 10^{20}$ & $9.8 \!\times\! 10^{19}$ & $1.5 \!\times\! 10^{19}$ & $1.8 \!\times\! 10^{18}$ & $1.0 \!\times\! 10^{20}$ & $69$ & $81$ & $100$ & $100$ & $100$  \\
       & $15$ & $3.0$ & $2.3 \!\times\! 10^{20}$ & $9.9 \!\times\! 10^{19}$ & $2.2 \!\times\! 10^{19}$ & $6.7 \!\times\! 10^{18}$ & $7.8 \!\times\! 10^{19}$ & $57$ & $76$ & $100$ & $100$ & $100$  \\
       & $25$ & $3.0$ & $2.5 \!\times\! 10^{20}$ & $6.8 \!\times\! 10^{19}$ & $3.2 \!\times\! 10^{19}$ & $1.0 \!\times\! 10^{19}$ & $4.4 \!\times\! 10^{19}$ & $45$ & $97$ & $100$ & $100$ & $100$  \\
       & $35$ & $3.0$ & $2.3 \!\times\! 10^{20}$ & $5.3 \!\times\! 10^{19}$ & $2.9 \!\times\! 10^{19}$ & $8.3 \!\times\! 10^{18}$ & $1.6 \!\times\! 10^{19}$ & $48$ & $80$ & $100$ & $100$ & $100$  \\
       & $45$ & $3.0$ & $8.7 \!\times\! 10^{20}$ & $3.4 \!\times\! 10^{19}$ & $1.8 \!\times\! 10^{19}$ & $2.4 \!\times\! 10^{18}$ & $4.1 \!\times\! 10^{18}$ & $32$ & $94$ & $100$ & $100$ & $100$  \\
       & $55$ & $3.0$ & $1.0 \!\times\! 10^{21}$ & $2.2 \!\times\! 10^{19}$ & $4.1 \!\times\! 10^{18}$ & $4.2 \!\times\! 10^{17}$ & $7.8 \!\times\! 10^{17}$ & $33$ & $100$ & $100$ & $100$ & $100$  \\
       & $65$ & $3.0$ & $1.1 \!\times\! 10^{21}$ & $4.2 \!\times\! 10^{18}$ & $4.8 \!\times\! 10^{17}$ & $3.9 \!\times\! 10^{16}$ & $6.0 \!\times\! 10^{16}$ & $38$ & $100$ & $100$ & $100$ & $100$  \\
       & $75$ & $3.0$ & $1.1 \!\times\! 10^{21}$ & $5.1 \!\times\! 10^{17}$ & $9.9 \!\times\! 10^{15}$ & $0.0$ & $0.0$ & $39$ & $100$ & $100$ & $0$ & $0$  \\
      \cline{0-0} \multirow{2}{*}{$N=10^{5.5}$} & $5$ & $2.0$ & $4.4 \!\times\! 10^{18}$ & $2.5 \!\times\! 10^{17}$ & $0.0$ & $0.0$ & $7.2 \!\times\! 10^{18}$ & $100$ & $100$ & $0$ & $0$ & $100$  \\
       & $15$ & $2.0$ & $4.0 \!\times\! 10^{19}$ & $9.7 \!\times\! 10^{18}$ & $2.2 \!\times\! 10^{18}$ & $2.6 \!\times\! 10^{17}$ & $9.4 \!\times\! 10^{18}$ & $100$ & $100$ & $100$ & $100$ & $100$  \\
       & $25$ & $2.0$ & $7.7 \!\times\! 10^{19}$ & $2.4 \!\times\! 10^{19}$ & $6.9 \!\times\! 10^{18}$ & $9.3 \!\times\! 10^{17}$ & $5.3 \!\times\! 10^{18}$ & $95$ & $100$ & $100$ & $100$ & $100$  \\
       & $35$ & $2.0$ & $1.1 \!\times\! 10^{20}$ & $3.6 \!\times\! 10^{19}$ & $8.9 \!\times\! 10^{18}$ & $7.7 \!\times\! 10^{17}$ & $2.0 \!\times\! 10^{18}$ & $72$ & $100$ & $100$ & $100$ & $100$  \\
       & $45$ & $2.0$ & $9.4 \!\times\! 10^{20}$ & $3.0 \!\times\! 10^{19}$ & $3.5 \!\times\! 10^{18}$ & $2.7 \!\times\! 10^{17}$ & $5.2 \!\times\! 10^{17}$ & $34$ & $100$ & $100$ & $100$ & $100$  \\
       & $55$ & $2.0$ & $1.0 \!\times\! 10^{21}$ & $1.1 \!\times\! 10^{19}$ & $7.4 \!\times\! 10^{17}$ & $4.2 \!\times\! 10^{16}$ & $9.0 \!\times\! 10^{16}$ & $35$ & $100$ & $100$ & $100$ & $100$  \\
       & $65$ & $2.0$ & $1.1 \!\times\! 10^{21}$ & $1.6 \!\times\! 10^{18}$ & $5.3 \!\times\! 10^{16}$ & $0.0$ & $0.0$ & $38$ & $100$ & $100$ & $0$ & $0$  \\
       & $75$ & $2.0$ & $1.1 \!\times\! 10^{21}$ & $2.1 \!\times\! 10^{16}$ & $0.0$ & $0.0$ & $0.0$ & $39$ & $100$ & $0$ & $0$ & $0$  \\
       & $5$ & $3.0$ & $2.2 \!\times\! 10^{20}$ & $9.5 \!\times\! 10^{19}$ & $1.6 \!\times\! 10^{19}$ & $2.1 \!\times\! 10^{18}$ & $1.0 \!\times\! 10^{20}$ & $69$ & $85$ & $100$ & $100$ & $100$  \\
       & $15$ & $3.0$ & $2.3 \!\times\! 10^{20}$ & $9.4 \!\times\! 10^{19}$ & $2.2 \!\times\! 10^{19}$ & $6.8 \!\times\! 10^{18}$ & $7.8 \!\times\! 10^{19}$ & $56$ & $83$ & $100$ & $100$ & $100$  \\
       & $25$ & $3.0$ & $2.3 \!\times\! 10^{20}$ & $6.7 \!\times\! 10^{19}$ & $3.2 \!\times\! 10^{19}$ & $1.1 \!\times\! 10^{19}$ & $4.4 \!\times\! 10^{19}$ & $48$ & $99$ & $100$ & $100$ & $100$  \\
       & $35$ & $3.0$ & $2.2 \!\times\! 10^{20}$ & $5.2 \!\times\! 10^{19}$ & $2.8 \!\times\! 10^{19}$ & $8.4 \!\times\! 10^{18}$ & $1.6 \!\times\! 10^{19}$ & $49$ & $84$ & $100$ & $100$ & $100$  \\
       & $45$ & $3.0$ & $8.7 \!\times\! 10^{20}$ & $3.2 \!\times\! 10^{19}$ & $1.8 \!\times\! 10^{19}$ & $2.6 \!\times\! 10^{18}$ & $4.6 \!\times\! 10^{18}$ & $31$ & $100$ & $100$ & $100$ & $100$  \\
       & $55$ & $3.0$ & $9.9 \!\times\! 10^{20}$ & $2.3 \!\times\! 10^{19}$ & $4.5 \!\times\! 10^{18}$ & $4.9 \!\times\! 10^{17}$ & $8.8 \!\times\! 10^{17}$ & $33$ & $100$ & $100$ & $100$ & $100$  \\
       & $65$ & $3.0$ & $1.1 \!\times\! 10^{21}$ & $5.0 \!\times\! 10^{18}$ & $5.8 \!\times\! 10^{17}$ & $5.3 \!\times\! 10^{16}$ & $7.7 \!\times\! 10^{16}$ & $37$ & $100$ & $100$ & $100$ & $100$  \\
       & $75$ & $3.0$ & $1.1 \!\times\! 10^{21}$ & $2.6 \!\times\! 10^{17}$ & $0.0$ & $0.0$ & $0.0$ & $39$ & $100$ & $0$ & $0$ & $0$  \\
      \cline{0-0} \multirow{2}{*}{$N=10^{5}$} & $5$ & $2.0$ & $3.6 \!\times\! 10^{18}$ & $2.7 \!\times\! 10^{17}$ & $0.0$ & $0.0$ & $4.1 \!\times\! 10^{18}$ & $100$ & $100$ & $0$ & $0$ & $100$  \\
       & $15$ & $2.0$ & $2.9 \!\times\! 10^{19}$ & $5.6 \!\times\! 10^{18}$ & $7.1 \!\times\! 10^{17}$ & $0.0$ & $3.4 \!\times\! 10^{18}$ & $100$ & $100$ & $100$ & $0$ & $100$  \\
       & $25$ & $2.0$ & $7.4 \!\times\! 10^{19}$ & $2.6 \!\times\! 10^{19}$ & $6.7 \!\times\! 10^{18}$ & $8.7 \!\times\! 10^{17}$ & $5.2 \!\times\! 10^{18}$ & $99$ & $100$ & $100$ & $100$ & $100$  \\
       & $35$ & $2.0$ & $1.2 \!\times\! 10^{20}$ & $3.5 \!\times\! 10^{19}$ & $8.4 \!\times\! 10^{18}$ & $7.8 \!\times\! 10^{17}$ & $1.9 \!\times\! 10^{18}$ & $70$ & $100$ & $100$ & $100$ & $100$  \\
       & $45$ & $2.0$ & $9.6 \!\times\! 10^{20}$ & $2.8 \!\times\! 10^{19}$ & $3.5 \!\times\! 10^{18}$ & $2.0 \!\times\! 10^{17}$ & $6.5 \!\times\! 10^{17}$ & $34$ & $100$ & $100$ & $100$ & $100$  \\
       & $55$ & $2.0$ & $1.0 \!\times\! 10^{21}$ & $8.9 \!\times\! 10^{18}$ & $6.1 \!\times\! 10^{17}$ & $3.4 \!\times\! 10^{16}$ & $3.1 \!\times\! 10^{16}$ & $35$ & $100$ & $100$ & $100$ & $100$  \\
       & $65$ & $2.0$ & $1.1 \!\times\! 10^{21}$ & $1.4 \!\times\! 10^{18}$ & $0.0$ & $0.0$ & $0.0$ & $38$ & $100$ & $0$ & $0$ & $0$  \\
       & $75$ & $2.0$ & $1.1 \!\times\! 10^{21}$ & $0.0$ & $0.0$ & $0.0$ & $0.0$ & $39$ & $0$ & $0$ & $0$ & $0$  \\
       & $5$ & $3.0$ & $2.3 \!\times\! 10^{20}$ & $8.9 \!\times\! 10^{19}$ & $1.8 \!\times\! 10^{19}$ & $2.2 \!\times\! 10^{18}$ & $9.9 \!\times\! 10^{19}$ & $66$ & $90$ & $100$ & $100$ & $100$  \\
       & $15$ & $3.0$ & $2.2 \!\times\! 10^{20}$ & $9.2 \!\times\! 10^{19}$ & $2.3 \!\times\! 10^{19}$ & $7.0 \!\times\! 10^{18}$ & $7.7 \!\times\! 10^{19}$ & $58$ & $86$ & $100$ & $100$ & $100$  \\
       & $25$ & $3.0$ & $2.4 \!\times\! 10^{20}$ & $6.6 \!\times\! 10^{19}$ & $3.3 \!\times\! 10^{19}$ & $1.0 \!\times\! 10^{19}$ & $4.3 \!\times\! 10^{19}$ & $47$ & $100$ & $100$ & $100$ & $100$  \\
       & $35$ & $3.0$ & $2.3 \!\times\! 10^{20}$ & $4.8 \!\times\! 10^{19}$ & $2.7 \!\times\! 10^{19}$ & $8.3 \!\times\! 10^{18}$ & $1.5 \!\times\! 10^{19}$ & $48$ & $92$ & $100$ & $100$ & $100$  \\
       & $45$ & $3.0$ & $8.9 \!\times\! 10^{20}$ & $3.1 \!\times\! 10^{19}$ & $1.6 \!\times\! 10^{19}$ & $2.6 \!\times\! 10^{18}$ & $4.2 \!\times\! 10^{18}$ & $31$ & $100$ & $100$ & $100$ & $100$  \\
       & $55$ & $3.0$ & $1.0 \!\times\! 10^{21}$ & $2.1 \!\times\! 10^{19}$ & $3.7 \!\times\! 10^{18}$ & $4.7 \!\times\! 10^{17}$ & $6.5 \!\times\! 10^{17}$ & $34$ & $100$ & $100$ & $100$ & $100$  \\
       & $65$ & $3.0$ & $1.1 \!\times\! 10^{21}$ & $4.0 \!\times\! 10^{18}$ & $5.4 \!\times\! 10^{17}$ & $0.0$ & $0.0$ & $38$ & $100$ & $100$ & $0$ & $0$  \\
       & $75$ & $3.0$ & $1.1 \!\times\! 10^{21}$ & $3.4 \!\times\! 10^{16}$ & $0.0$ & $0.0$ & $0.0$ & $39$ & $100$ & $0$ & $0$ & $0$  \\
      \cline{0-0} \multirow{2}{*}{p-Rhea} & $5$ & $2.0$ & $2.2 \!\times\! 10^{20}$ & $3.8 \!\times\! 10^{19}$ & $4.1 \!\times\! 10^{18}$ & $3.7 \!\times\! 10^{17}$ & $7.5 \!\times\! 10^{18}$ & $81$ & $100$ & $100$ & $100$ & $100$  \\
       & $15$ & $2.0$ & $2.1 \!\times\! 10^{20}$ & $5.3 \!\times\! 10^{19}$ & $1.6 \!\times\! 10^{19}$ & $1.7 \!\times\! 10^{18}$ & $9.6 \!\times\! 10^{18}$ & $85$ & $100$ & $100$ & $100$ & $100$  \\
       & $25$ & $2.0$ & $2.0 \!\times\! 10^{20}$ & $7.5 \!\times\! 10^{19}$ & $1.9 \!\times\! 10^{19}$ & $1.7 \!\times\! 10^{18}$ & $8.3 \!\times\! 10^{18}$ & $97$ & $100$ & $100$ & $100$ & $100$  \\
       & $35$ & $2.0$ & $7.8 \!\times\! 10^{20}$ & $7.5 \!\times\! 10^{19}$ & $1.0 \!\times\! 10^{19}$ & $7.5 \!\times\! 10^{17}$ & $3.4 \!\times\! 10^{18}$ & $60$ & $100$ & $100$ & $100$ & $100$  \\
       & $45$ & $2.0$ & $2.1 \!\times\! 10^{21}$ & $2.4 \!\times\! 10^{19}$ & $2.6 \!\times\! 10^{18}$ & $2.3 \!\times\! 10^{17}$ & $9.1 \!\times\! 10^{17}$ & $52$ & $100$ & $100$ & $100$ & $100$  \\
       & $55$ & $2.0$ & $2.2 \!\times\! 10^{21}$ & $5.0 \!\times\! 10^{18}$ & $4.1 \!\times\! 10^{17}$ & $5.1 \!\times\! 10^{16}$ & $2.0 \!\times\! 10^{17}$ & $55$ & $100$ & $100$ & $100$ & $100$  \\
       & $65$ & $2.0$ & $2.3 \!\times\! 10^{21}$ & $8.1 \!\times\! 10^{17}$ & $5.0 \!\times\! 10^{16}$ & $1.1 \!\times\! 10^{16}$ & $3.3 \!\times\! 10^{16}$ & $57$ & $100$ & $100$ & $100$ & $100$  \\
       & $75$ & $2.0$ & $2.3 \!\times\! 10^{21}$ & $6.9 \!\times\! 10^{16}$ & $5.5 \!\times\! 10^{15}$ & $3.4 \!\times\! 10^{14}$ & $2.0 \!\times\! 10^{15}$ & $57$ & $100$ & $100$ & $100$ & $100$  \\
       & $5$ & $3.0$ & $4.2 \!\times\! 10^{20}$ & $2.7 \!\times\! 10^{20}$ & $1.2 \!\times\! 10^{20}$ & $3.6 \!\times\! 10^{19}$ & $9.7 \!\times\! 10^{19}$ & $63$ & $71$ & $99$ & $100$ & $100$  \\
       & $15$ & $3.0$ & $4.1 \!\times\! 10^{20}$ & $2.3 \!\times\! 10^{20}$ & $9.4 \!\times\! 10^{19}$ & $3.4 \!\times\! 10^{19}$ & $7.6 \!\times\! 10^{19}$ & $64$ & $79$ & $99$ & $100$ & $100$  \\
       & $25$ & $3.0$ & $3.2 \!\times\! 10^{20}$ & $1.7 \!\times\! 10^{20}$ & $7.8 \!\times\! 10^{19}$ & $2.5 \!\times\! 10^{19}$ & $5.6 \!\times\! 10^{19}$ & $70$ & $95$ & $100$ & $100$ & $100$  \\
       & $35$ & $3.0$ & $5.4 \!\times\! 10^{20}$ & $1.3 \!\times\! 10^{20}$ & $5.2 \!\times\! 10^{19}$ & $1.1 \!\times\! 10^{19}$ & $2.8 \!\times\! 10^{19}$ & $79$ & $98$ & $100$ & $100$ & $100$  \\
       & $45$ & $3.0$ & $1.8 \!\times\! 10^{21}$ & $8.4 \!\times\! 10^{19}$ & $1.7 \!\times\! 10^{19}$ & $2.5 \!\times\! 10^{18}$ & $7.5 \!\times\! 10^{18}$ & $51$ & $100$ & $100$ & $100$ & $100$  \\
       & $55$ & $3.0$ & $2.2 \!\times\! 10^{21}$ & $2.1 \!\times\! 10^{19}$ & $3.7 \!\times\! 10^{18}$ & $3.7 \!\times\! 10^{17}$ & $1.1 \!\times\! 10^{18}$ & $54$ & $100$ & $100$ & $100$ & $100$  \\
       & $65$ & $3.0$ & $2.3 \!\times\! 10^{21}$ & $3.7 \!\times\! 10^{18}$ & $3.6 \!\times\! 10^{17}$ & $4.2 \!\times\! 10^{16}$ & $1.2 \!\times\! 10^{17}$ & $57$ & $100$ & $100$ & $100$ & $100$  \\
       & $75$ & $3.0$ & $2.3 \!\times\! 10^{21}$ & $3.6 \!\times\! 10^{17}$ & $2.8 \!\times\! 10^{16}$ & $2.5 \!\times\! 10^{15}$ & $8.1 \!\times\! 10^{15}$ & $57$ & $100$ & $100$ & $100$ & $100$  \\
      \cline{0-0} \multirow{2}{*}{ANEOS} & $25$ & $2.0$ & $1.5 \!\times\! 10^{20}$ & $2.5 \!\times\! 10^{19}$ & $1.9 \!\times\! 10^{18}$ & $8.9 \!\times\! 10^{16}$ & $5.1 \!\times\! 10^{18}$ & $100$ & $99$ & $100$ & $100$ & $100$  \\
      \multirow{2}{*}{\& AQUA} & $35$ & $2.0$ & $9.0 \!\times\! 10^{19}$ & $9.4 \!\times\! 10^{18}$ & $7.8 \!\times\! 10^{17}$ & $5.0 \!\times\! 10^{15}$ & $1.7 \!\times\! 10^{18}$ & $100$ & $99$ & $100$ & $100$ & $100$  \\
       & $45$ & $2.0$ & $2.3 \!\times\! 10^{19}$ & $1.4 \!\times\! 10^{18}$ & $1.3 \!\times\! 10^{17}$ & $2.6 \!\times\! 10^{15}$ & $3.3 \!\times\! 10^{17}$ & $100$ & $99$ & $100$ & $100$ & $100$  \\
       & $25$ & $3.0$ & $2.3 \!\times\! 10^{20}$ & $1.0 \!\times\! 10^{20}$ & $3.3 \!\times\! 10^{19}$ & $5.1 \!\times\! 10^{18}$ & $5.5 \!\times\! 10^{19}$ & $94$ & $100$ & $99$ & $100$ & $100$  \\
       & $35$ & $3.0$ & $6.9 \!\times\! 10^{20}$ & $1.3 \!\times\! 10^{20}$ & $2.7 \!\times\! 10^{19}$ & $3.1 \!\times\! 10^{18}$ & $9.7 \!\times\! 10^{18}$ & $83$ & $100$ & $100$ & $99$ & $99$  \\
       & $45$ & $3.0$ & $1.9 \!\times\! 10^{21}$ & $4.1 \!\times\! 10^{19}$ & $5.4 \!\times\! 10^{18}$ & $4.6 \!\times\! 10^{17}$ & $1.4 \!\times\! 10^{18}$ & $61$ & $100$ & $99$ & $99$ & $99$  \\
      \cline{0-0} $M_{\rm pD} = 1$ & $15$ & $3.0$ & $2.3 \!\times\! 10^{20}$ & $5.5 \!\times\! 10^{19}$ & $2.2 \!\times\! 10^{19}$ & $5.9 \!\times\! 10^{18}$ & $3.8 \!\times\! 10^{19}$ & $80$ & $100$ & $100$ & $100$ & $100$  \\
       & $25$ & $3.0$ & $1.7 \!\times\! 10^{20}$ & $6.9 \!\times\! 10^{19}$ & $2.6 \!\times\! 10^{19}$ & $6.4 \!\times\! 10^{18}$ & $1.8 \!\times\! 10^{19}$ & $79$ & $100$ & $100$ & $100$ & $100$  \\
       & $35$ & $3.0$ & $1.7 \!\times\! 10^{20}$ & $7.2 \!\times\! 10^{19}$ & $1.9 \!\times\! 10^{19}$ & $2.8 \!\times\! 10^{18}$ & $5.2 \!\times\! 10^{18}$ & $77$ & $99$ & $100$ & $100$ & $100$  \\
      $M_{\rm pD} = 2$ & $15$ & $3.0$ & $3.9 \!\times\! 10^{20}$ & $1.0 \!\times\! 10^{20}$ & $4.2 \!\times\! 10^{19}$ & $7.9 \!\times\! 10^{18}$ & $6.8 \!\times\! 10^{19}$ & $83$ & $100$ & $100$ & $100$ & $100$  \\
       & $25$ & $3.0$ & $2.9 \!\times\! 10^{20}$ & $1.3 \!\times\! 10^{20}$ & $5.0 \!\times\! 10^{19}$ & $8.6 \!\times\! 10^{18}$ & $2.9 \!\times\! 10^{19}$ & $86$ & $100$ & $100$ & $100$ & $100$  \\
       & $35$ & $3.0$ & $3.9 \!\times\! 10^{20}$ & $1.3 \!\times\! 10^{20}$ & $3.2 \!\times\! 10^{19}$ & $3.3 \!\times\! 10^{18}$ & $7.7 \!\times\! 10^{18}$ & $81$ & $99$ & $100$ & $100$ & $100$  \\
      $M_{\rm pD} = 4$ & $15$ & $3.0$ & $6.7 \!\times\! 10^{20}$ & $2.0 \!\times\! 10^{20}$ & $8.2 \!\times\! 10^{19}$ & $1.7 \!\times\! 10^{19}$ & $1.3 \!\times\! 10^{20}$ & $89$ & $100$ & $100$ & $100$ & $100$  \\
       & $25$ & $3.0$ & $4.9 \!\times\! 10^{20}$ & $2.4 \!\times\! 10^{20}$ & $9.2 \!\times\! 10^{19}$ & $2.0 \!\times\! 10^{19}$ & $5.8 \!\times\! 10^{19}$ & $91$ & $100$ & $100$ & $100$ & $100$  \\
       & $35$ & $3.0$ & $7.7 \!\times\! 10^{20}$ & $2.4 \!\times\! 10^{20}$ & $5.5 \!\times\! 10^{19}$ & $6.6 \!\times\! 10^{18}$ & $1.5 \!\times\! 10^{19}$ & $86$ & $100$ & $100$ & $100$ & $100$  \\
      $M_{\rm pD} = 8$ & $15$ & $3.0$ & $9.7 \!\times\! 10^{20}$ & $3.8 \!\times\! 10^{20}$ & $1.6 \!\times\! 10^{20}$ & $4.0 \!\times\! 10^{19}$ & $2.6 \!\times\! 10^{20}$ & $97$ & $100$ & $100$ & $100$ & $100$  \\
       & $25$ & $3.0$ & $8.1 \!\times\! 10^{20}$ & $4.6 \!\times\! 10^{20}$ & $1.8 \!\times\! 10^{20}$ & $4.5 \!\times\! 10^{19}$ & $1.2 \!\times\! 10^{20}$ & $96$ & $100$ & $100$ & $100$ & $100$  \\
       & $35$ & $3.0$ & $1.6 \!\times\! 10^{21}$ & $4.1 \!\times\! 10^{20}$ & $1.0 \!\times\! 10^{20}$ & $1.6 \!\times\! 10^{19}$ & $3.1 \!\times\! 10^{19}$ & $88$ & $100$ & $100$ & $100$ & $100$  \\
  \hline
  \end{longtable}
  }
\end{center}
\twocolumngrid


\newpage


\bibliographystyle{aasjournal}
\bibliography{./gihr.bib}{}

\begin{thebibliography}{}
\expandafter\ifx\csname natexlab\endcsname\relax\def\natexlab#1{#1}\fi
\providecommand{\url}[1]{\href{#1}{#1}}
\providecommand{\dodoi}[1]{doi:~\href{http://doi.org/#1}{\nolinkurl{#1}}}
\providecommand{\doeprint}[1]{\href{http://ascl.net/#1}{\nolinkurl{http://ascl.net/#1}}}
\providecommand{\doarXiv}[1]{\href{https://arxiv.org/abs/#1}{\nolinkurl{https://arxiv.org/abs/#1}}}

\bibitem[{{Agnor} \& {Hamilton}(2006)}]{Agnor+Hamilton2006}
{Agnor}, C.~B., \& {Hamilton}, D.~P. 2006, \nat, 441, 192,
  \dodoi{10.1038/nature04792}

\bibitem[{{Asphaug} \& {Reufer}(2013)}]{Asphaug+Reufer2013}
{Asphaug}, E., \& {Reufer}, A. 2013, \icarus, 223, 544,
  \dodoi{10.1016/j.icarus.2012.12.009}

\bibitem[{{Balsara}(1995)}]{Balsara1995}
{Balsara}, D.~S. 1995, J. Comput. Phys., 121, 357,
  \dodoi{10.1016/S0021-9991(95)90221-X}

\bibitem[{{Canup}(2010)}]{Canup2010}
{Canup}, R.~M. 2010, Nature, 468, 943, \dodoi{10.1038/nature09661}

\bibitem[{{Charnoz} {et~al.}(2009){Charnoz}, {Dones}, {Esposito}, {Estrada}, \&
  {Hedman}}]{Charnoz+2009}
{Charnoz}, S., {Dones}, L., {Esposito}, L.~W., {Estrada}, P.~R., \& {Hedman},
  M.~M. 2009, in Saturn from Cassini-Huygens, ed. M.~K. {Dougherty}, L.~W.
  {Esposito}, \& S.~M. {Krimigis}, 537, \dodoi{10.1007/978-1-4020-9217-6_17}

\bibitem[{{Chen} {et~al.}(2014){Chen}, {Nimmo}, \& {Glatzmaier}}]{Chen+2014}
{Chen}, E.~M.~A., {Nimmo}, F., \& {Glatzmaier}, G.~A. 2014, \icarus, 229, 11,
  \dodoi{10.1016/j.icarus.2013.10.024}

\bibitem[{{Crida} \& {Charnoz}(2012)}]{Crida+Charnoz2012}
{Crida}, A., \& {Charnoz}, S. 2012, Science, 338, 1196,
  \dodoi{10.1126/science.1226477}

\bibitem[{{Crida} \& {Charnoz}(2014)}]{Crida+Charnoz2014}
{Crida}, A., \& {Charnoz}, S. 2014, in Complex Planetary Systems, Proceedings
  of the International Astronomical Union, Vol. 310, 182--189,
  \dodoi{10.1017/S1743921314008229}

\bibitem[{{Crida} {et~al.}(2019){Crida}, {Charnoz}, {Hsu}, \&
  {Dones}}]{Crida+2019}
{Crida}, A., {Charnoz}, S., {Hsu}, H.-W., \& {Dones}, L. 2019, Nature
  Astronomy, 3, 967, \dodoi{10.1038/s41550-019-0876-y}

\bibitem[{{{\'C}uk} {et~al.}(2016){{\'C}uk}, {Dones}, \&
  {Nesvorn{\'y}}}]{Cuk+2016b}
{{\'C}uk}, M., {Dones}, L., \& {Nesvorn{\'y}}, D. 2016, \apj, 820, 97,
  \dodoi{10.3847/0004-637X/820/2/97}

\bibitem[{{{\'C}uk} {et~al.}(2020){{\'C}uk}, {El Moutamid}, \&
  {Tiscareno}}]{Cuk+2020}
{{\'C}uk}, M., {El Moutamid}, M., \& {Tiscareno}, M.~S. 2020, The Planetary
  Science Journal, 1, 22, \dodoi{10.3847/PSJ/ab9748}

\bibitem[{Cuzzi \& Estrada(1998)}]{Cuzzi+Estrada1998}
Cuzzi, J.~N., \& Estrada, P.~R. 1998, \icarus, 132, 1

\bibitem[{{Dones} {et~al.}(2007){Dones}, {Agnor}, \& {Asphaug}}]{Dones+2007}
{Dones}, H.~C., {Agnor}, C.~B., \& {Asphaug}, E. 2007, in Bulletin of the
  American Astronomical Society, Vol.~39, AAS/Division for Planetary Sciences
  Meeting Abstracts \#39, 420

\bibitem[{{Dones}(1991)}]{Dones1991}
{Dones}, L. 1991, \icarus, 92, 194, \dodoi{10.1016/0019-1035(91)90045-U}

\bibitem[{{Doyle} {et~al.}(1989){Doyle}, {Dones}, \& {Cuzzi}}]{Doyle+1989}
{Doyle}, L.~R., {Dones}, L., \& {Cuzzi}, J.~N. 1989, \icarus, 80, 104,
  \dodoi{10.1016/0019-1035(89)90163-2}

\bibitem[{{Dubinski}(2019)}]{Dubinski2019}
{Dubinski}, J. 2019, \icarus, 321, 291, \dodoi{10.1016/j.icarus.2018.11.034}

\bibitem[{{Durisen} \& {Estrada}(2023)}]{Durisen+Estrada2023}
{Durisen}, R.~H., \& {Estrada}, P.~R. 2023, \icarus, 400, 115221,
  \dodoi{10.1016/j.icarus.2022.115221}

\bibitem[{{Estrada} \& {Durisen}(2023)}]{Estrada+Durisen2023}
{Estrada}, P.~R., \& {Durisen}, R.~H. 2023, \icarus, 400, 115296,
  \dodoi{10.1016/j.icarus.2022.115296}

\bibitem[{{Estrada} {et~al.}(2015){Estrada}, {Durisen}, {Cuzzi}, \&
  {Morgan}}]{Estrada+2015}
{Estrada}, P.~R., {Durisen}, R.~H., {Cuzzi}, J.~N., \& {Morgan}, D.~A. 2015,
  \icarus, 252, 415, \dodoi{10.1016/j.icarus.2015.02.005}

\bibitem[{{Estrada} {et~al.}(2018){Estrada}, {Durisen}, \&
  {Latter}}]{Estrada+2018}
{Estrada}, P.~R., {Durisen}, R.~H., \& {Latter}, H.~N. 2018, in Planetary Ring
  Systems.~Properties, Structure, and Evolution, ed. M.~S. {Tiscareno} \& C.~D.
  {Murray}, 198--224, \dodoi{10.1017/9781316286791.009}

\bibitem[{{Ferguson} {et~al.}(2022){Ferguson}, {Rhoden}, \&
  {Kirchoff}}]{Ferguson+2022}
{Ferguson}, S.~N., {Rhoden}, A.~R., \& {Kirchoff}, M.~R. 2022, Journal of
  Geophysical Research (Planets), 127, e07204, \dodoi{10.1029/2022JE007204}

\bibitem[{{Fuller} {et~al.}(2016){Fuller}, {Luan}, \& {Quataert}}]{Fuller+2016}
{Fuller}, J., {Luan}, J., \& {Quataert}, E. 2016, \mnras, 458, 3867,
  \dodoi{10.1093/mnras/stw609}

\bibitem[{{Genda} {et~al.}(2015){Genda}, {Fujita}, {Kobayashi}, {Tanaka}, \&
  {Abe}}]{Genda+2015}
{Genda}, H., {Fujita}, T., {Kobayashi}, H., {Tanaka}, H., \& {Abe}, Y. 2015,
  \icarus, 262, 58, \dodoi{10.1016/j.icarus.2015.08.029}

\bibitem[{{Gingold} \& {Monaghan}(1977)}]{Gingold+Monaghan1977}
{Gingold}, R.~A., \& {Monaghan}, J.~J. 1977, \mnras, 181, 375,
  \dodoi{10.1093/mnras/181.3.375}

\bibitem[{{Goldreich} {et~al.}(1989){Goldreich}, {Murray}, {Longaretti}, \&
  {Banfield}}]{Goldreich+1989}
{Goldreich}, P., {Murray}, N., {Longaretti}, P.~Y., \& {Banfield}, D. 1989,
  Science, 245, 500, \dodoi{10.1126/science.245.4917.500}

\bibitem[{{Goldreich} \& {Soter}(1966)}]{Goldreich+Soter1966}
{Goldreich}, P., \& {Soter}, S. 1966, \icarus, 5, 375,
  \dodoi{10.1016/0019-1035(66)90051-0}

\bibitem[{{Haldemann} {et~al.}(2020){Haldemann}, {Alibert}, {Mordasini}, \&
  {Benz}}]{Haldemann+2020}
{Haldemann}, J., {Alibert}, Y., {Mordasini}, C., \& {Benz}, W. 2020, \aap, 643,
  A105, \dodoi{10.1051/0004-6361/202038367}

\bibitem[{{Harris}(1984)}]{Harris1984}
{Harris}, A.~W. 1984, in IAU Colloq. 75: Planetary Rings, ed. R.~{Greenberg} \&
  A.~{Brahic}, 641--659

\bibitem[{{Hosono} {et~al.}(2017){Hosono}, {Iwasawa}, {Tanikawa}, {Nitadori},
  {Muranushi}, \& {Makino}}]{Hosono+2017}
{Hosono}, N., {Iwasawa}, M., {Tanikawa}, A., {et~al.} 2017, Publ. Astron. Soc.
  Jpn., 69, 26, \dodoi{10.1093/pasj/psw131}

\bibitem[{{Hsu} {et~al.}(2018){Hsu}, {Schmidt}, {Kempf}, {Postberg},
  {Moragas-Klostermeyer}, {Sei{\ss}}, {Hoffmann}, {Burton}, {Ye}, {Kurth},
  {Hor{\'a}nyi}, {Khawaja}, {Spahn}, {Schirdewahn}, {O'Donoghue}, {Moore},
  {Cuzzi}, {Jones}, \& {Srama}}]{Hsu+2018}
{Hsu}, H.-W., {Schmidt}, J., {Kempf}, S., {et~al.} 2018, Science, 362, aat3185,
  \dodoi{10.1126/science.aat3185}

\bibitem[{{Hyodo} \& {Charnoz}(2017)}]{Hyodo+Charnoz2017}
{Hyodo}, R., \& {Charnoz}, S. 2017, \aj, 154, 34,
  \dodoi{10.3847/1538-3881/aa74c9}

\bibitem[{{Hyodo} {et~al.}(2017){Hyodo}, {Charnoz}, {Ohtsuki}, \&
  {Genda}}]{Hyodo+2017c}
{Hyodo}, R., {Charnoz}, S., {Ohtsuki}, K., \& {Genda}, H. 2017, \icarus, 282,
  195, \dodoi{10.1016/j.icarus.2016.09.012}

\bibitem[{{Iess} {et~al.}(2019){Iess}, {Militzer}, {Kaspi}, {Nicholson},
  {Durante}, {Racioppa}, {Anabtawi}, {Galanti}, {Hubbard}, {Mariani},
  {Tortora}, {Wahl}, \& {Zannoni}}]{Iess+2019}
{Iess}, L., {Militzer}, B., {Kaspi}, Y., {et~al.} 2019, Science, 364, aat2965,
  \dodoi{10.1126/science.aat2965}

\bibitem[{{Ip}(1988)}]{Ip1988}
{Ip}, W.-H. 1988, \aap, 199, 340

\bibitem[{{Jacobson}(2022)}]{Jacobson2022}
{Jacobson}, R.~A. 2022, \aj, 164, 199, \dodoi{10.3847/1538-3881/ac90c9}

\bibitem[{{Kegerreis} {et~al.}(2019){Kegerreis}, {Eke}, {Gonnet}, {Korycansky},
  {Massey}, {Schaller}, \& {Teodoro}}]{Kegerreis+2019}
{Kegerreis}, J.~A., {Eke}, V.~R., {Gonnet}, P., {et~al.} 2019, \mnras, 487,
  1536, \dodoi{10.1093/mnras/stz1606}

\bibitem[{{Kegerreis} {et~al.}(2020){Kegerreis}, {Eke}, {Massey}, \&
  {Teodoro}}]{Kegerreis+2020}
{Kegerreis}, J.~A., {Eke}, V.~R., {Massey}, R.~J., \& {Teodoro}, L.~F.~A. 2020,
  \apj, 897, 161, \dodoi{10.3847/1538-4357/ab9810}

\bibitem[{{Kegerreis} {et~al.}(2022){Kegerreis}, {Ruiz-Bonilla}, {Eke},
  {Massey}, {Sandnes}, \& {Teodoro}}]{Kegerreis+2022}
{Kegerreis}, J.~A., {Ruiz-Bonilla}, S., {Eke}, V.~R., {et~al.} 2022, \apjl,
  937, L40, \dodoi{10.3847/2041-8213/abb5fb}

\bibitem[{Kempf {et~al.}(2023)Kempf, Altobelli, Cuzzi, Estrada, \&
  Srama}]{Kempf+2023}
Kempf, S., Altobelli, N.~Schmidt, J., Cuzzi, J.~N., Estrada, P.~R., \& Srama,
  R. 2023, Science Advances, 9, eadf8537, \dodoi{10.1126/sciadv.adf8537}

\bibitem[{{Lainey} {et~al.}(2020){Lainey}, {Casajus}, {Fuller}, {Zannoni},
  {Tortora}, {Cooper}, {Murray}, {Modenini}, {Park}, {Robert}, \&
  {Zhang}}]{Lainey+2020}
{Lainey}, V., {Casajus}, L.~G., {Fuller}, J., {et~al.} 2020, Nature Astronomy,
  4, 1053, \dodoi{10.1038/s41550-020-1120-5}

\bibitem[{{Laplace} {et~al.}(1829){Laplace}, {Bowditch}, \&
  {Bowditch}}]{Laplace+1829}
{Laplace}, P.~S., {Bowditch}, N., \& {Bowditch}, N.~I. 1829, {M{\'e}canique
  c{\'e}leste}

\bibitem[{{Leinhardt} \& {Stewart}(2012)}]{Leinhardt+Stewart2012}
{Leinhardt}, Z.~M., \& {Stewart}, S.~T. 2012, \apj, 745, 79,
  \dodoi{10.1088/0004-637X/745/1/79}

\bibitem[{{Lindal} {et~al.}(1985){Lindal}, {Sweetnam}, \&
  {Eshleman}}]{Lindal+1985}
{Lindal}, G.~F., {Sweetnam}, D.~N., \& {Eshleman}, V.~R. 1985, \aj, 90, 1136,
  \dodoi{10.1086/113820}

\bibitem[{{Lissauer}(1995)}]{Lissauer1995}
{Lissauer}, J.~J. 1995, \icarus, 114, 217, \dodoi{10.1006/icar.1995.1057}

\bibitem[{{Lissauer} {et~al.}(1988){Lissauer}, {Squyres}, \&
  {Hartmann}}]{Lissauer+1988}
{Lissauer}, J.~J., {Squyres}, S.~W., \& {Hartmann}, W.~K. 1988, \jgr, 93,
  13776, \dodoi{10.1029/JB093iB11p13776}

\bibitem[{{Lucy}(1977)}]{Lucy1977}
{Lucy}, L.~B. 1977, \aj, 82, 1013, \dodoi{10.1086/112164}

\bibitem[{{Melosh}(2007)}]{Melosh2007}
{Melosh}, H.~J. 2007, Meteorit. Planet. Sci, 42, 2079,
  \dodoi{10.1111/j.1945-5100.2007.tb01009.x}

\bibitem[{{Meyer} \& {Wisdom}(2007)}]{Meyer+Wisdom2007}
{Meyer}, J., \& {Wisdom}, J. 2007, \icarus, 188, 535,
  \dodoi{10.1016/j.icarus.2007.03.001}

\bibitem[{{Mosqueira} \&
  {Estrada}(2003{\natexlab{a}})}]{Mosqueira+Estrada2003b}
{Mosqueira}, I., \& {Estrada}, P.~R. 2003{\natexlab{a}}, \icarus, 163, 232,
  \dodoi{10.1016/S0019-1035(03)00077-0}

\bibitem[{{Mosqueira} \& {Estrada}(2003{\natexlab{b}})}]{Mosqueira+Estrada2003}
---. 2003{\natexlab{b}}, \icarus, 163, 198,
  \dodoi{10.1016/S0019-1035(03)00076-9}

\bibitem[{{Murray} \& {Dermott}(1999)}]{Murray+Dermott1999}
{Murray}, C.~D., \& {Dermott}, S.~F. 1999, {Solar system dynamics}

\bibitem[{{Nixon} {et~al.}(2018){Nixon}, {Lorenz}, {Achterberg}, {Buch},
  {Coll}, {Clark}, {Courtin}, {Hayes}, {Iess}, {Johnson}, {Lopes},
  {Mastrogiuseppe}, {Mandt}, {Mitchell}, {Raulin}, {Rymer}, {Todd Smith},
  {Solomonidou}, {Sotin}, {Strobel}, {Turtle}, {Vuitton}, {West}, \&
  {Yelle}}]{Nixon+2018}
{Nixon}, C.~A., {Lorenz}, R.~D., {Achterberg}, R.~K., {et~al.} 2018, \planss,
  155, 50, \dodoi{10.1016/j.pss.2018.02.009}

\bibitem[{{O'Donoghue} {et~al.}(2019){O'Donoghue}, {Moore}, {Connerney},
  {Melin}, {Stallard}, {Miller}, \& {Baines}}]{ODonoghue+2019}
{O'Donoghue}, J., {Moore}, L., {Connerney}, J., {et~al.} 2019, \icarus, 322,
  251, \dodoi{10.1016/j.icarus.2018.10.027}

\bibitem[{{Pollack} {et~al.}(1976){Pollack}, {Grossman}, {Moore}, \&
  {Graboske}}]{Pollack+1976}
{Pollack}, J.~B., {Grossman}, A.~S., {Moore}, R., \& {Graboske}, Jr., H.~C.
  1976, \icarus, 29, 35, \dodoi{10.1016/0019-1035(76)90100-7}

\bibitem[{{Ruiz-Bonilla} {et~al.}(2022){Ruiz-Bonilla}, {Borrow}, {Eke},
  {Kegerreis}, {Massey}, {Sandnes}, \& {Teodoro}}]{RuizBonilla+2022}
{Ruiz-Bonilla}, S., {Borrow}, J., {Eke}, V.~R., {et~al.} 2022, \mnras, 512,
  4660, \dodoi{10.1093/mnras/stac857}

\bibitem[{{Ruiz-Bonilla} {et~al.}(2021){Ruiz-Bonilla}, {Eke}, {Kegerreis},
  {Massey}, \& {Teodoro}}]{RuizBonilla+2021}
{Ruiz-Bonilla}, S., {Eke}, V.~R., {Kegerreis}, J.~A., {Massey}, R.~J., \&
  {Teodoro}, L.~F.~A. 2021, \mnras, 500, 2861, \dodoi{10.1093/mnras/staa3385}

\bibitem[{{Sandnes} {et~al.}(2023){Sandnes}, {Eke}, {Kegerreis}, {Massey},
  {Ruiz-Bonilla}, \& {Teodoro}}]{Sandnes+2023}
{Sandnes}, T.~D., {Eke}, V.~R., {Kegerreis}, J.~A., {et~al.} 2023, in prep.

\bibitem[{{Schaller} {et~al.}(2018){Schaller}, {Gonnet}, {Chalk}, \&
  {Draper}}]{Schaller+2018}
{Schaller}, M., {Gonnet}, P., {Chalk}, A. B.~G., \& {Draper}, P.~W. 2018,
  {SWIFT: SPH With Inter-dependent Fine-grained Tasking}, Astrophysics Source
  Code Library.
\newblock \doeprint{1805.020}

\bibitem[{{Schaller} {et~al.}(2023){Schaller}, {Borrow}, {Draper}, {Ivkovic},
  {McAlpine}, {Vandenbroucke}, {Bah{\'e}}, {Chaikin}, {Chalk}, {Keung Chan},
  {Correa}, {van Daalen}, {Elbers}, {Gonnet}, {Hausammann}, {Helly},
  {Hu{\v{s}}ko}, {Kegerreis}, {Nobels}, {Ploeckinger}, {Revaz}, {Roper},
  {Ruiz-Bonilla}, {Sandnes}, {Uyttenhove}, {Willis}, \&
  {Xiang}}]{Schaller+2023}
{Schaller}, M., {Borrow}, J., {Draper}, P.~W., {et~al.} 2023, arXiv e-prints,
  arXiv:2305.13380, \dodoi{10.48550/arXiv.2305.13380}

\bibitem[{{Stewart} {et~al.}(2020){Stewart}, {Davies}, {Duncan}, {Lock},
  {Root}, {Townsend}, {Kraus}, {Caracas}, \& {Jacobsen}}]{Stewart+2020}
{Stewart}, S., {Davies}, E., {Duncan}, M., {et~al.} 2020, in American Institute
  of Physics Conference Series, Vol. 2272, American Institute of Physics
  Conference Series, 080003, \dodoi{10.1063/12.0000946}

\bibitem[{Tillotson(1962)}]{Tillotson1962}
Tillotson, J.~H. 1962, General Atomic Report, GA-3216, 141

\bibitem[{{Tiscareno} {et~al.}(2013){Tiscareno}, {Hedman}, {Burns}, \&
  {Castillo-Rogez}}]{Tiscareno+2013}
{Tiscareno}, M.~S., {Hedman}, M.~M., {Burns}, J.~A., \& {Castillo-Rogez}, J.
  2013, \apjl, 765, L28, \dodoi{10.1088/2041-8205/765/2/L28}

\bibitem[{{Waite} {et~al.}(2018){Waite}, {Perryman}, {Perry}, {Miller}, {Bell},
  {Cravens}, {Glein}, {Grimes}, {Hedman}, {Cuzzi}, {Brockwell}, {Teolis},
  {Moore}, {Mitchell}, {Persoon}, {Kurth}, {Wahlund}, {Morooka}, {Hadid},
  {Chocron}, {Walker}, {Nagy}, {Yelle}, {Ledvina}, {Johnson}, {Tseng},
  {Tucker}, \& {Ip}}]{Waite+2018}
{Waite}, J.~H., {Perryman}, R.~S., {Perry}, M.~E., {et~al.} 2018, Science, 362,
  aat2382, \dodoi{10.1126/science.aat2382}

\bibitem[{{White} {et~al.}(2017){White}, {Schenk}, {Bellagamba}, {Grimm},
  {Dombard}, \& {Bray}}]{White+2017}
{White}, O.~L., {Schenk}, P.~M., {Bellagamba}, A.~W., {et~al.} 2017, \icarus,
  288, 37, \dodoi{10.1016/j.icarus.2017.01.025}

\bibitem[{{Wisdom}(1980)}]{Wisdom1980}
{Wisdom}, J. 1980, \aj, 85, 1122, \dodoi{10.1086/112778}

\bibitem[{{Wisdom} {et~al.}(2022){Wisdom}, {Dbouk}, {Militzer}, {Hubbard},
  {Nimmo}, {Downey}, \& {French}}]{Wisdom+2022}
{Wisdom}, J., {Dbouk}, R., {Militzer}, B., {et~al.} 2022, Science, 377, 1285,
  \dodoi{10.1126/science.abn1234}

\bibitem[{{Zahnle} {et~al.}(2003){Zahnle}, {Schenk}, {Levison}, \&
  {Dones}}]{Zahnle+2003}
{Zahnle}, K., {Schenk}, P., {Levison}, H., \& {Dones}, L. 2003, \icarus, 163,
  263, \dodoi{10.1016/S0019-1035(03)00048-4}

\bibitem[{{Zhang} {et~al.}(2017{\natexlab{a}}){Zhang}, {Hayes}, {Janssen},
  {Nicholson}, {Cuzzi}, {de Pater}, \& {Dunn}}]{Zhang+2017b}
{Zhang}, Z., {Hayes}, A.~G., {Janssen}, M.~A., {et~al.} 2017{\natexlab{a}},
  \icarus, 294, 14, \dodoi{10.1016/j.icarus.2017.04.008}

\bibitem[{{Zhang} {et~al.}(2017{\natexlab{b}}){Zhang}, {Hayes}, {Janssen},
  {Nicholson}, {Cuzzi}, {de Pater}, {Dunn}, {Estrada}, \&
  {Hedman}}]{Zhang+2017a}
---. 2017{\natexlab{b}}, \icarus, 281, 297,
  \dodoi{10.1016/j.icarus.2016.07.020}

\end{thebibliography}




\end{document}